\documentclass[a4paper]{article}
\usepackage[english]{babel}
\usepackage[utf8x]{inputenc}
\usepackage[T1]{fontenc}
\usepackage{mathtools}
\usepackage{amsfonts}
\usepackage{kpfonts}
\usepackage{multirow}
\usepackage{adjustbox}

\def\multiset#1#2{\ensuremath{\left(\kern-.3em\left(\genfrac{}{}{0pt}{}{#1}{#2}\right)\kern-.3em\right)}}
\usepackage[ruled,vlined]{algorithm2e}
\usepackage[a4paper,top=3cm,bottom=2cm,left=3cm,right=3cm,marginparwidth=1.75cm]{geometry}

\usepackage{amsmath}
\usepackage{graphicx}
\usepackage[colorinlistoftodos]{todonotes}
\usepackage[colorlinks=true, allcolors=blue]{hyperref}

\title{Nonparametric inference of higher order interaction patterns in networks}
\author{Anatol E. Wegner$^{1*}$ \and Sofia C. Olhede$^2$}
\date{%
    $^1$\small{Chair of Machine Learning for Complex Networks,\\ Center for Artificial Intelligence and Data Science,\\
University of Würzburg, \\97070 Würzburg, Germany\\[2ex]%
    $^2$Institute of Mathematics,\\ École Polytechnique Fédérale de Lausanne,\\
 1015 Lausanne, Switzerland\\[2ex]%
 $^*$ anatol.wegner@uni-wuerzburg.de\\[2ex]
    \today
}}
\begin{document}
\maketitle

\begin{abstract}  

We propose a method for obtaining parsimonious decompositions of networks into higher order interactions which can take the form of arbitrary motifs. 
The method is based on a class of analytically solvable generative models, where vertices are connected via explicit copies of motifs, which in combination with non-parametric priors allow us to infer higher order interactions from dyadic graph data without any prior knowledge on the types or frequencies of such interactions. Crucially, we also consider 'degree--corrected' models that correctly reflect the degree distribution of the network and consequently prove to be a better fit for many real world--networks compared to non-degree corrected models. We test the presented approach on simulated data for which we recover the set of underlying higher order interactions to a high degree of accuracy. For empirical networks the method identifies concise sets of atomic subgraphs from within thousands of candidates that cover a large fraction of edges and include higher order interactions of known structural and functional significance. The method not only produces an explicit higher order representation of the network but also a fit of the network to analytically tractable models opening new avenues for the systematic study of higher order network structures.

\end{abstract}
The reduction of large complex systems to elementary units and their interactions is one of the primary modes of operation in science. Much of our understanding of the natural world stems from our knowledge on such basic building blocks; ranging from elementary particles, to atoms and molecules, to living cells and organisms. Typically elementary units at smaller scales combine in specific patterns to form larger units of increasing complexity and variety.

There are also many indications that higher order structures play an important role in the structural and functional organization of complex networks. For instance, many real--world networks contain certain small connectivity patterns, known as network motifs, in much larger numbers than expected in random graphs with conditionally independent edges \cite{Milo2002NetworkNetworks}. In biological and technological networks these small recurring circuit elements appear to contribute to their function by performing modular tasks which is further supported by the observation that network motifs are broadly shared among networks representing systems with similar functions \cite{Milo2004SuperfamiliesNetworks}.

One of the major difficulties in quantifying higher order network structures such as network motifs is the sheer number of potential interaction patterns that can exist in groups of even moderate size. For instance in undirected networks there are 11117 different ways of connecting 8 vertices and in directed networks there are 9364 such motifs on just 5 vertices. As a result counting subgraphs quickly becomes an ineffective way of describing local network structures as the size of the subgraphs is increased. Moreover, subgraphs are coupled through a complex web of dependencies which further complicates the problem of formulating concise yet informative descriptors of the local structure of networks. As the size and hence potential variety of the patterns included in descriptions of local network structures is increased one is further faced with computational challenges  due to the graph isomorphism problem and the complexity of subgraph enumeration.

In this article we introduce a nonparametric method that is based on the assumption that networks are made of higher order interactions that can take the form of any simply connected motif. Consequently, we consider generative models where networks are constructed  using not only edges but also explicit copies of higher order subgraphs \cite{Bollobas2011SparseClustering,Karrer2010RandomSubgraphs,Wegner2021AtomicNetworks}. More specifically we focus on maximum entropy models that result from constraining the types and distributions of atomic subgraphs used to construct the network \cite{Wegner2021AtomicNetworks}. In contrast to exponential random graphs \cite{2012ExponentialNetworks} that result from maximizing the entropy under constraints on expected subgraph counts, these model can be studied analytically in terms of their properties \cite{Bollobas2011SparseClustering,Karrer2010RandomSubgraphs,Wegner2021AtomicNetworks}. Our approach differs from other approaches to motifs analysis \cite{Milo2002NetworkNetworks} in that it seeks to find explicit decompositions of networks into recurring subgraphs rather than comparing counts of subgraphs in the network against expected counts under a null model. In other words, instead of defining network motifs in terms of deviations from a null model of the network, we treat motifs as statistically significant higher order interaction patterns that are to be inferred from the data. For this we follow a nonparametric Bayesian approach which naturally balances goodness of fit and parametric complexity and, does not require ad-hoc assumptions regarding the types or frequencies of higher order motifs present in the data. This approach, being based on the principle of parsimony, naturally balances goodness of fit and model complexity making it possible to infer concise sets of higher order interactions from within large sets of candidate motifs. 

Our approach is also closely related to other representations of higher order interactions namely hypergraphs and simplicial complexes. Recent extensive research efforts in the area of hypergraphs and simplicial complexes \cite{Battiston2020NetworksDynamics} have yielded methods and models for analysing the structural and dynamical properties of networked systems that comprise higher order interactions. For instance, recent studies have shown that the presence of higher order interactions can lead to markedly different behaviour in dynamical systems \cite{Skardal2020HigherSwitching,Millan2020ExplosiveComplexes,Iacopini2019SimplicialContagion}. Despite this recent surge in interest in higher order networks most empirical data sets only contain pairwise interactions. Consequently, methods for extracting higher order interactions from graph data \cite{Young2021HypergraphData} are of much current interest.

In hypergraphs and simplicial complexes higher order interactions are clique-like i.e. nodes participating in interactions are assumed to be interacting uniformly with all other nodes participating in such interactions. Although restricting higher order interactions to cliques greatly simplifies models of the higher order interactions, from an inference perspective, allowing only cliques is problematic since in principle higher order interactions could take the form of any motif. This is especially relevant, for instance, for directed networks as these usually do not contain large numbers of fully connected subgraphs. In order to circumvent these restrictions we consider higher order network representations, called subgraph configurations, that can be understood as generalized hypergraphs where higher order interactions can take the form of any simply connected subgraph/motif. Nevertheless the method will in many networks identify cliques as higher order interactions as it implicitly selects motifs based on their density, symmetry and frequency making cliques primary candidates for higher order interactions if present in the network under consideration. 

From a methodological standpoint the presented method is similar to inference based methods for community detection that use the Stochastic Block Models (SBMs) as generative models. The SBM \cite{Holland1983StochasticSteps} and its degree corrected variant \cite{Karrer2011StochasticNetworks} have been at the center of much of the progress in the area of community detection and approaches based on the statistical inference of SBMs \cite{Karrer2011StochasticNetworks, peixoto2017nonparametric,peixoto2013parsimonious,Newman2015GeneralizedNetworks,Newman2016EstimatingNetwork}. Statistical inference methods have also been applied to time--dependent and multilayer networks \cite{Peixoto2015InferringNetworks} and have found use in methods for the reconstruction of networks from noisy data \cite{Newman2018NetworkData,Peixoto2018ReconstructingErrors} and dynamical processes on graphs \cite{Peixoto2019NetworkDynamics}. 
Although inference--based methods have mostly concentrated on the SBM and its variants more recently methods that consider alternative generative models have started to emerge. For instance \cite{Young2021HypergraphData} introduced a method for inferring hypergraph representations from graph data that relies on random hypergraphs as generative models. 
The presented approach also builds upon previous work \cite{wegner2014subgraph}. In this article we formulate the problem in a more formal Bayesian framework and use more realistic degree corrected models that not only constrain the types and frequencies atomic subgraphs but also their distributions across the nodes of the network. As shall be demonstrated later, degree corrected models are in general a better fit for real world networks and hence result in a significantly improved method.  

Our results demonstrate that many empirical networks can be represented more parsimoniously by including higher order interaction in their representations. Moreover, being based on statistical inference the method also allow us to fit networks to generative models that more accurately reflect their higher order structure. These models are higher order generalizations of the configuration models and are amenable to analytic and numeric studies of the topological and dynamical implications of higher order structures.

\section*{Results}
\subsection*{Generative models for networks with higher--order motifs}\label{models}
\begin{figure}
    \centering
    \includegraphics[width=0.7\textwidth]{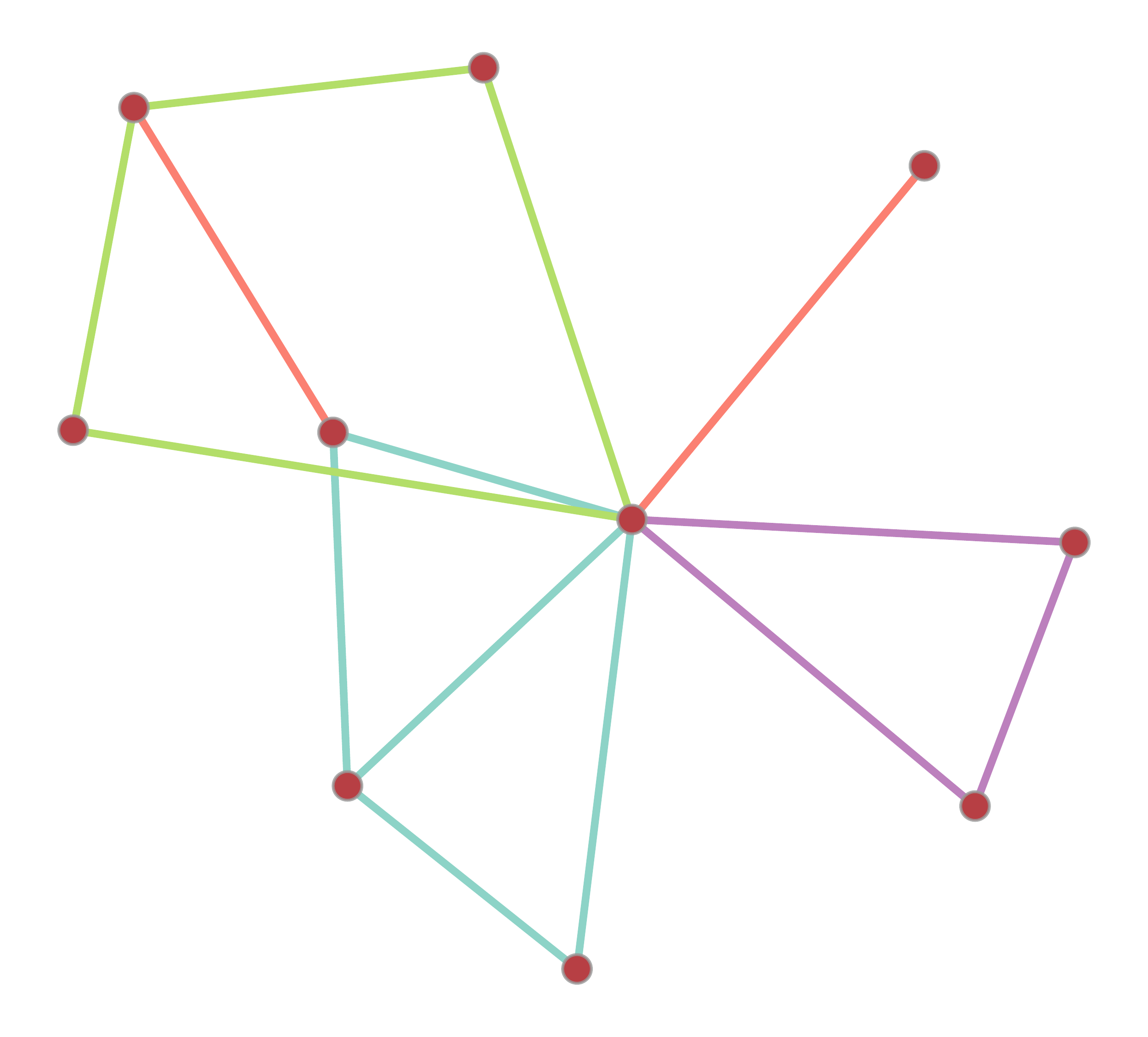}
    \caption{A subgraph configuration consisting two single edges, a triangle, a 4-cycle and a diamond.}
    \label{fig:Conf}
\end{figure}

We base our approach on generative models where vertices are connected not only by single edges but also copies of higher order subgraphs, that we call ``atoms''. Assuming such a generative model leads us to consider latent spaces that correspond the atomic subgraphs added to the graph during the generation process. We call these latent states {\em subgraph configurations}. Subgraph configurations are decompositions of a network into smaller subgraphs and can be thought of as generalized hypergraphs where hyper edges have the form of any connected motif.  More formally a subgraph configuration $C$ on a set of vertices $V$ is a set of subgraphs of the complete graph $K_V$ on $V$. Given such a $C$ we define the atoms of $C$, $M(C)$,  as the set of motifs occurring in $C$. For a graphical illustration of a subgraph configuration see Fig. \ref{fig:Conf}. Note that any subgraph configuration $C$ can be transformed into a graph $G$ by simply taking the union of the edges contained in the subgraphs in $C$ i.e. $E(G)=\bigcup_{S\in C}E(S)$. We says a subgraph configuration $C$ covers $G$ if $E(G)=\bigcup_{S\in C}E(S)$. 

In the formulation of our method we use maximum entropy ensembles of subgraph configurations that results from imposing hard constraints on the counts and distributions of atomic subgraphs used to construct the network \cite{Wegner2021AtomicNetworks}. 
In these models, which are also known as microcanonical ensembles,  all subgraph configurations that satisfy the given constraints are equi--probable. Consequently, the likelihood of configurations can be computed by simply counting the number of configurations that are compatible with the given constraints. Another advantage of using microcanonical models is that every configuration is compatible with a single set of parameters which allows marginal probabilities to be computed in closed form without computing costly integrals. 

\subsubsection*{Motifs, automorphisms and orbits}
\begin{figure}
    \centering
    \includegraphics[width=0.15\textwidth]{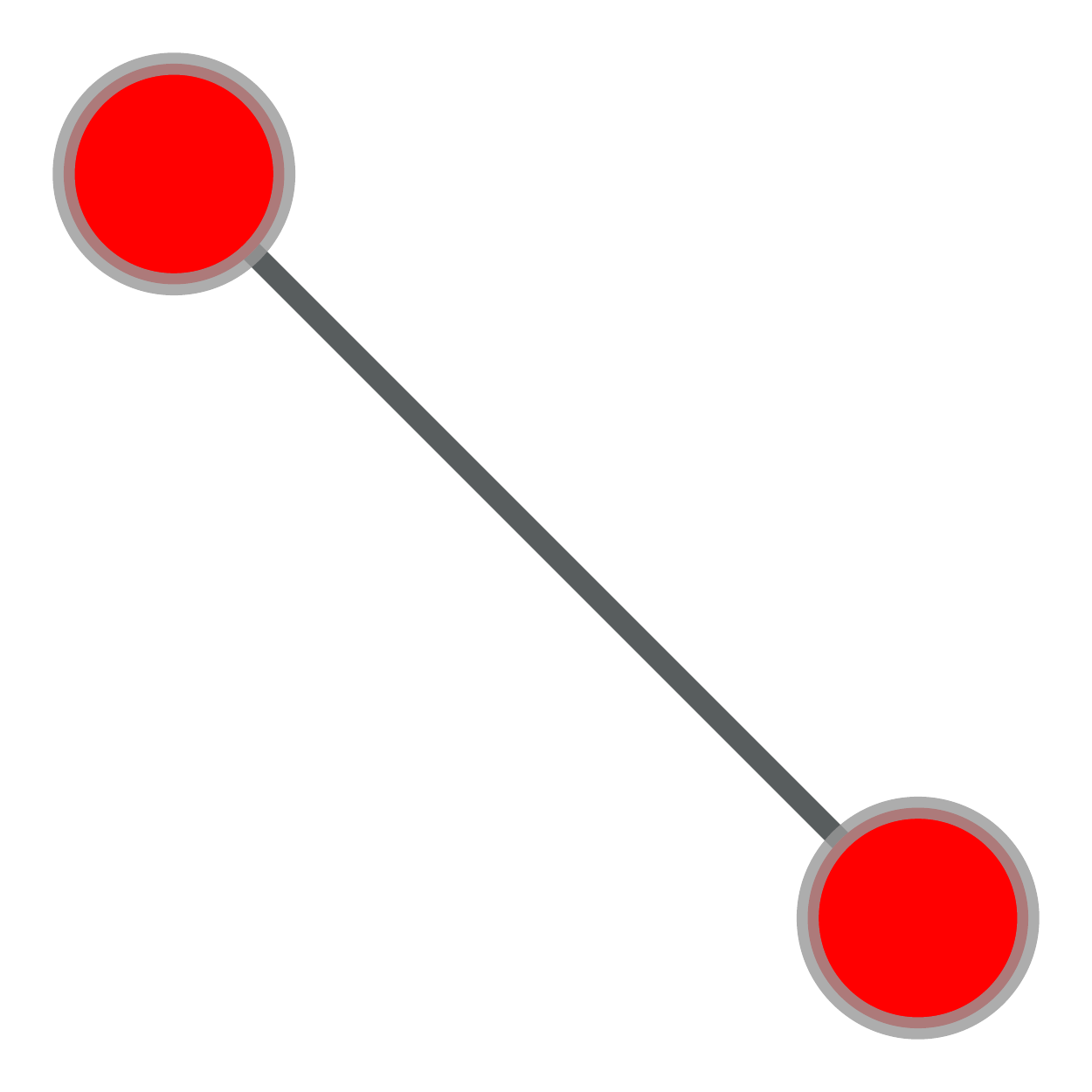}
    \includegraphics[width=0.15\textwidth]{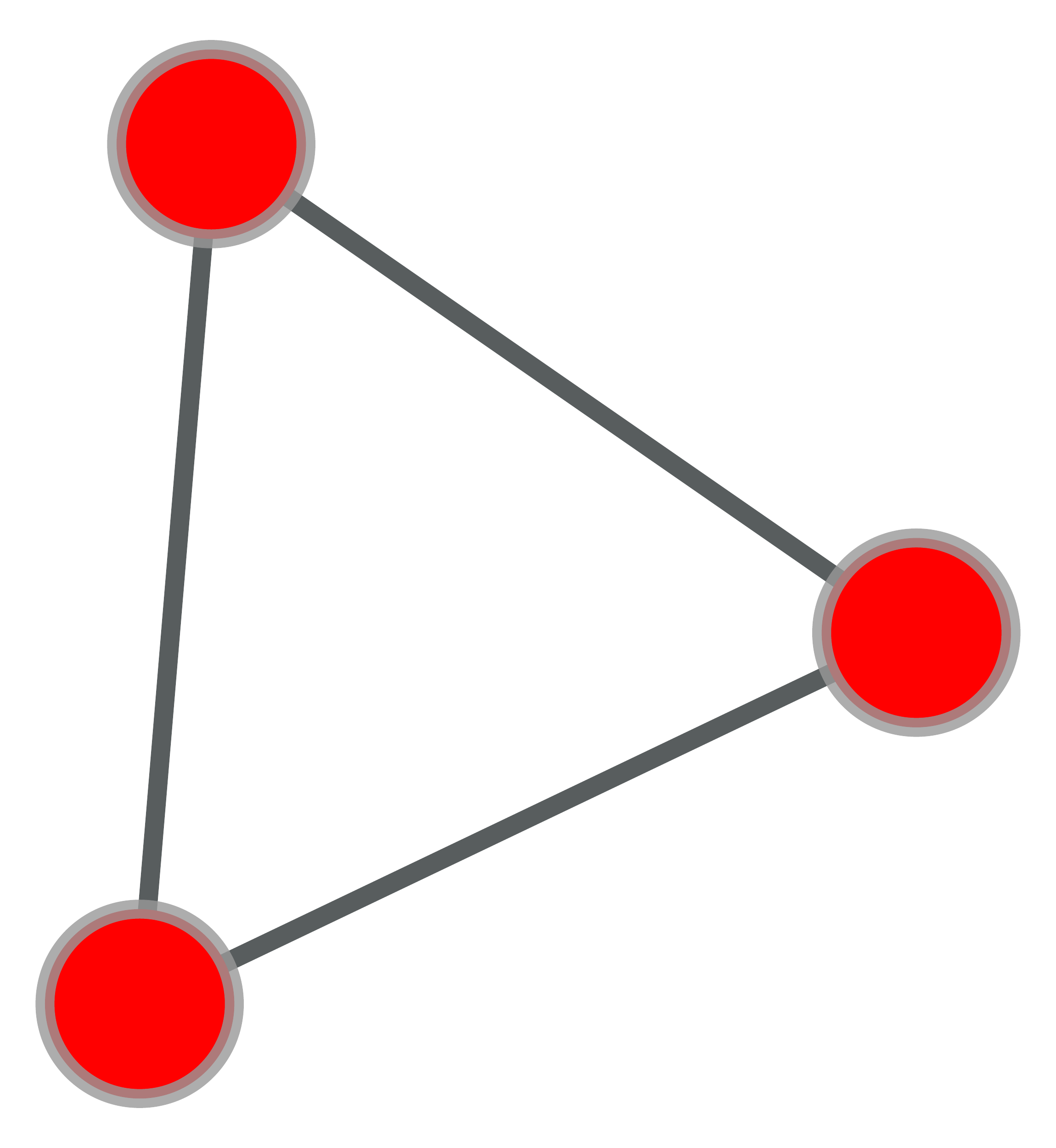}
    \includegraphics[width=0.15\textwidth]{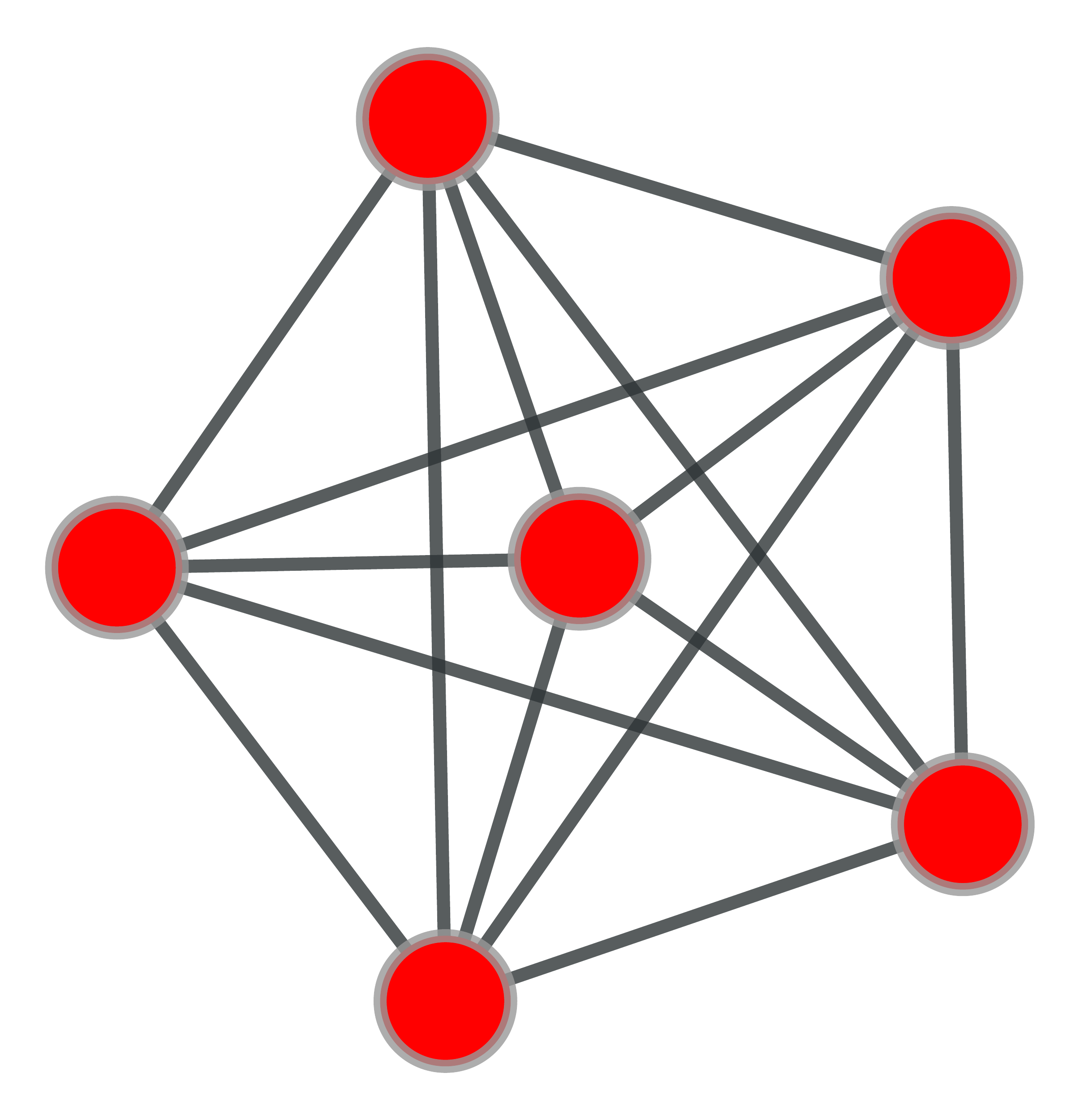}
    \includegraphics[width=0.15\textwidth]{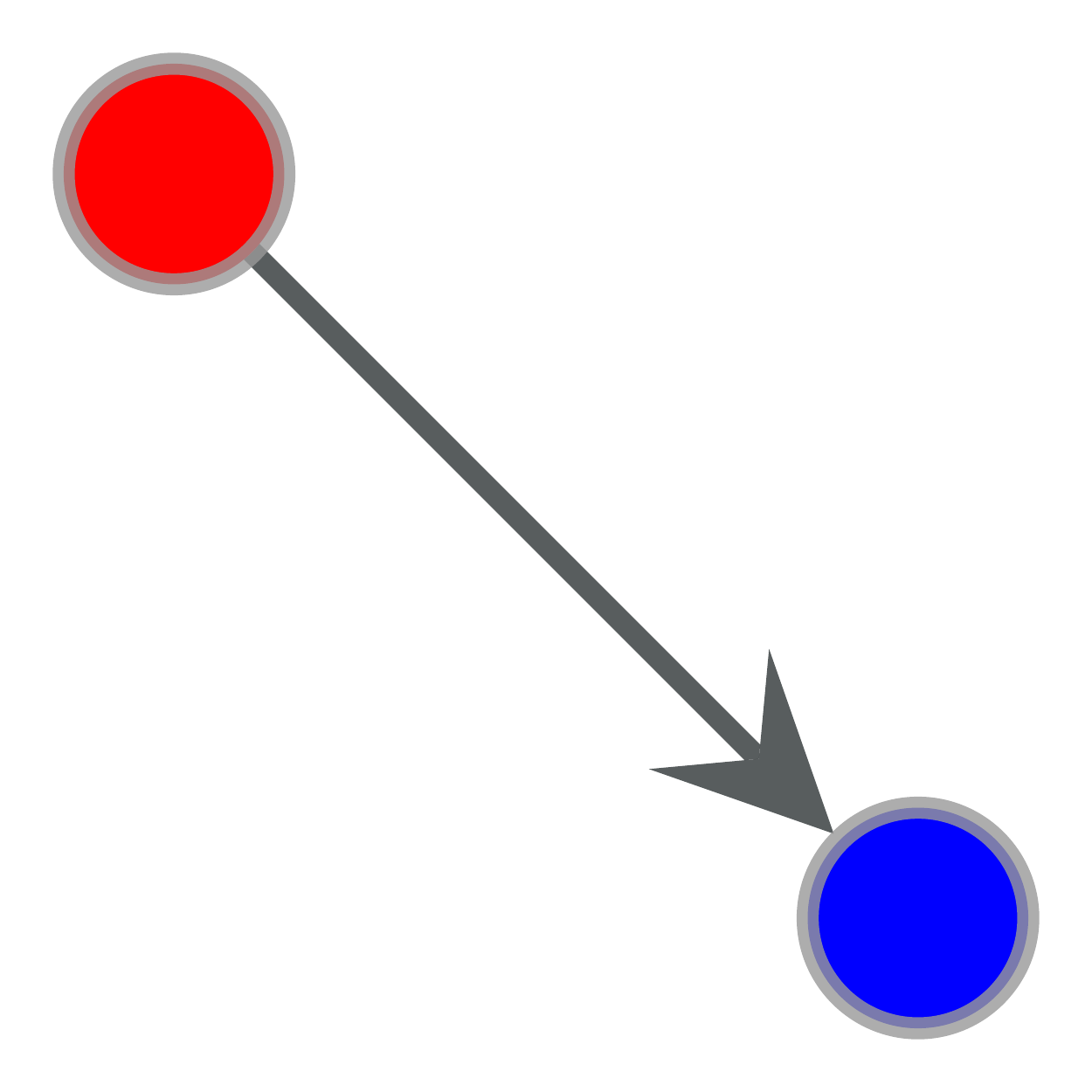}
    \includegraphics[width=0.15\textwidth]{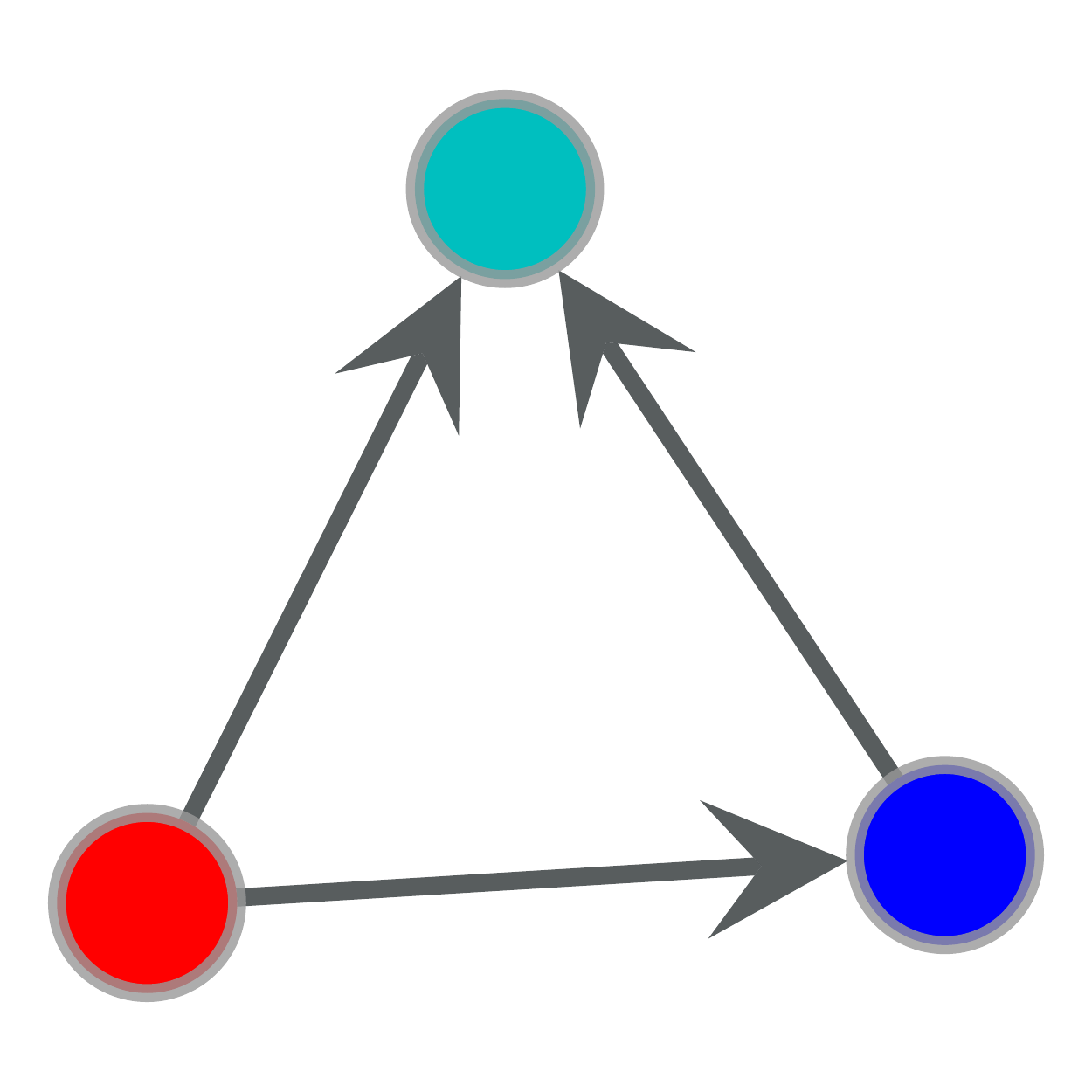}
    \includegraphics[width=0.15\textwidth]{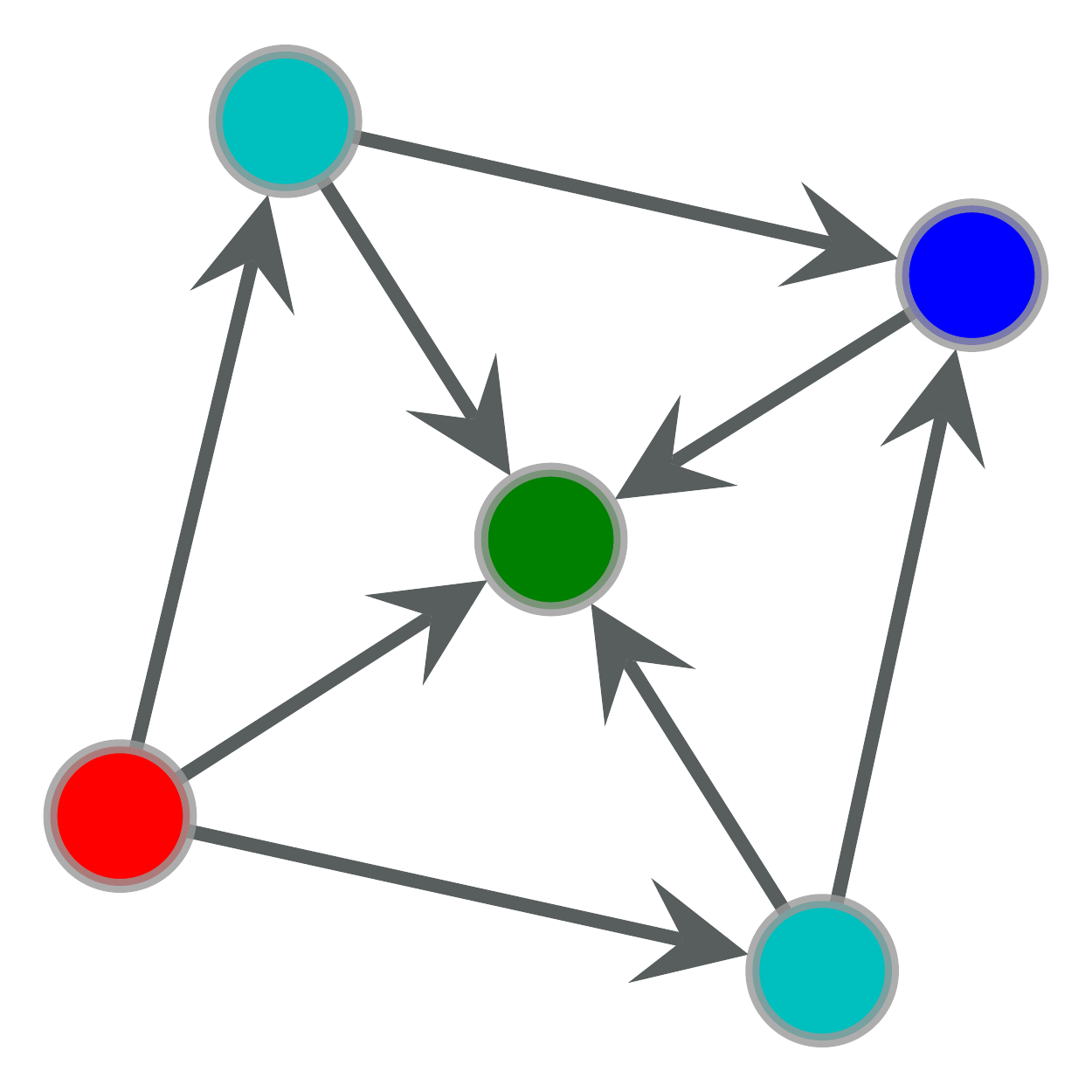}
    \caption{Examples of undirected and directed motifs with vertex colours indicating orbits.}
    \label{fig:orbits}
\end{figure}

First, we review some basic concepts that are essential in formulating the generative model starting with graph isomorphism. Two graphs $G$ and $H$, are said to be isomorphic if there exists a bijection $\phi: V(G) \rightarrow V(H)$ such that $(v,v')\in E(G) \iff (\phi(v),\phi(v')) \in E(H)$. Being isomorphic is an equivalence relation of which the equivalence classes correspond to unlabelled graphs also known as motifs. We denote motifs using lower-case letters and write $G \simeq g$. Automorphisms are special types of isomorphisms which map a \textbf{graph} to itself. Automorphisms are essentially vertex permutations that leave the structure of the graph unchanged. The set of all automorphisms of $G$ form a group which we denote as $Aut(G)$. Finally, the orbits of a graph are subsets of vertices that can be mapped onto each other by $Aut(G)$ and correspond to classes of structurally identical vertices of $G$ (see Fig.\ref{fig:orbits}).  The $i^{th}$ orbit of $G$ is denoted as $O_{G,i}$. 

In the remainder of the text we shall not explicitly distinguish between undirected and directed graphs because in the context of subgraph configurations this distinction can be reduced to the set of atomic subgraphs and the character of their symmetries. The only modification required for the directed case is in the definition of graph isomorphisms which for directed  graphs are required to also conserve edge directions. 

\subsubsection*{Microcanonical models with fixed subgraph counts}

The simplest type of subgraph configuration model (SGCM) on $N$ vertices with atoms $M=\{m\}$ is the ensemble of all subgraph configurations that contain exactly $n_m$ subgraphs of type $m\in M$. In the microcanonical ensemble all such configurations have equal probability which can be computed by simply counting the total number of such configurations:   
\begin{equation}\label{mcP}
P(C|\mathbf{n}_m,M)=\Big(\prod_{m\in M} {|\mathcal{H}_{N,m}| \choose n_m}\Big)^{-1},
\end{equation}
where $\mathcal{H}_{N,m}$ is set of all $m$-subgraphs of the complete graph $K_N$. It follows from the definition of the automorphism group that: 
\begin{equation}
|\mathcal{H}_{N,m}|=\frac{N!}{(N-|m|)!|Aut(m)|}.
\end{equation}

In such models atomic subgraphs are distributed uniformly over the vertices and hence we shall refer to these types of models as homogeneous SGCMs. When $M$ contains only the single edge motif, homogeneous models reduce to the classical Erd\"os R\'enyi random graph $G_{N,e}$ where all graphs with $N$ vertices and $e$ edges are equi--probable. Throughout this article we assume that networks are sparse i.e. $n_m=O(N)$ for which $\log(P)$ in Eq.\ref{mcP} can be approximated using Stirling's formula \cite{wegner2014subgraph}.

\subsubsection*{Subgraph configuration models and random graphs}
Distributions over subgraph configurations such as the one introduced above can be turned into distributions over graphs by projecting subgraph configurations onto graphs. In general we shall assume that the graphs under consideration are simple i.e.  that they do not contain parallel edges or self loops. In this case subgraph configurations can be projected onto graphs by simply taking the union of the edges of the subgraphs in a given configuration $C$. Using this projection any distribution over subgraph configurations $\{C\}$, denoted $P(C)$, can be mapped onto a distribution over graphs where the probability of graph $G$ is given by the sum of the probabilities of the subgraph configurations of which the projection is $G$, i.e.:
\begin{equation}\label{pc}
 P(G)=\sum\limits_{C|\bigcup_{s\in C}E(s)=E(G)} P(C).
\end{equation}
Note that subgraph configurations differ from other types of latent states, such as community assignments in the SBM, in that each subgraph configuration maps to a unique graph. Therefore, subgraph configurations can be seen as a general class of exact/lossless graph representations that includes many known graph representations including edge lists, adjacency list and bipartite representations.

\subsubsection*{Degree--corrected models}\label{sgcm}
In homogeneous SGCMs atomic subgraphs are distributed uniformly over the vertices which in turn results in graphs with Poisson type degree distributions \cite{Bollobas2011SparseClustering}. Therefore homogeneous models are in general not a good fit for empirical networks that have highly heterogeneous or power-law type degree distributions. Thus we consider degree--corrected subgraph configuration models where one not only constrains the counts of atomic subgraphs but also the number of atomic subgraphs attached to each vertex. For this we consider orbit degree sequences of configurations where
$d_{m,i}(C)(v)$ specifies the number $m$-subgraphs in $C$ that are attached to $v$ at orbit $i$. For instance the central vertex of the configuration in Fig.\ref{fig:Conf} has edge degree 1, triangle degree 1, 4-cycle degree 1. Whereas these motifs all have a single orbit the diamond has two orbits i.e. one corresponding to the vertices with degree 2 and one corresponding to the vertices with degree 3 for which the central vertex of the configuration in Fig.\ref{fig:Conf} has degrees 0 and 1, respectively. Note that the orbit degree sequence is defined with respect to a subgraph configuration $C$ and should not be confused with the number of subgraphs of type $m$ containing a certain vertex $v$ in the graph. As in the case of fixed subgraph counts the microcanonical ensemble of subgraph configurations having orbit degree sequence $d_{m,i}(v)$ is defined as the uniform ensemble over all configurations having orbit degree sequence $d_{m,i}(C)(v)$. As a result the probability of configurations can be calculated by counting the number of configurations with orbit degree sequence $d_{m,i}(C)(v)$ \cite{Wegner2021AtomicNetworks}. A brief derivation of the likelihood can be found in the Methods section. The degree--corrected subgraph configuration model is closely related to the model by Karrer and Newman introduced in~\cite{Karrer2010RandomSubgraphs} which is defined in terms of a stub-matching process similar to the (edge) configuration model.

Note that the degree of a vertex $v$ (counting parallel edges) is fully determined by the orbit degree:
\begin{equation}
d(v)=\sum_{m,i}d_{m,i}(v) d(O_{m,i}),
\end{equation}
where $d(O_{m,i})$ is the degree of vertices in orbit $O_{m,i}$ and hence the degree corrected model can model networks with heterogeneous degree distributions. Although we discard parallel-edges in our formulation the expected number of such multi-edges is $O(1)$ for sparse models. 

\subsubsection*{Coarse grained degree corrected models}
Constraining the atomic degrees at the level of orbits in certain cases significantly increases the model complexity of DC-SGCMs due to the fact that the model requires a degree sequence of length $|V|$ to be specified for every orbit of each atom in $M$. Although the Bayesian approach safeguards against over--fitting by balancing goodness of fit and parametric complexity, using models with high parametric complexity can inhibit the detection of  regularities, i.e. lead to under-fitting, due to the high  complexity cost associated to including such regularities in the model. Thus, in the context of statistical inference degree--corrected models that have lower model complexity for the same set of atoms need to be considered. It should however be iterated that the final choice should be made on the basis of how well each model fits the data, for instance as we shall do in this paper, via the posterior odds ratio.

The parametric complexity of DC-SGCMs can be reduced by aggregating components of the orbit degree sequence at different levels of granularity. One such option is to constrain atomic degrees at the level of motifs i.e. by the fixing the sum $d_m(v)=\sum_i d_{m,i}(v)$ instead of each $d_{m,i}(v)$ separately which results in a model that requires $|M|$ degree sequences. The parametric complexity of the model can be even further reduced by only constraining the total number of atomic subgraphs attached to each vertex i.e. $d_{t}(v)=\sum_{m}\sum_{i}d_{m,i}(v)$ which results in a model that only requires a single degree sequence. In the case of directed graphs we also consider a model where orbits are grouped according to the their in-- and out--degrees because grouping orbits according to atoms will in general lead to a mixing of in-- and out--degrees which in turn results in models where edge directions are random. For this we group orbits in to three groups corresponding to orbits that only have incoming edges ($d_{in}(v)=\sum_{m,i|d_{out}(O_{m,i})=0}d_{m,i}$), only out going edges and have both in-- and out--going edges, respectively. In this model the atomic degree sequence has 3 components and hence can be seen as a generalization of the configuration model where the in--, out-- and mutual--edge degrees are conserved separately \cite{Milo2002NetworkNetworks}. The likelihood of coarse grained ensembles can be obtained in closed form by the same procedure as the orbit degree model \cite{Wegner2021AtomicNetworks}. 

\subsection*{Inference}
Our goal is to infer a subgraph configuration $C$ for given the graph $G$. Hence, we consider $P(C|G)$ which can be obtained via Bayes' theorem: 
\begin{equation}
P(C|G)=\frac{P(G|C)P(C)}{P(G)},
\end{equation}
where $P(G|C)=1$ if $\bigcup_{s\in C}E(s)=E(G)$ and 0 otherwise, and $P(C)$ is given by the prior over subgraph configurations. Note that subgraph configurations are latent structures that fully determine the graph and therefore differ from other latent spaces such as node partitions in the SBM that do not uniquely determine the graph. 

We first consider the case of degree-corrected subgraph configuration models. For an atomic degree sequence $\mathbf {d_{m,o}}$ having motif set $M$ and count vectors $\mathbf{n_m}$ we assume a nested prior with a conditional dependence structure of the following form:
\begin{equation}
P(C)=P(C|\mathbf {d_{m,i}})P(\mathbf{d_{m,i}}|\mathbf{n_m})P(\mathbf{n_m}|M,e(G))P(M),
\end{equation}

where $e(G)$ is the number of edges in $G$. The above form  applies to all variants of the degree--corrected SGCM, the only difference being the number of components in the degree sequence. Similarly, in the case of the homogeneous SGCMs we consider a prior that has the following form:
$$P(C)=P(C|\mathbf{n_m})P(\mathbf{n_m}|M,e(G))P(M).$$

We seek to identify a subgraph configuration with maximum posterior probability (MAP) and will call such a configuration a MAP-configuration. Note that this is equivalent to inferring the latent state of the model rather than the parameters of the model. However, the two problems are closely related through Eq.\ref{pc}.

\subsubsection*{Model selection}
The model that is most suitable for describing a given network can be selected via posterior odds ratio. For this we first find the MAP-configuration for each model variant of the subgraph configuration model including homogeneous models. In the degree corrected case we consider models that constrain the distributions of atomic subgraphs at the level of orbits, atoms and total number of subgraphs. For directed networks in addition to these we also consider the directed orbit degree model. We then select the model-configuration combination that has highest posterior probability, which in turn is equivalent to selecting the configuration with the shortest description length. A more detail explanation of the model selection procedure is given in the methods section.

\subsubsection*{Minimum description length}
The Bayesian formulation outlined above is equivalent to finding a subgraph configuration that has Minimum Description Length (MDL)\cite{MDL}. Although the equivalence of MDL and Bayesian approaches holds more broadly, in our case it is more directly evident due to the discrete nature of the parameters.
\begin{equation}
\Sigma(C)=S(C|\mathbf {d},\mathbf{n_m},M)+\epsilon(\mathbf {d},\mathbf{n_m},M),
\end{equation}
where $\Sigma(C)$ is the description length (DL), $S(C)=-\log(P(C|\mathbf {d},\mathbf{n_m},M))$ the entropy which corresponds to the information required to specify the positions of the subgraphs given the model parameters and $\epsilon(\mathbf {d},\mathbf{n_m},M)=-\log(P(\mathbf{d},\mathbf{n_m},M))$ is the information required to describe the parameters of the model i.e. the model complexity. Consequently, finding a MAP-configuration is equivalent to finding a configuration with minimum description length.

\subsection*{The atomic structure of real world networks}\label{results}
Due to the computational complexity of the problem we restrict the set of motifs included in the analysis to connected motifs up to size 5 for directed networks and to motifs up to size 8 for undirected networks. There are  9578 such motifs in the directed case and 12112 in the undirected case. Despite the large number of potential motifs included in the analysis our method is able to identify concise sets of atoms (see $|M_{opt}|$ in Table \ref{Mselect}) that cover a large proportion of the edges in each of these networks. 

\subsubsection*{Model selection}
In this section we compare different subgraph configuration models, and their fit to data. For this, we consider different models introduced in the text so far, namely the homogeneous micro-canonical ensemble, the orbit degree--corrected model, the atomic degree model and the total degree model. For directed networks we also consider directed degree model where orbits are aggregated based on edge directions. The MAP configurations corresponding to different types of degree--corrected models tend to be quite similar both in terms of the atoms and subgraphs they contain. In general though, the number of atoms in the MAP configuration tends to increase as the level of detail at which the atomic degree sequence is conserved decreases since this decreases the model complexity cost of incorporating new atoms into the configuration.

Once we obtain the MAP-configuration for each model we select the optimal configuration-model pair by comparing their posterior odds ratio/description lengths, see Table \ref{Mselect}.  We find that for undirected networks the total degree model results in the shortest description length, except for the Malaria Genes network. For the directed networks we find that in general the directed degree model or total degree model result in the shortest description lengths. 

Our results show that the inclusion of higher order interactions in representations of real world networks leads to much more parsimonious representations when compared to their counterparts that only include dyadic interactions (i.e. single edges) with reductions in DL ranging from 297 to 13,007 nats. Note that this reduction in description length directly relates to the posterior odds ratio of the MAP configuration $C_{M}$ and the configuration consisting of all single edges i.e. $E(G)$ :  
$$\frac{P(C_{M}|G)}{P(E(G)|G)}= \exp(\Sigma(E(G))-\Sigma(C_{M})),$$
where $\Sigma(E(G))$ is the DL of $E(G)$ as given by the edge only configuration model. In other words we find that the networks in Table \ref{Mselect} are exponentially more likely to have been generated by the higher order models corresponding to their MAP configurations than by the edge only configuration model. Similarly, for real world networks we analysed, we find that degree corrected models in general result in significantly shorter description lengths than non--degree corrected models indicating that degree corrected models are an overall better fit to the data.

In the following section we provide a more detailed discussion of the MAP configurations for some of the networks in Table \ref{Mselect}. The results for the remaining networks are discussed in the SI. 

\begin{table}[]
\begin{adjustbox}{max width=\textwidth}
    \centering\footnotesize
    \begin{tabular}{c|c|c|c|c|c|c|c|c|c|c}
         Network&$|V(G)|$&$|E(G)|$&$\Sigma_o$&$\Sigma_m$&$\Sigma_t$&
         $\Sigma_d$&$\Sigma_h$&$\Sigma_e$&$\min_{}(\Delta\Sigma) $&$|M_{opt}|$ \\
         \hline
         
         Malaria Genes \cite{larremore2013network}&307&2,812&7,371&7,382&7,374&-&\textbf{7,336}&10,112&35&26\\
         
         CE gap junctions (h) \cite{cook2019whole}&469 &1,433&6,910& 6,915&\textbf{6,859}&-&6,887&7,186&28&17\\
         CE gap junctions (m) \cite{cook2019whole}&585&1,724&8,318& 8,320&\textbf{8,233}&-&8,275&8,766&42&19\\
         Reference connectome \cite{Szalkai2015TheV2.0}&1,015&4,275&16,569& 16,577&\textbf{16,150}&-&16,625&20,761&419&38\\
         Network Science\cite{Newman2006FindingMatrices} &1,589&2,742&11,700&11,702&\textbf{11,344}&-&11,707&18,916&356&16\\
         Yeast Proteins \cite{collins2007toward}&1,622&9,070&31,670&31,677&\textbf{30,377}&-&31,838&43,404&1,300&45\\
         CE Neural \cite{white1986structure}&297&2,345&9,421&9,641& 9,511&\textbf{9,339}&9,684&9,626&82&13\\
         CE Synapses (h) \cite{cook2019whole}&452&3,666&14,306&14,435&\textbf{14,023}&14,074&14,731&15,168&41&24\\
         CE Synapses (m)\cite{cook2019whole} &575&5,246&21,405&21,544&21,213&\textbf{20,997}&21,606&23,737&216&32\\
         Directed  connectome \cite{Szalkai2015TheV2.0}&1,015&4,008&21,733&21,863&21,311&\textbf{21,178}&21,918&22,796&133&21\\
         E.coli Metabolic \cite{meta}&2,275&5,793&33,933&33,854&\textbf{33,335}&33,790&36,137&37,958&455&12\\
         
    \end{tabular}
    \end{adjustbox}
    \caption{Description lengths (DL) of MAP configurations corresponding to degree corrected models that constrain the atomic degrees at different levels of granularity:  $\Sigma_o$:orbit degree,  $\Sigma_m$:motif degree,  $\Sigma_t$:total atomic degree, $\Sigma_d$:directed degree model. We also include the DL for the edge only configuration model ($\Sigma_e$) and homogeneous/non-degree corrected models ($\Sigma_h$) as baselines.  Description lengths are rounded to the nearest integer and given in nats. $\Delta\Sigma$ represents the difference in DL between the two models having the shortest DLs and $|M_{opt}|$ corresponds to the number of atoms in the MAP configuration with the shortest DL. CE stands for the worm {\em C. elegans} and, (h) and (m) indicate the hermaphrodite and male versions. }
    \label{Mselect}
    
\end{table}

\subsubsection*{Co-authorship network of network scientists}
We first consider a widely studied co-authorship network between network scientists \cite{Newman2006FindingMatrices}. We find that the MAP configuration of this network contains 15 nontrivial atoms other than the single edge that cover over 85\% of the edges (see Table \ref{NetsciM}). As expected in a collaboration network the MAP configuration contains many cliques corresponding to publications with more than two authors. In addition to cliques we also find higher order interactions where two high degree coauthors collaborate with the same group of researchers (Table \ref{NetsciM} Atoms 3 and 12) as well as patterns where two high degree authors collaborate with the same group of lower degree nodes but not with each other (Table \ref{NetsciM} Atoms 5,7,8,13 and 14).  

\begin{table}[h!]
\begin{center}

\begin{adjustbox}{max width=\textwidth}
\begin{tabular}{|c|c|c|c|c|c|c|c|c|}
\hline

$m$&\includegraphics[height=0.09\textwidth]{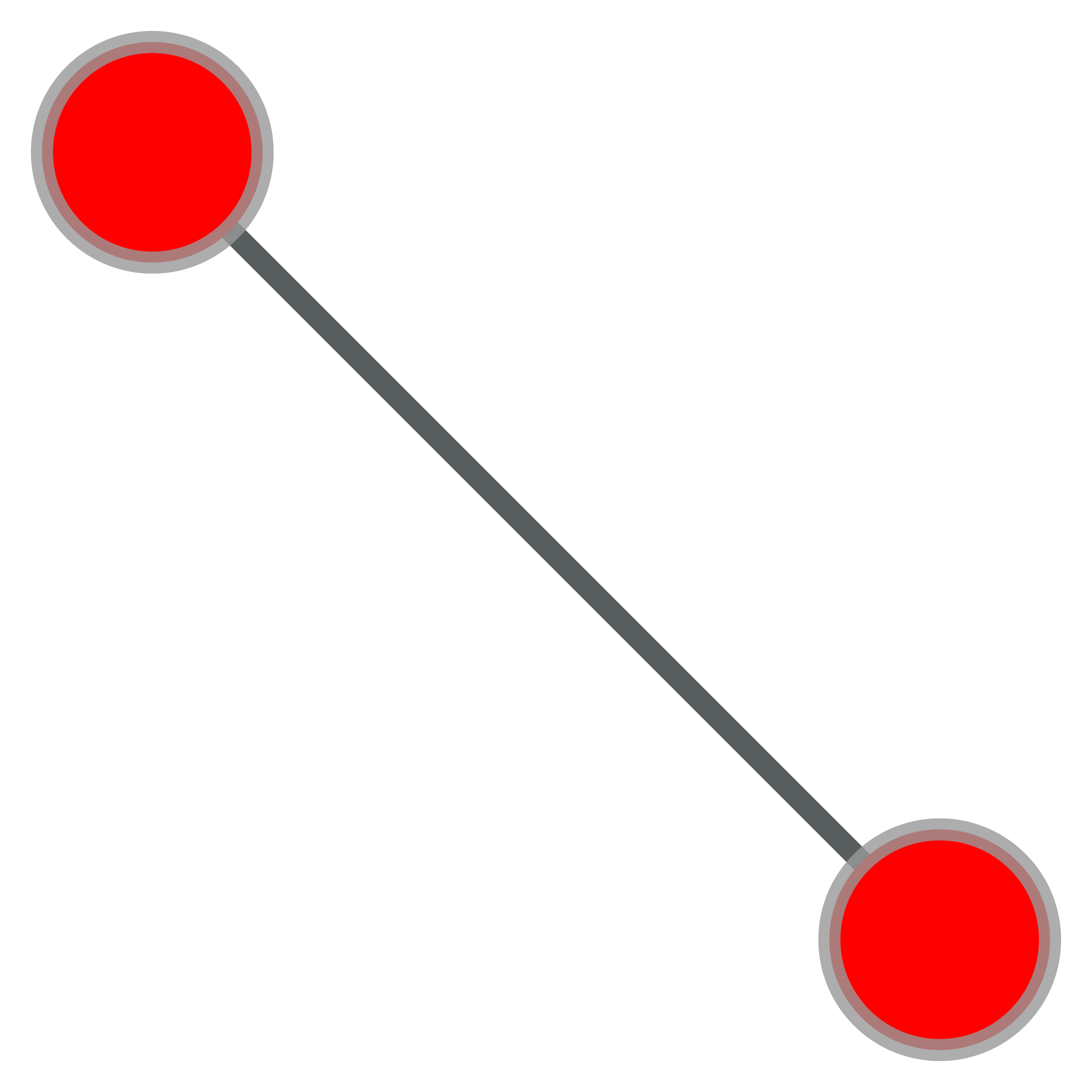}& \includegraphics[height=0.09\textwidth]{Netsci/M1.pdf}& \includegraphics[height=0.09\textwidth]{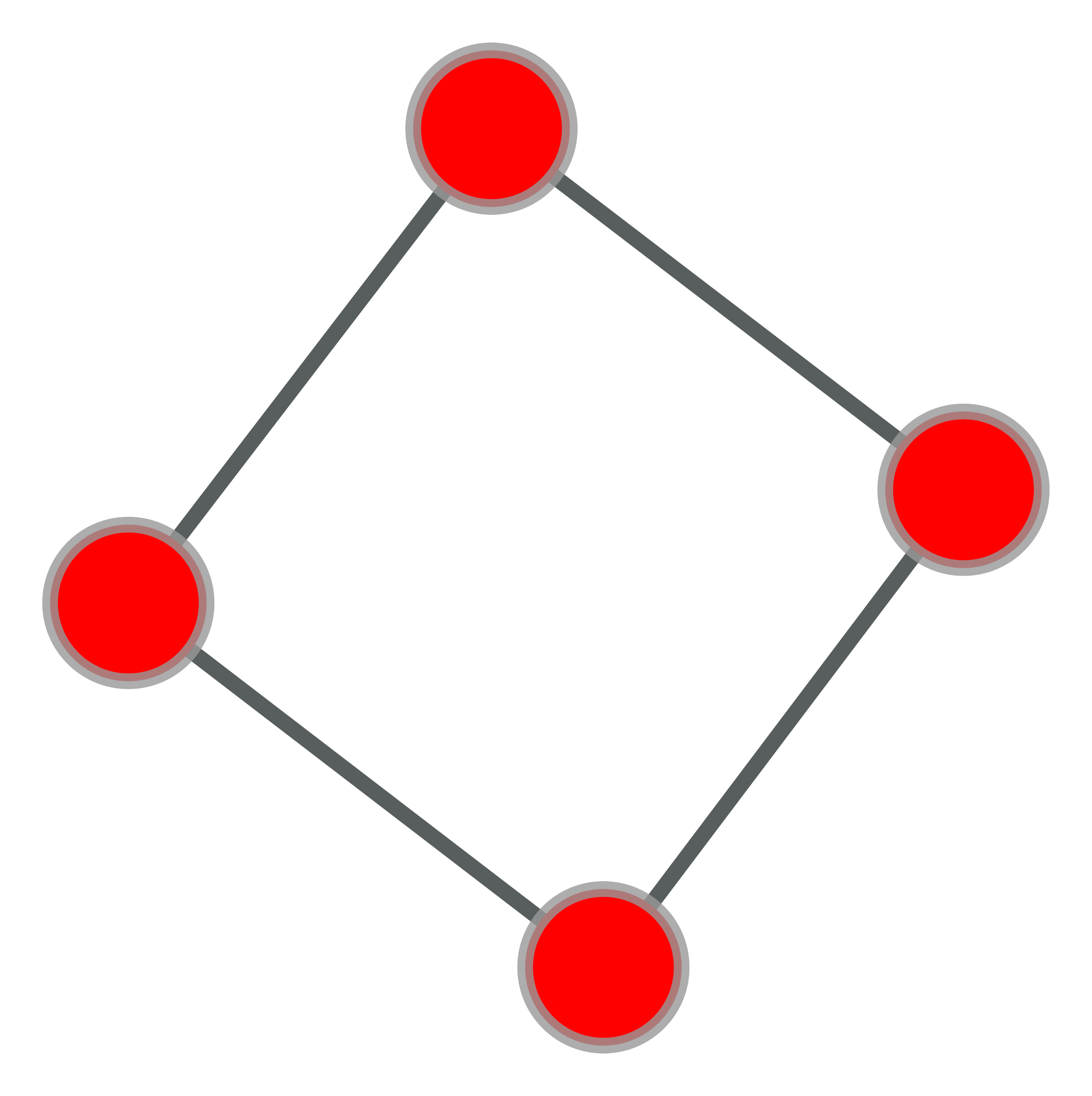}& \includegraphics[height=0.09\textwidth]{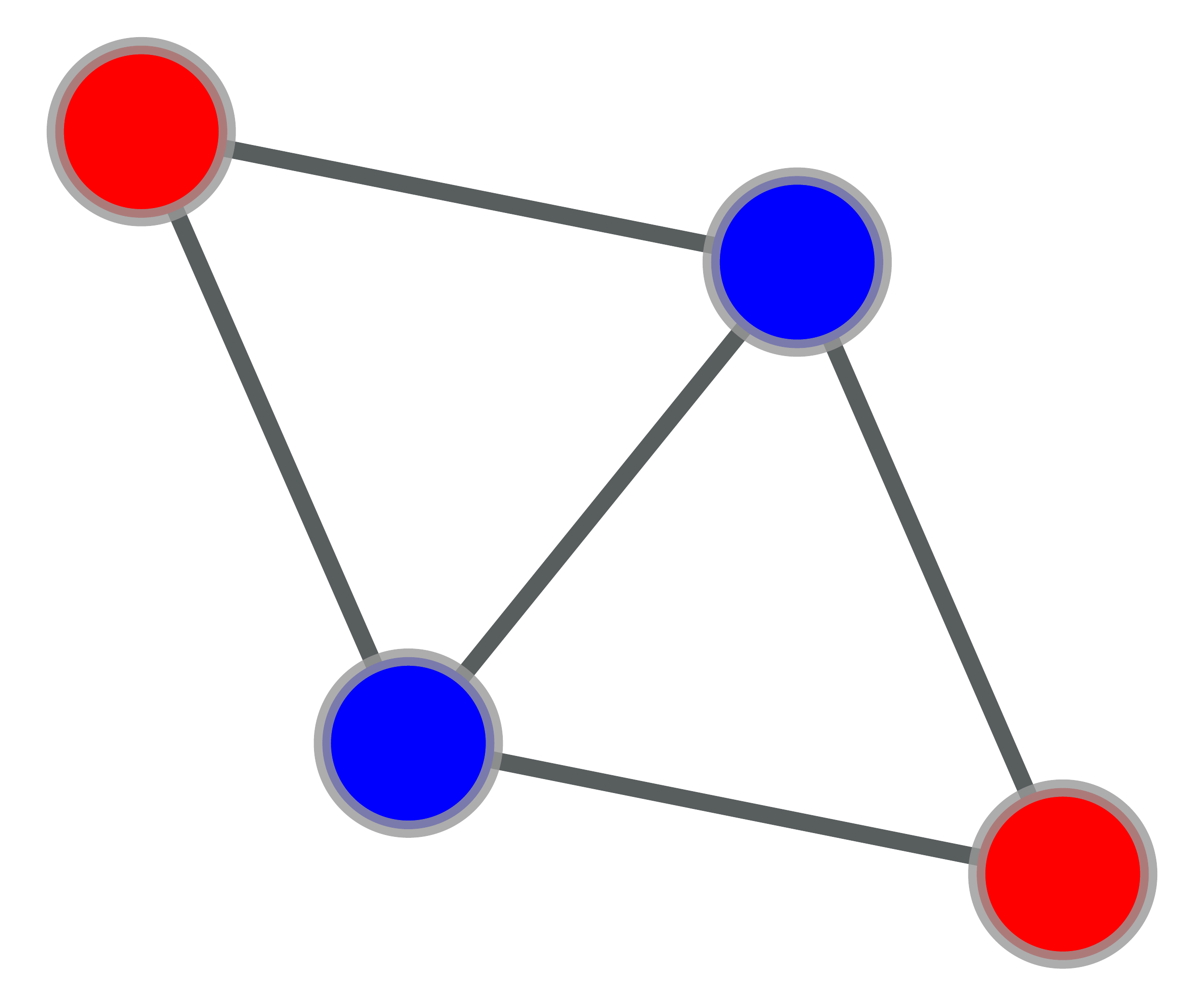}& \includegraphics[height=0.09\textwidth]{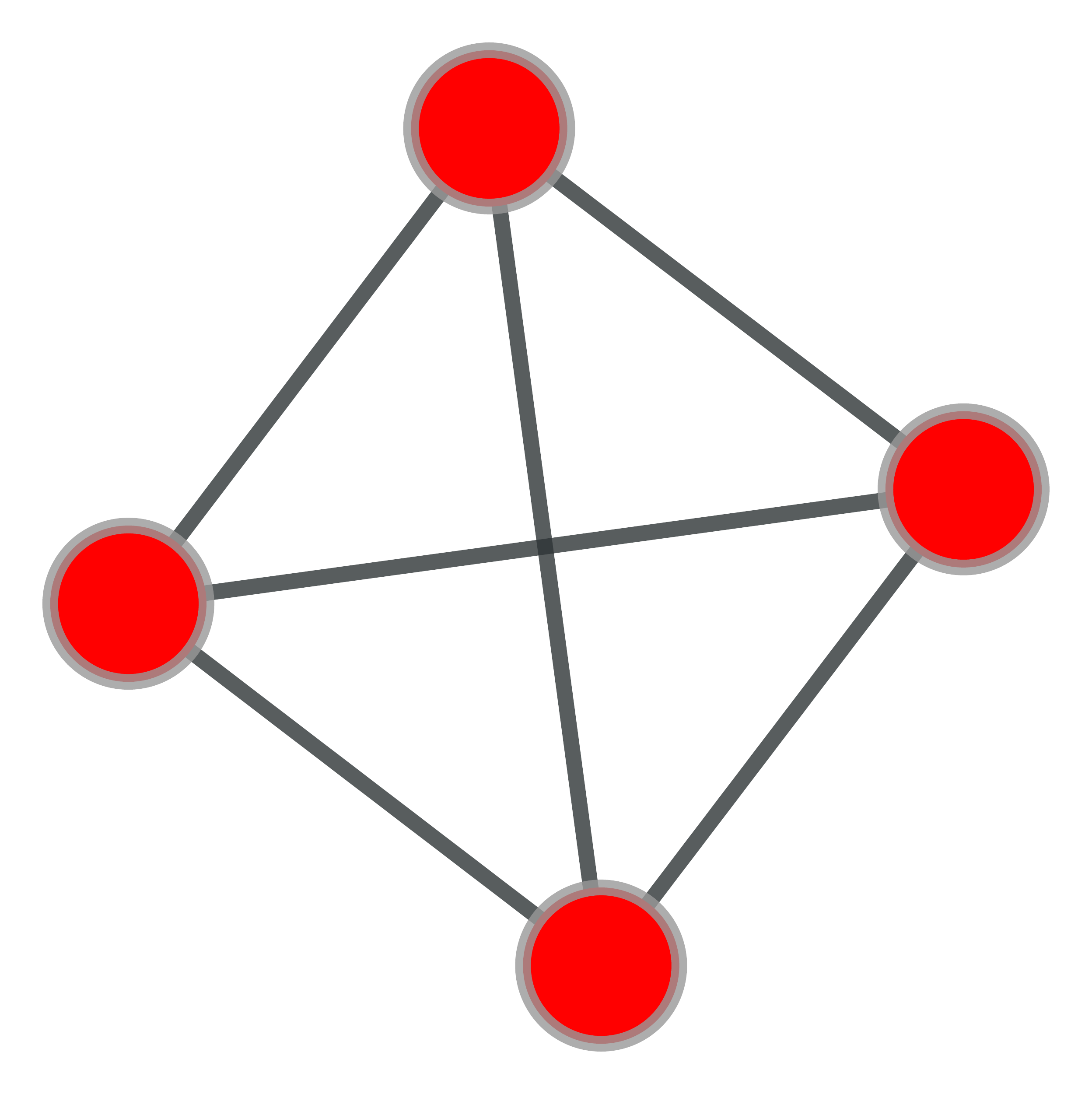} &\includegraphics[height=0.09\textwidth]{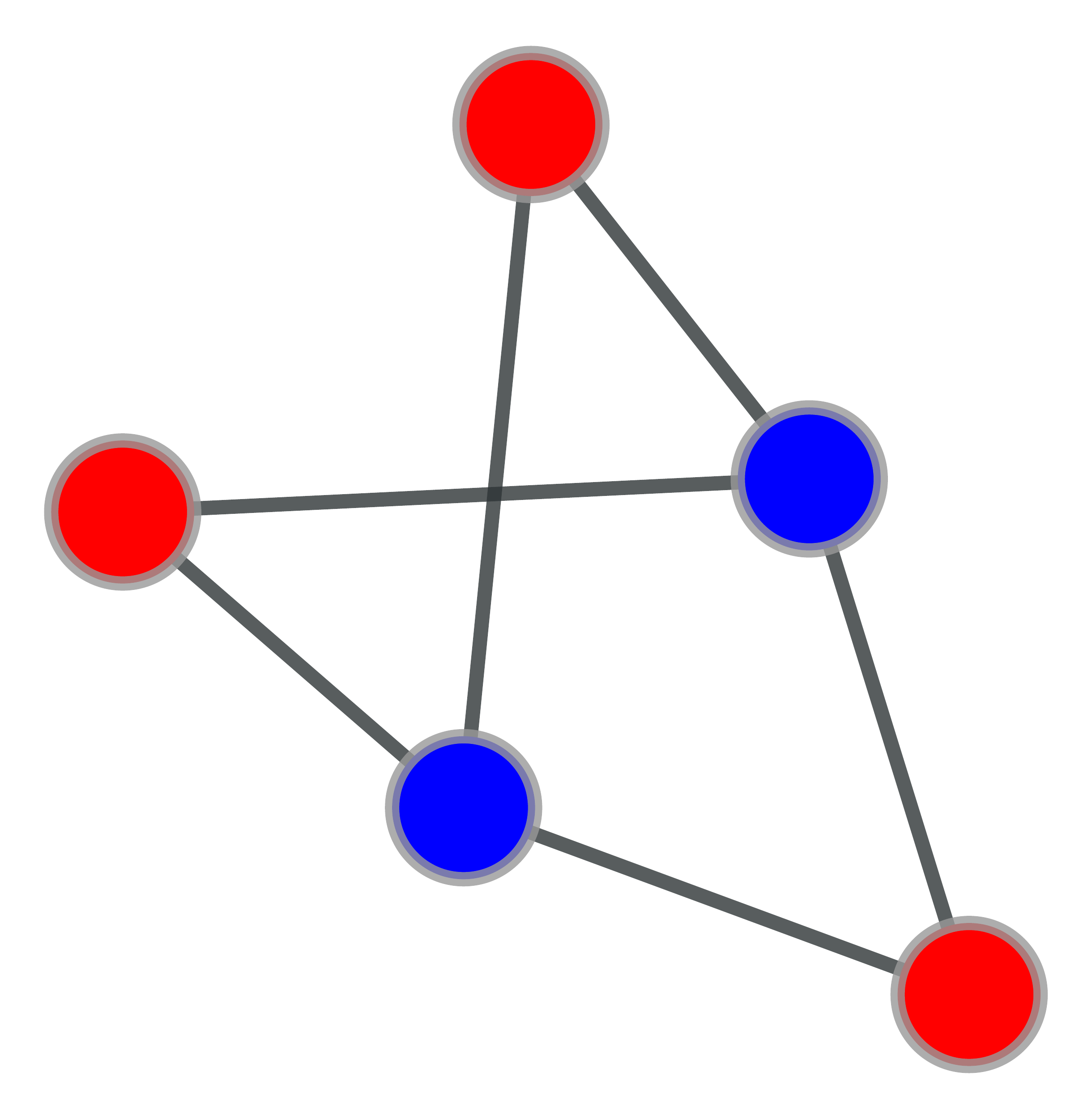}& \includegraphics[height=0.09\textwidth]{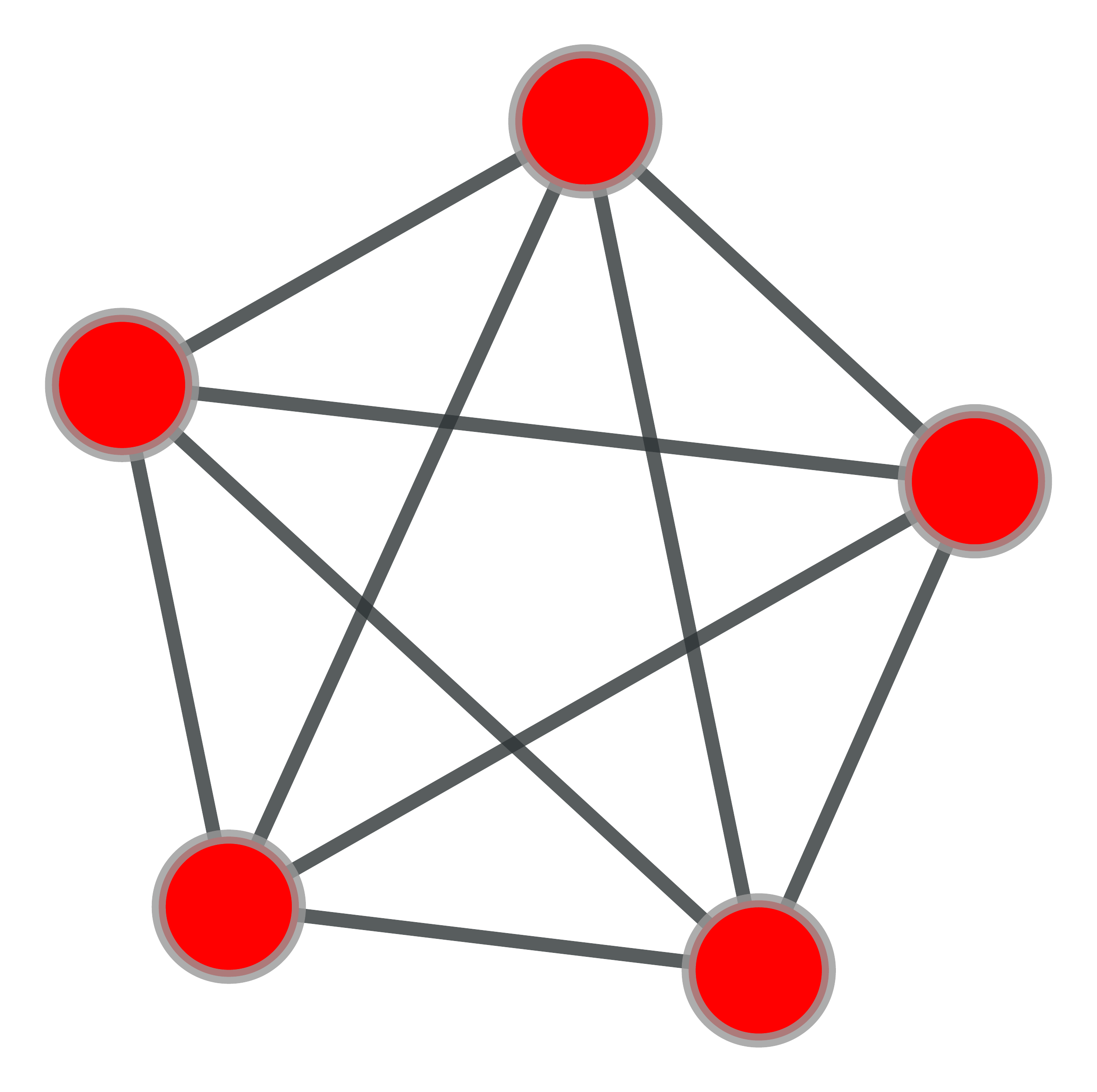}&\includegraphics[height=0.09\textwidth]{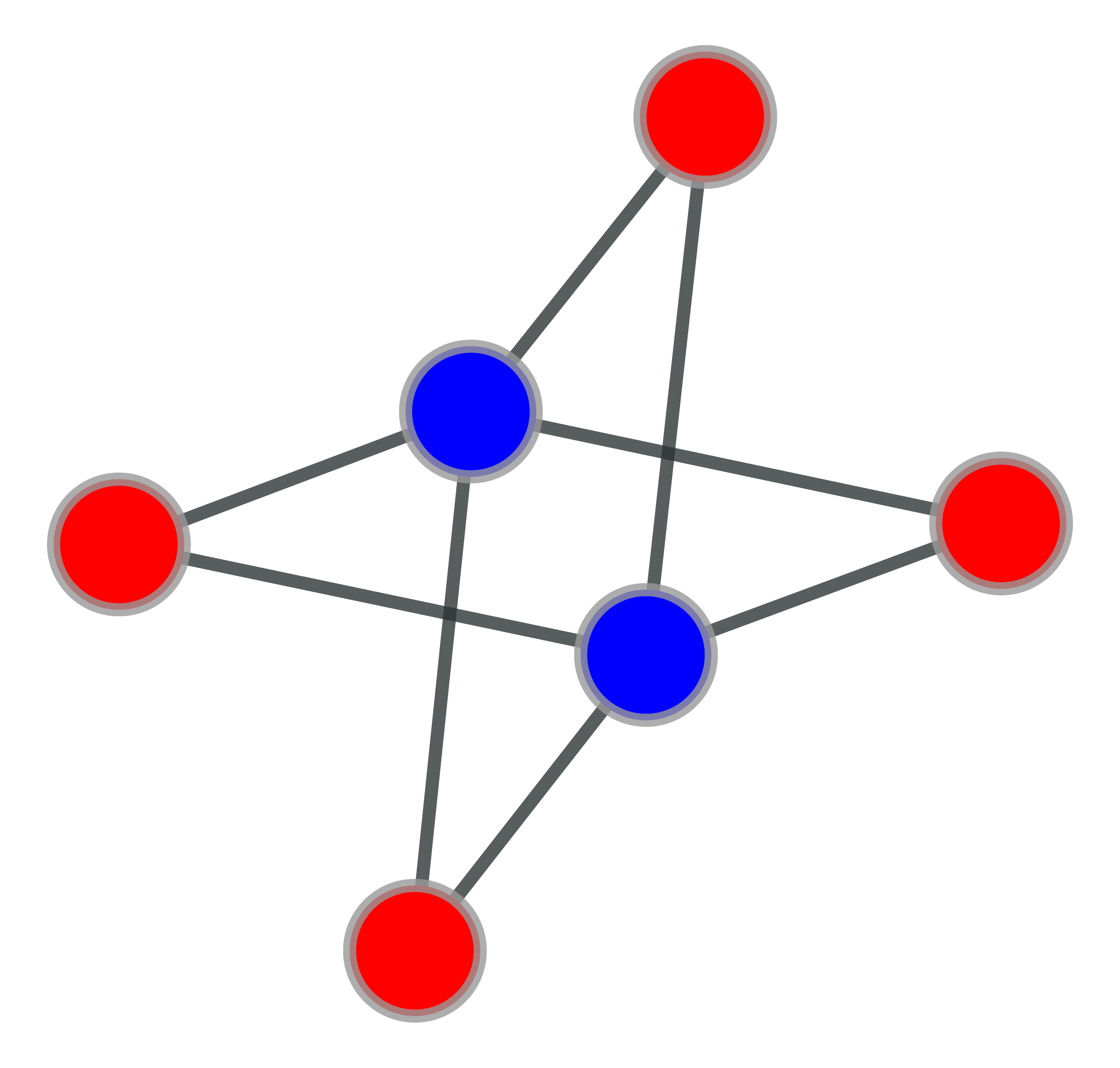}\\
\hline
Id&0&1&2&3&4&5&6&7\\
\hline
$n_m$&357& 130& 11& 14& 72& 4& 43& 4\\

\hline
$m$& \includegraphics[height=0.09\textwidth]{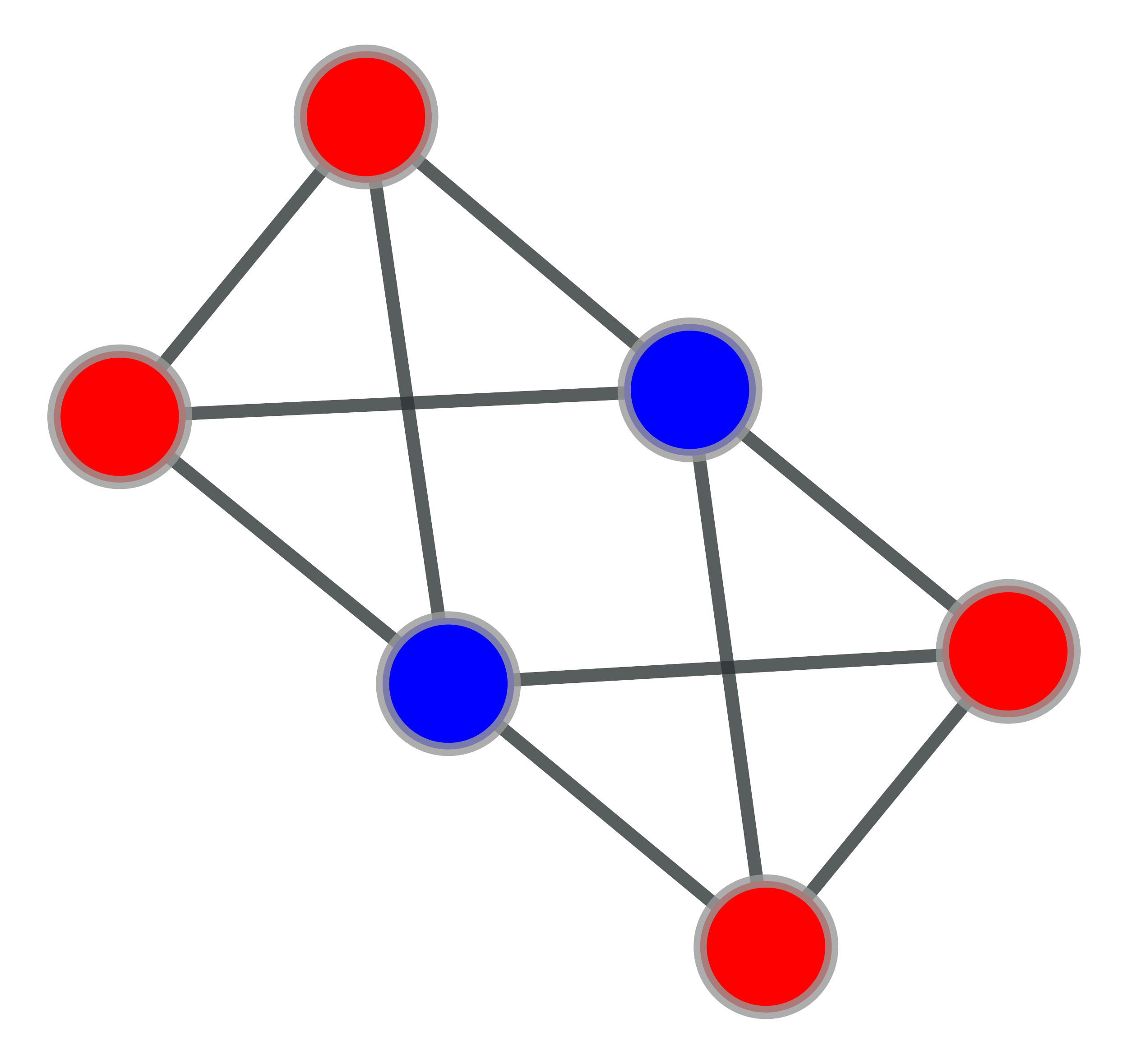}& \includegraphics[height=0.09\textwidth]{Netsci/M9.pdf}&\includegraphics[height=0.08\textwidth]{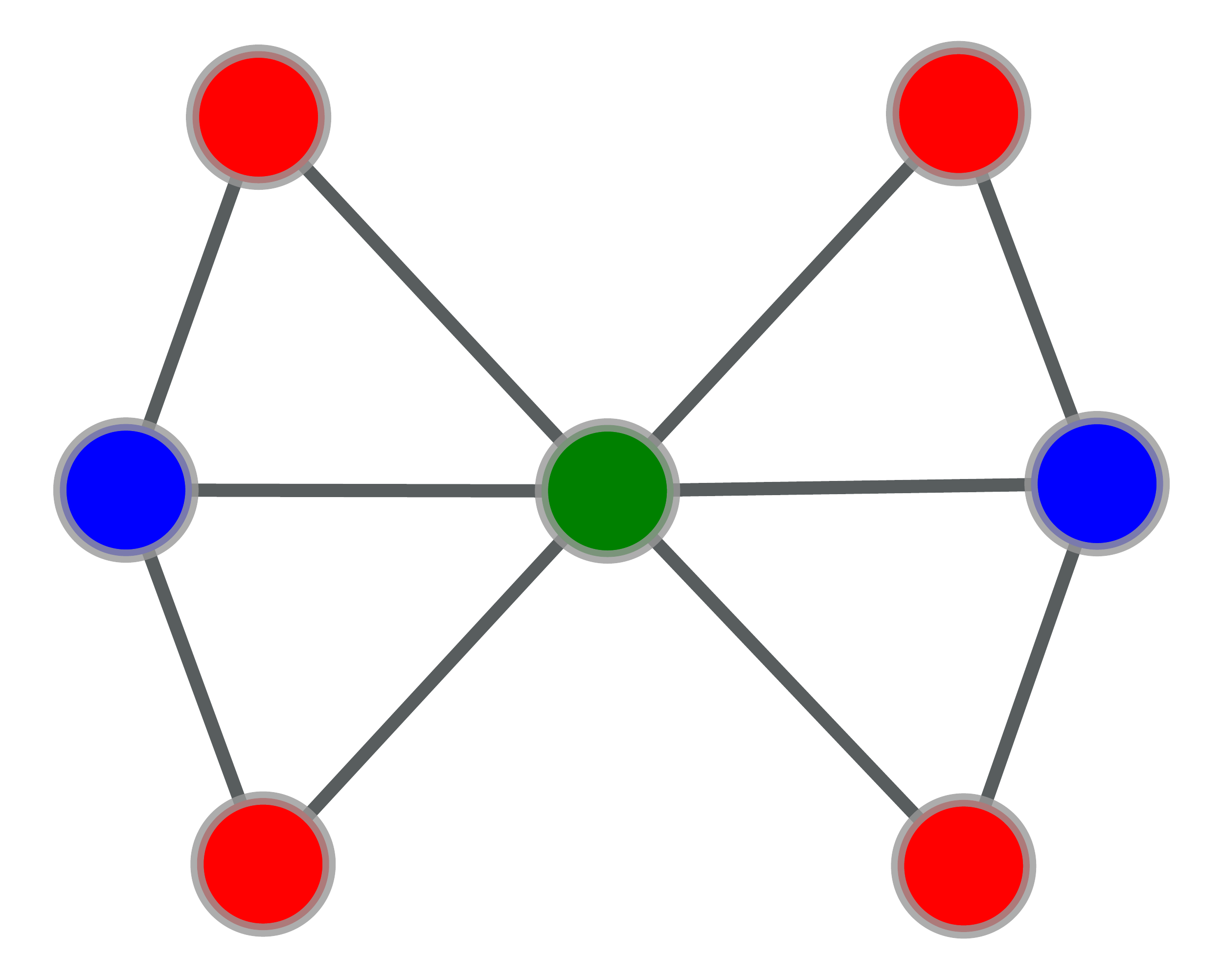}& \includegraphics[height=0.09\textwidth]{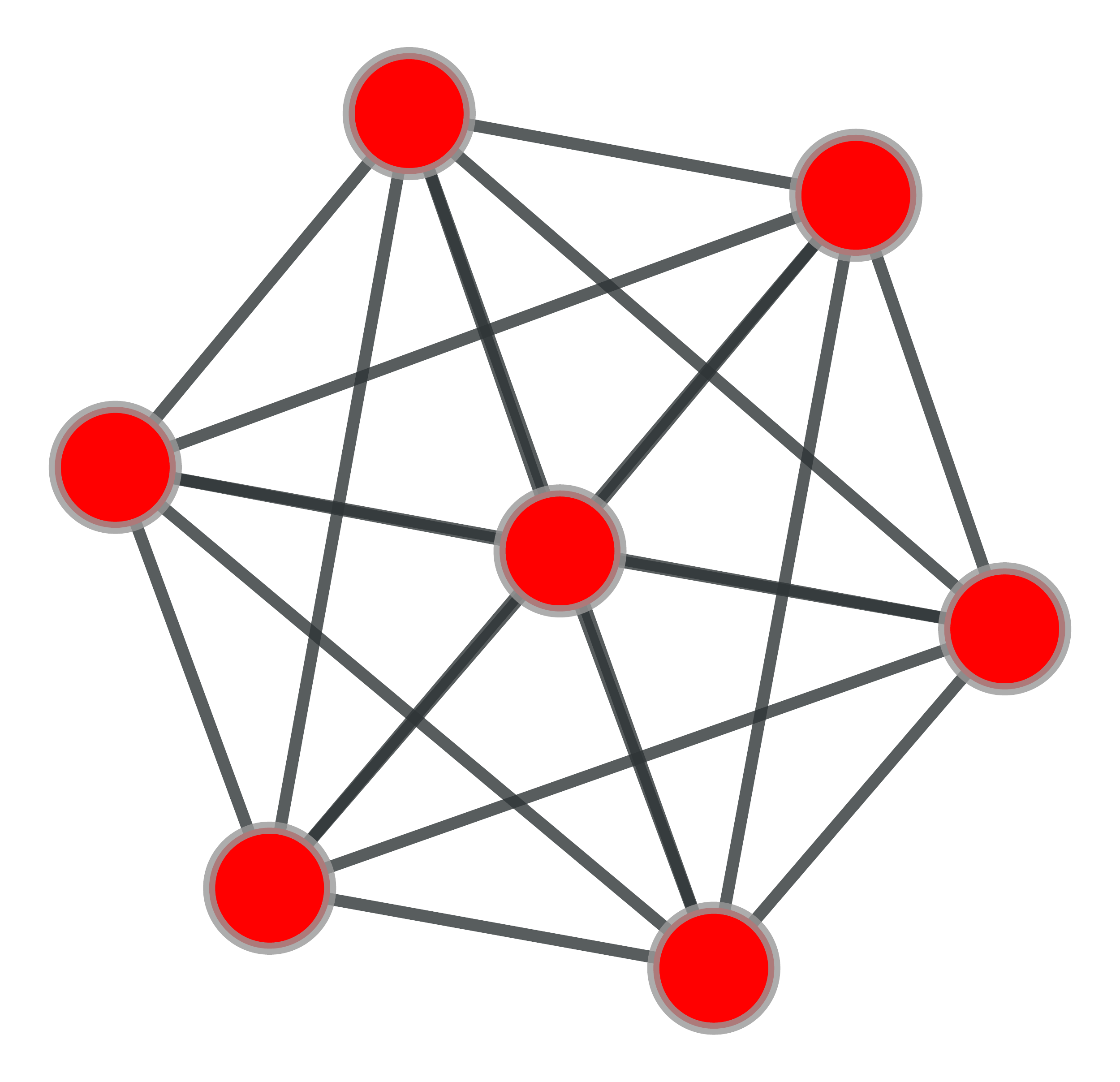}&\includegraphics[height=0.09\textwidth]{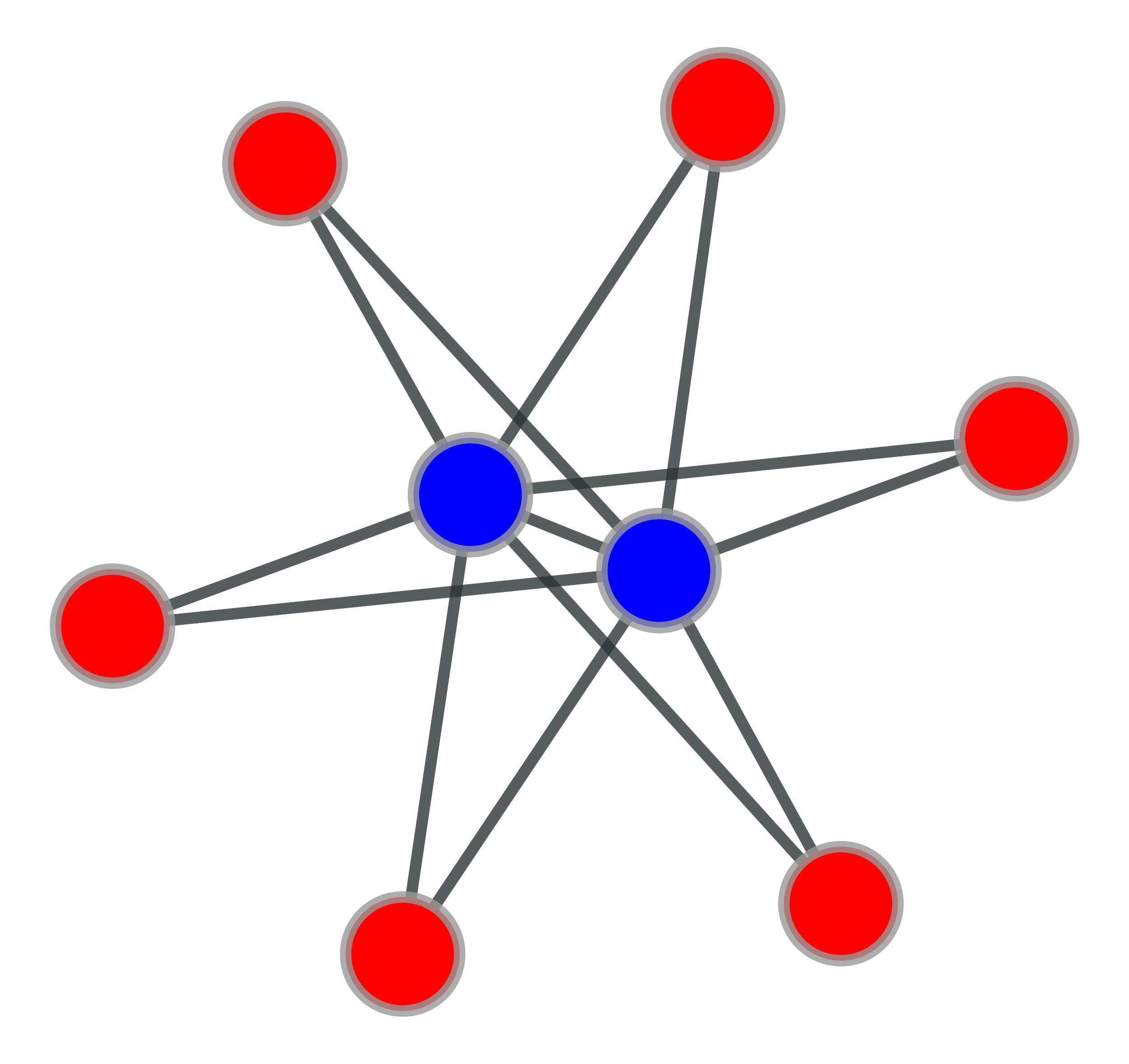}& \includegraphics[height=0.09\textwidth]{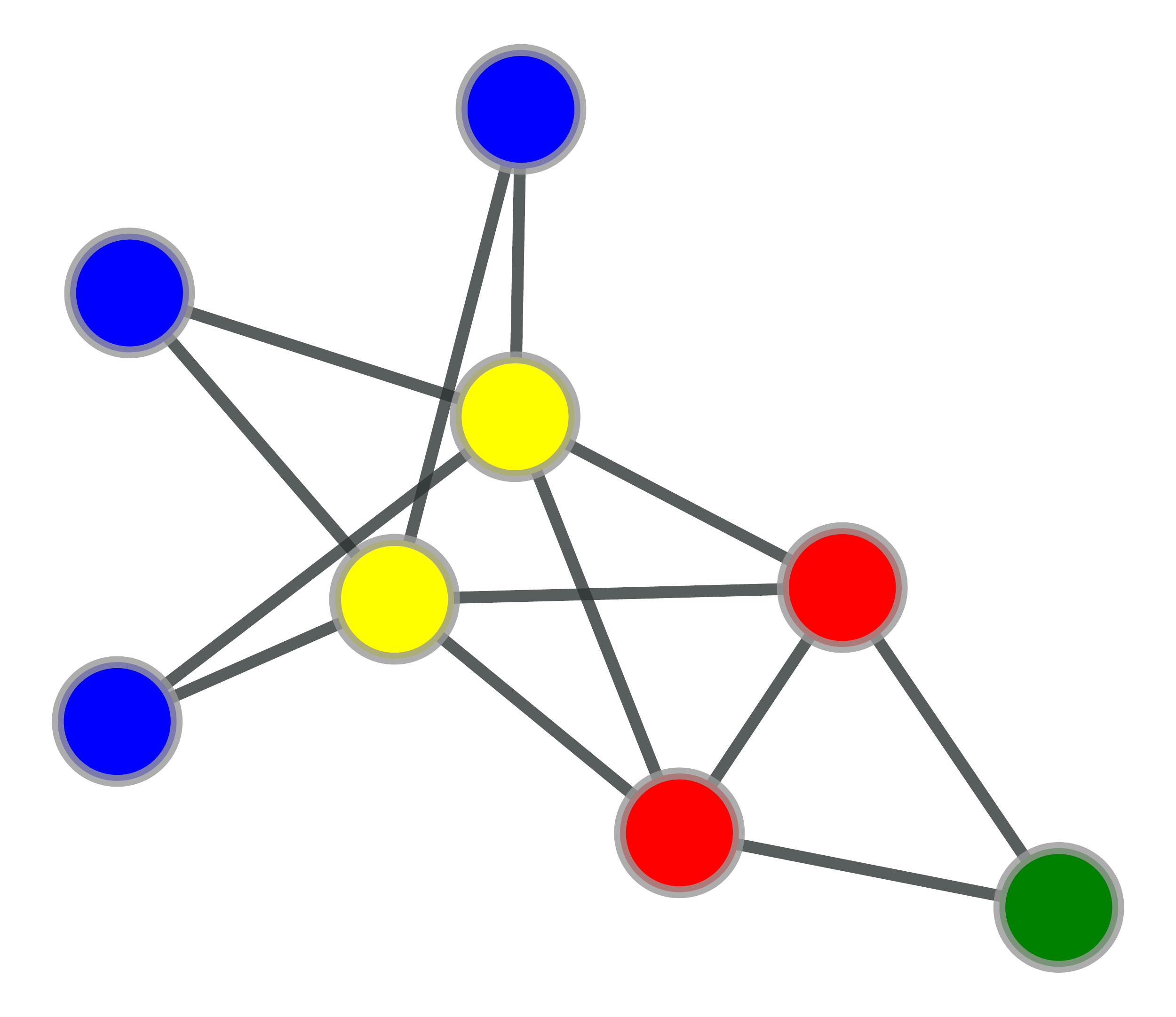}&\includegraphics[height=0.09\textwidth]{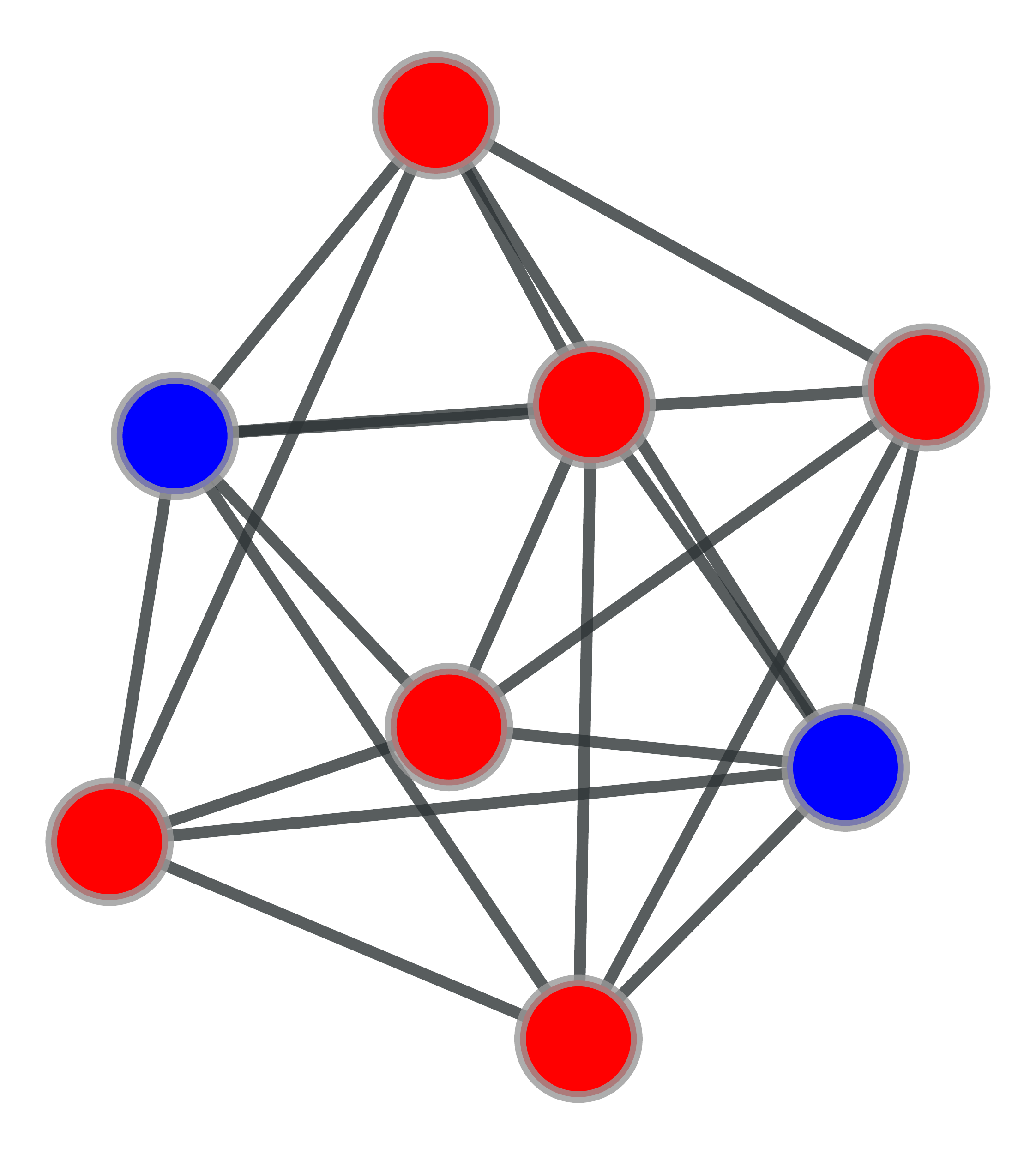}&\includegraphics[height=0.09\textwidth]{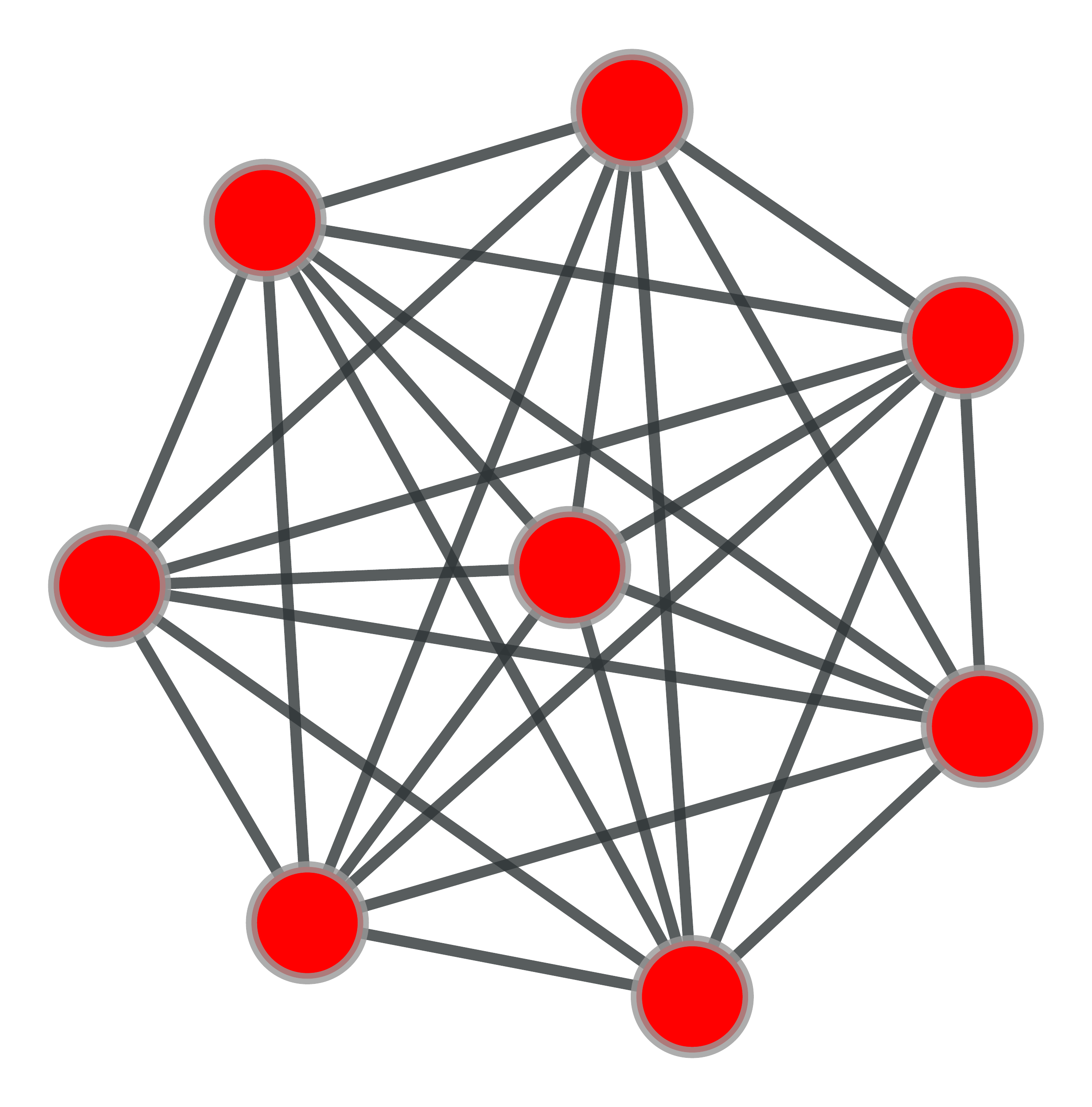} \\
\hline
Id&8&9&10&11&12&13&14&15\\
\hline
$n_m$&5& 14& 5& 4& 3& 2& 4& 15\\

\hline
\end{tabular}
\end{adjustbox}
\end{center}

\caption{Atoms found in the MAP configurations of the Network science collaboration network together with their respective counts in the MAP configuration ($n_m$).}
\label{NetsciM}

\end{table}

\begin{figure}[h!]
\includegraphics[width=\textwidth]{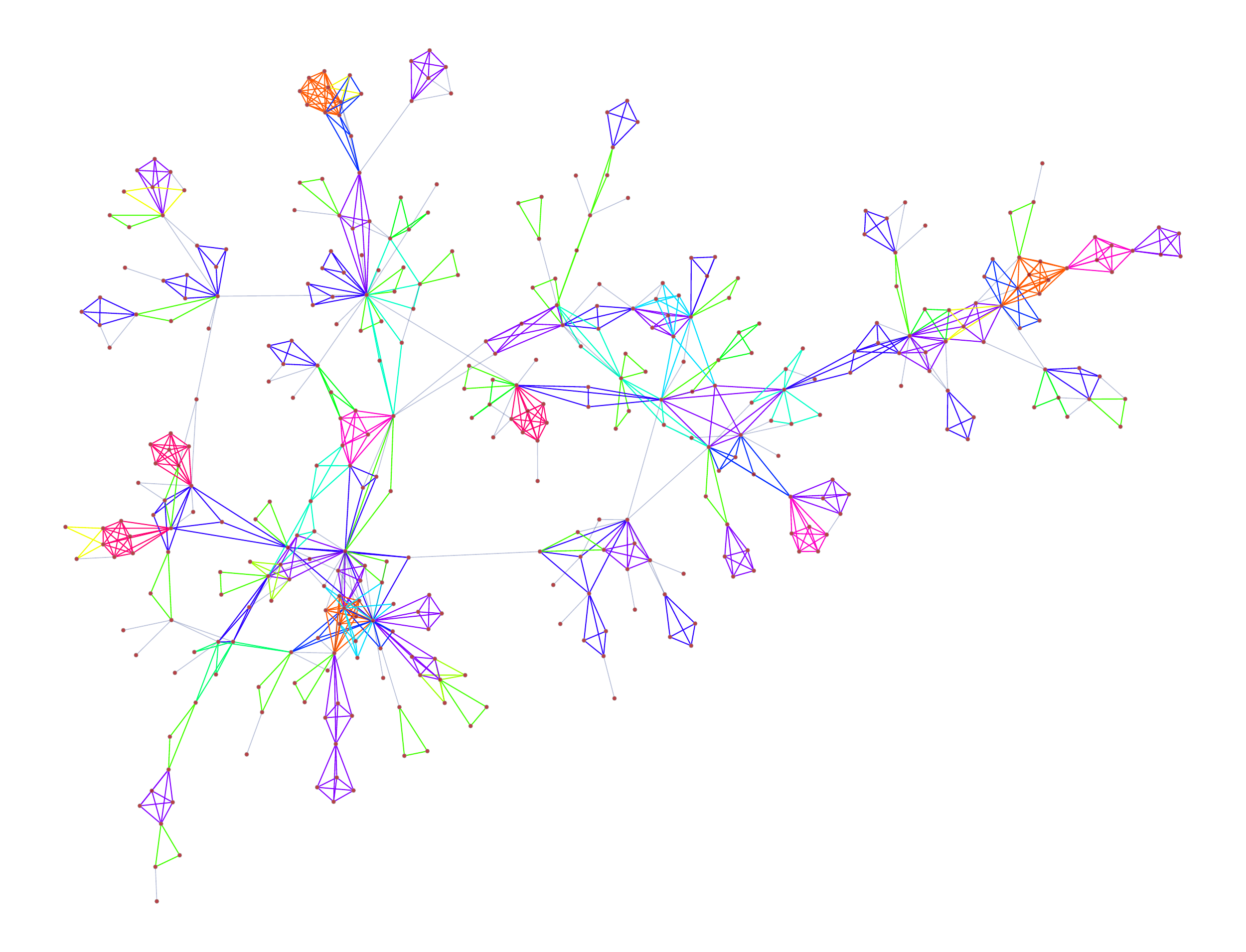}
\caption{Visualization of the largest connected component of collaborations between network scientists. Colours represent different atomic subgraph types in the MAP-configuration. A high resolution plot of the full network can be found in the SI.}
\label{netsciLCC}
\end{figure}

\subsubsection*{Human connectome}
Next we consider a directed connectome of the human brain from \cite{Szalkai2015TheV2.0}. The network has 1015 vertices and 4008 edges. We find that the MAP configuration contains 20 non-trivial patterns which cover approximately 75\% of the edges. 
The MAP configuration contains a large number of feed-forward loops (FFL) (Atom 1 in Table \ref{dibrain}) and bi--fan motifs (Atom 5 in Table \ref{dibrain}) usually associated with neuronal networks. We also find a large number of acyclic square shaped motifs (Atoms 2 and 3 in Table \ref{dibrain}) that can be interpreted as 4-node generalizations of the FFL, where an input node is connected to an output node via both direct connections and indirect connections in the form of longer directed paths. In addition to these basic types we also recover more complex atoms that can be interpreted as various combinations of these lower order atoms. Such larger arrangements of smaller motifs have recently been studied in \cite{adler2022emergence}. However, our result indicate that in the connectome lower order motifs combine in denser and more complex patterns than the pairwise combinations of lower order motifs studied in \cite{adler2022emergence} (see Atoms 12-20 in Table \ref{dibrain}). Interestingly, all atoms of the directed connectome contain at least one output node in the form of a node that has only incoming links. 

\begin{table}[h!]
\begin{center}
\begin{adjustbox}{max width=\textwidth}

\begin{tabular}{|c|c|c|c|c|c|c|c|}
\hline

$m$&\includegraphics[height=0.09\textwidth]{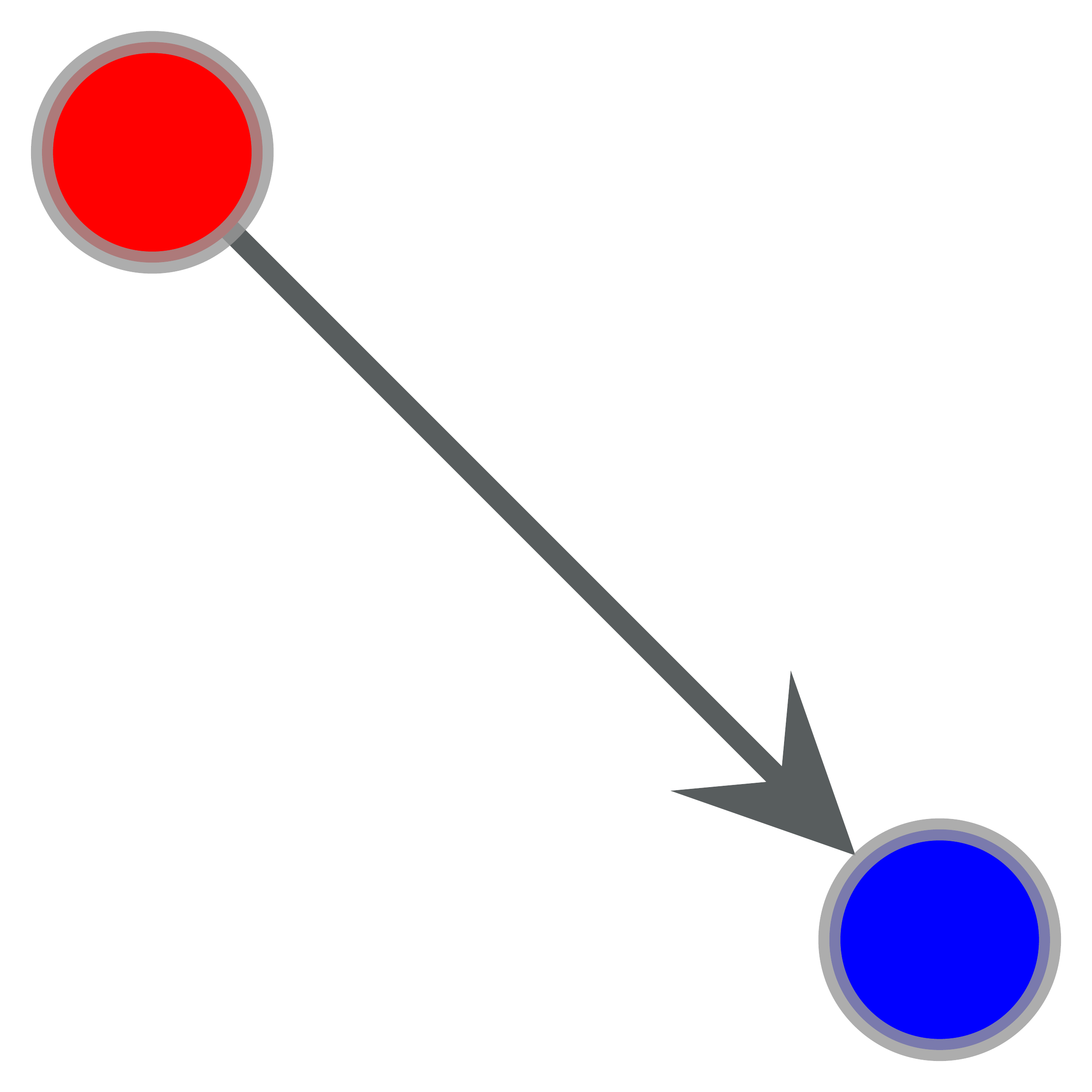}& \includegraphics[height=0.09\textwidth]{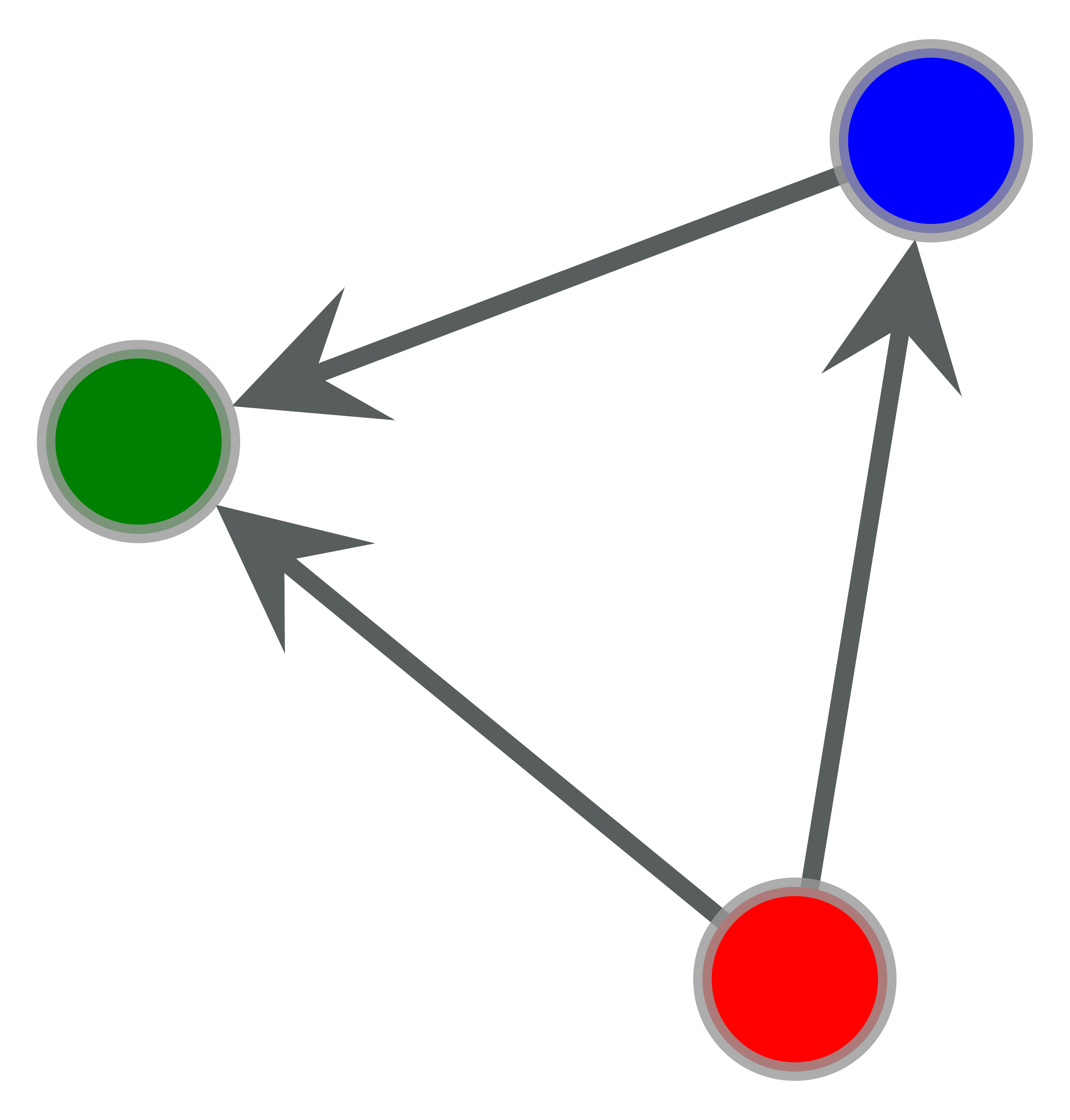}& \includegraphics[height=0.09\textwidth]{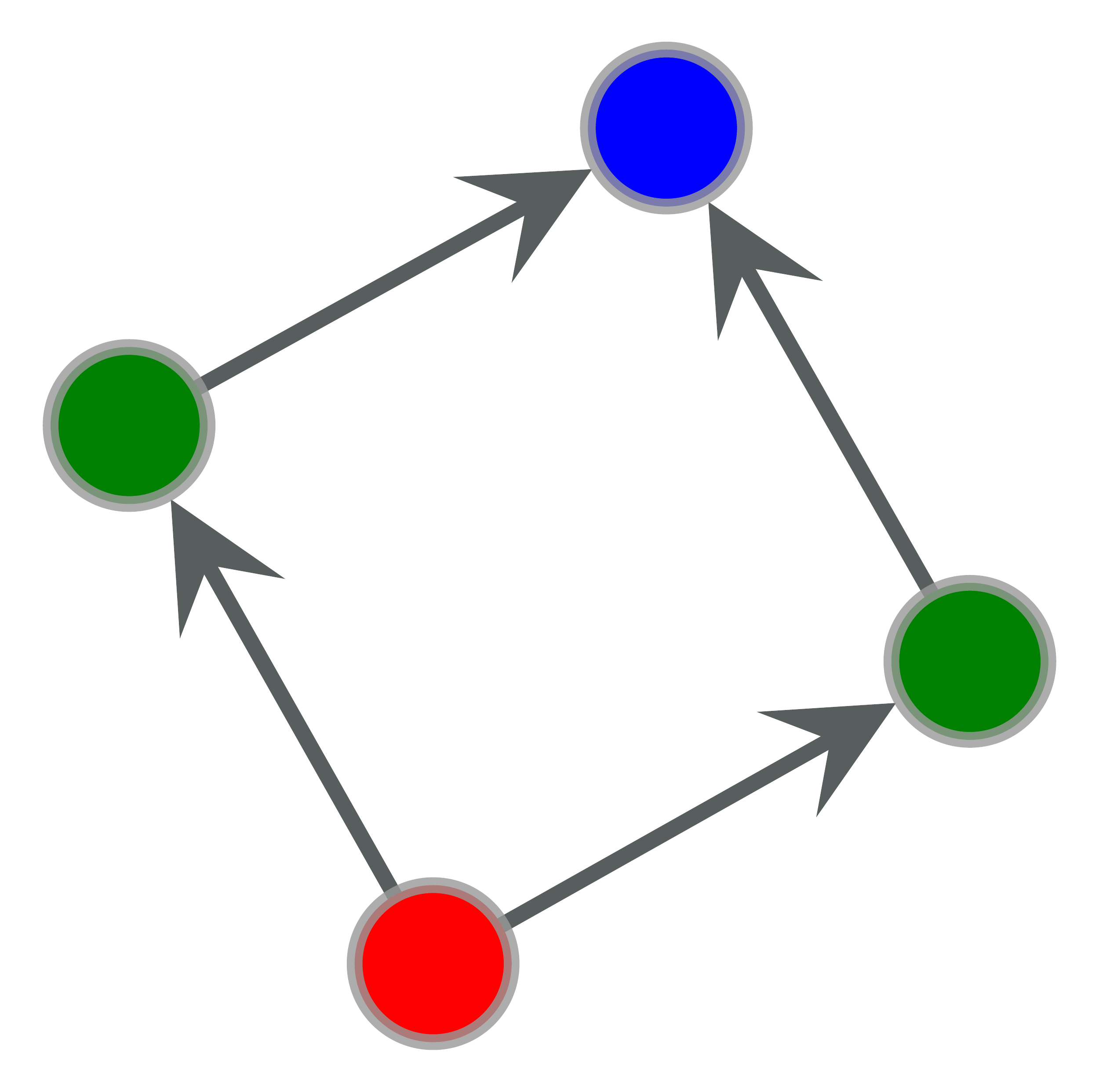}& \includegraphics[height=0.09\textwidth]{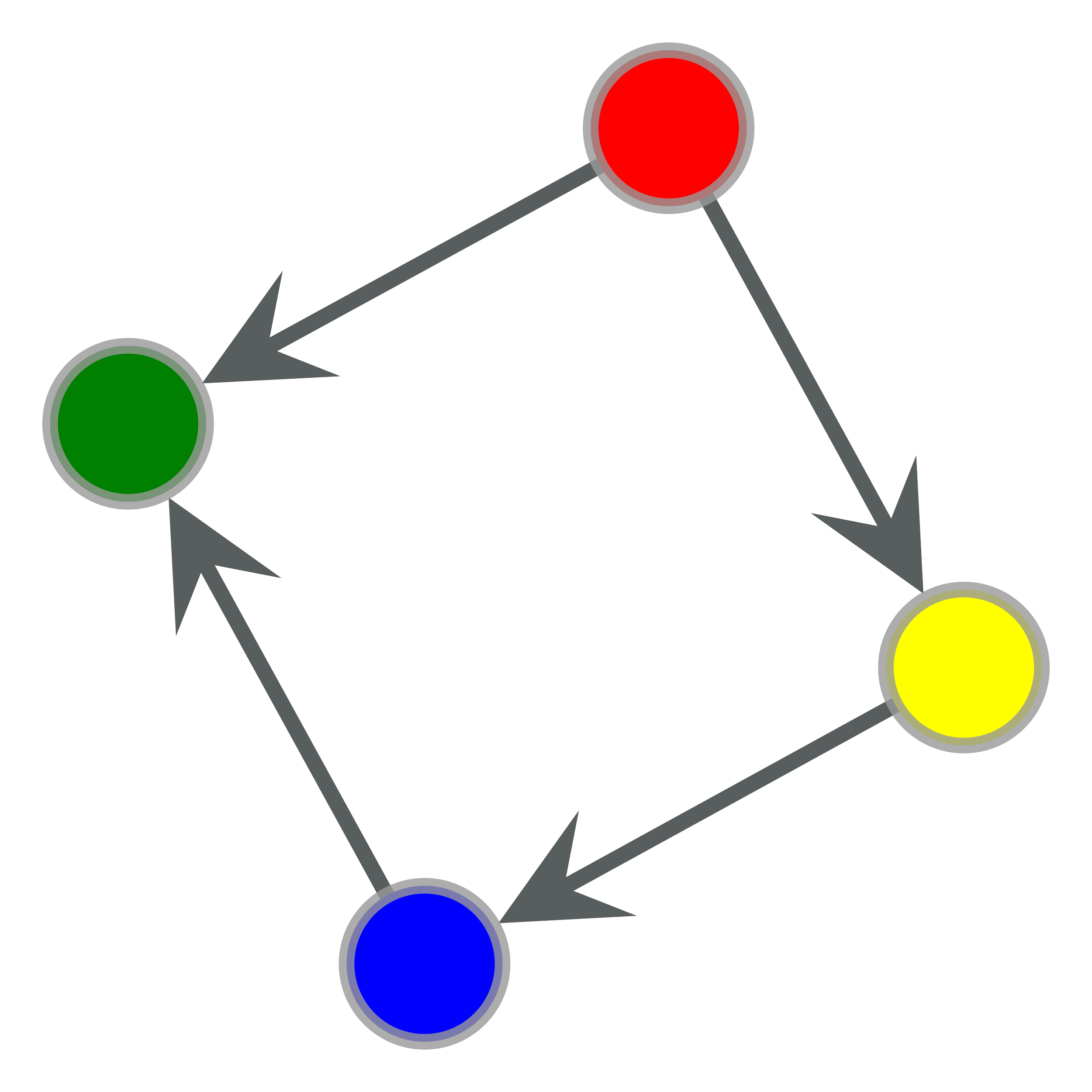}& \includegraphics[height=0.09\textwidth]{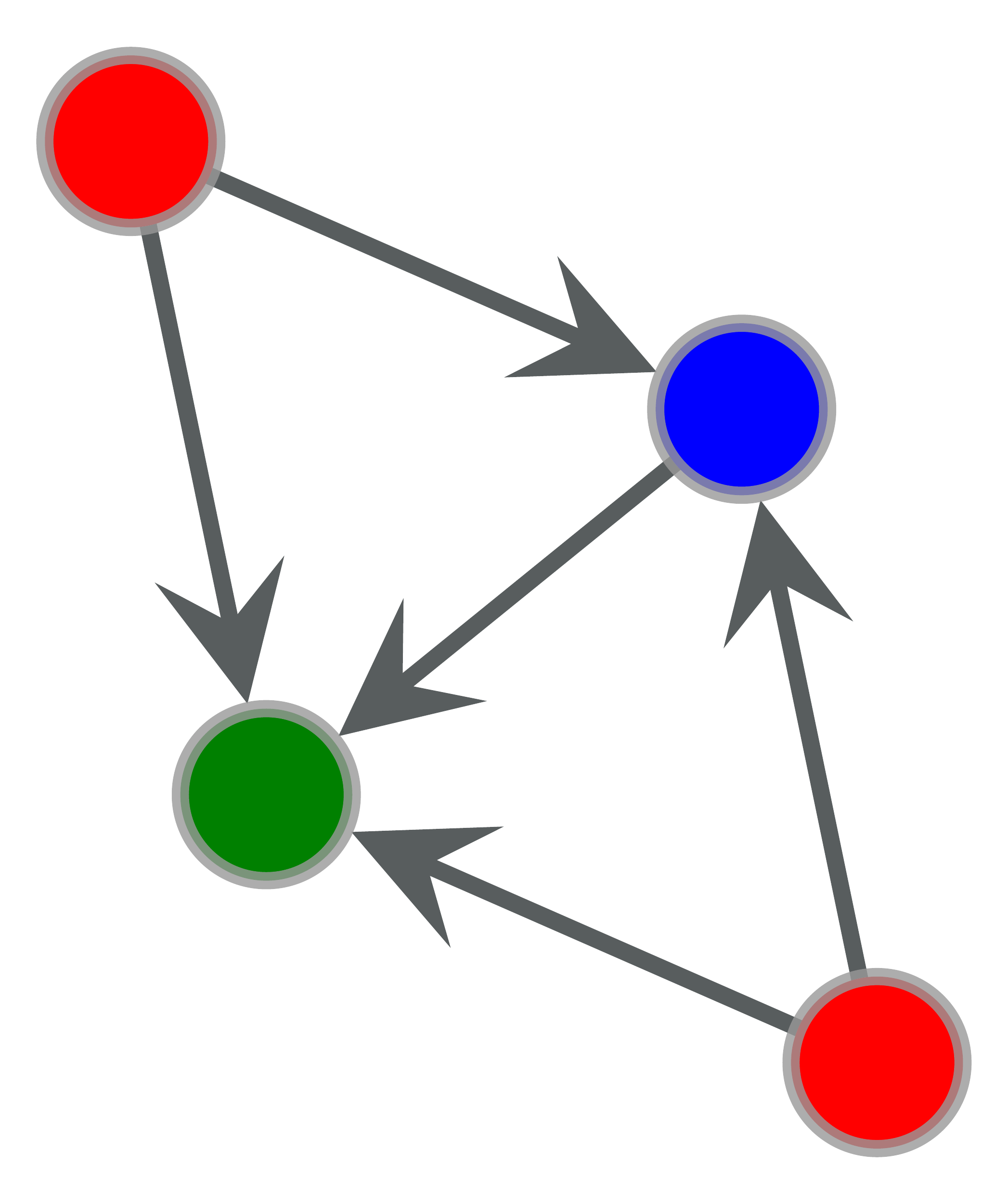}& \includegraphics[height=0.09\textwidth]{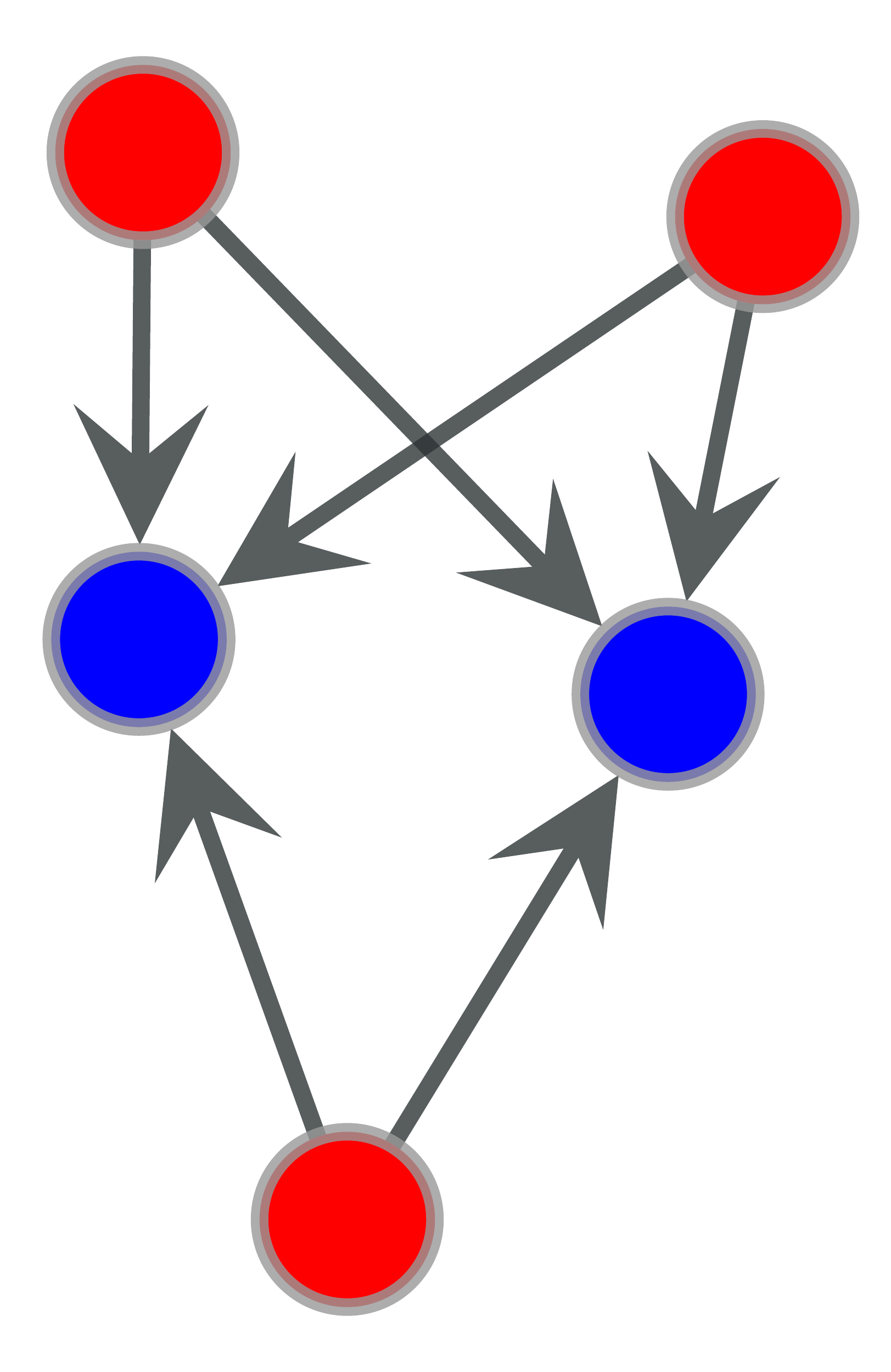}&
\includegraphics[height=0.09\textwidth]{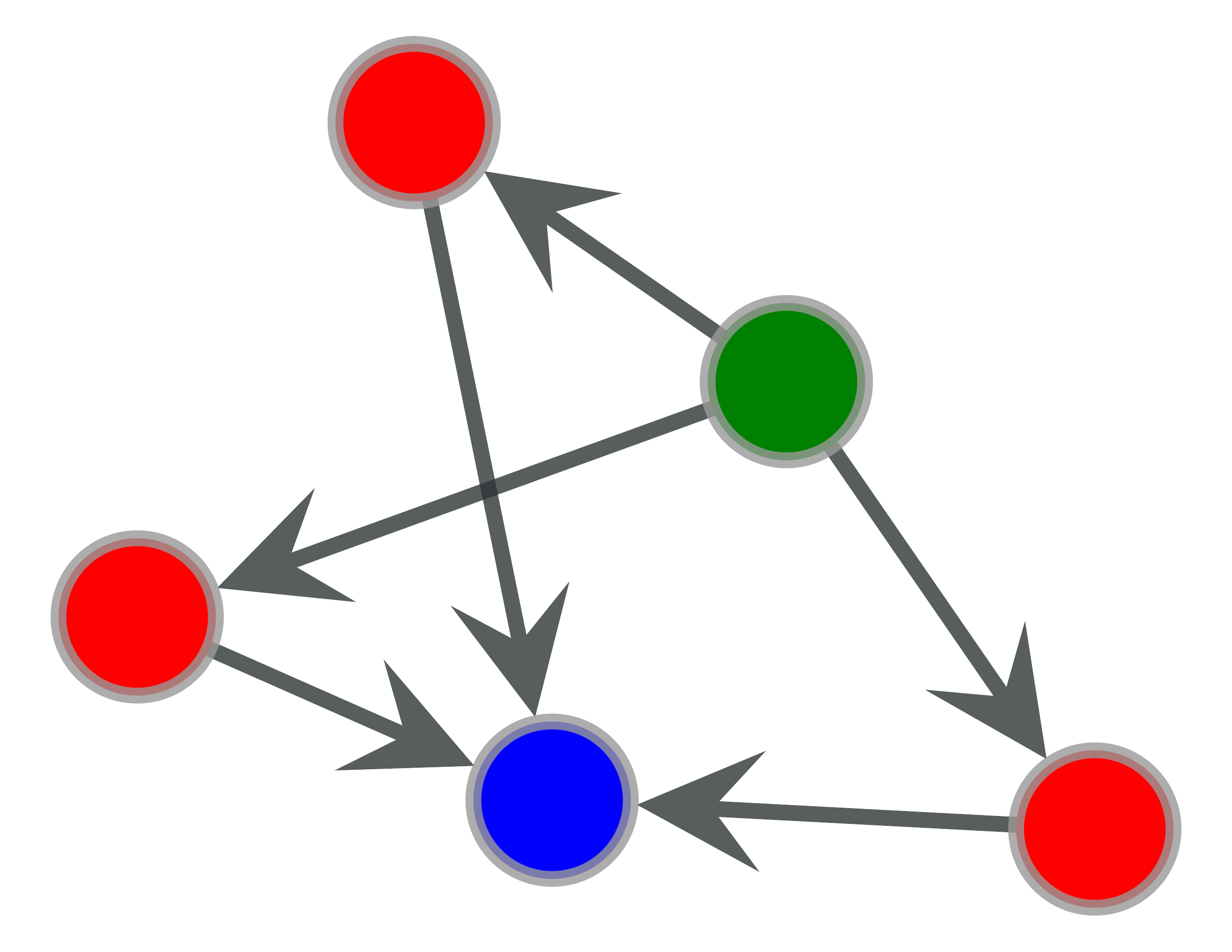}\\

\hline
Id&0& 1& 2& 3& 4& 5& 6\\
\hline
$n_m$&1086& 47& 37& 24& 27& 30& 25\\

\hline
$m$&\includegraphics[height=0.09\textwidth]{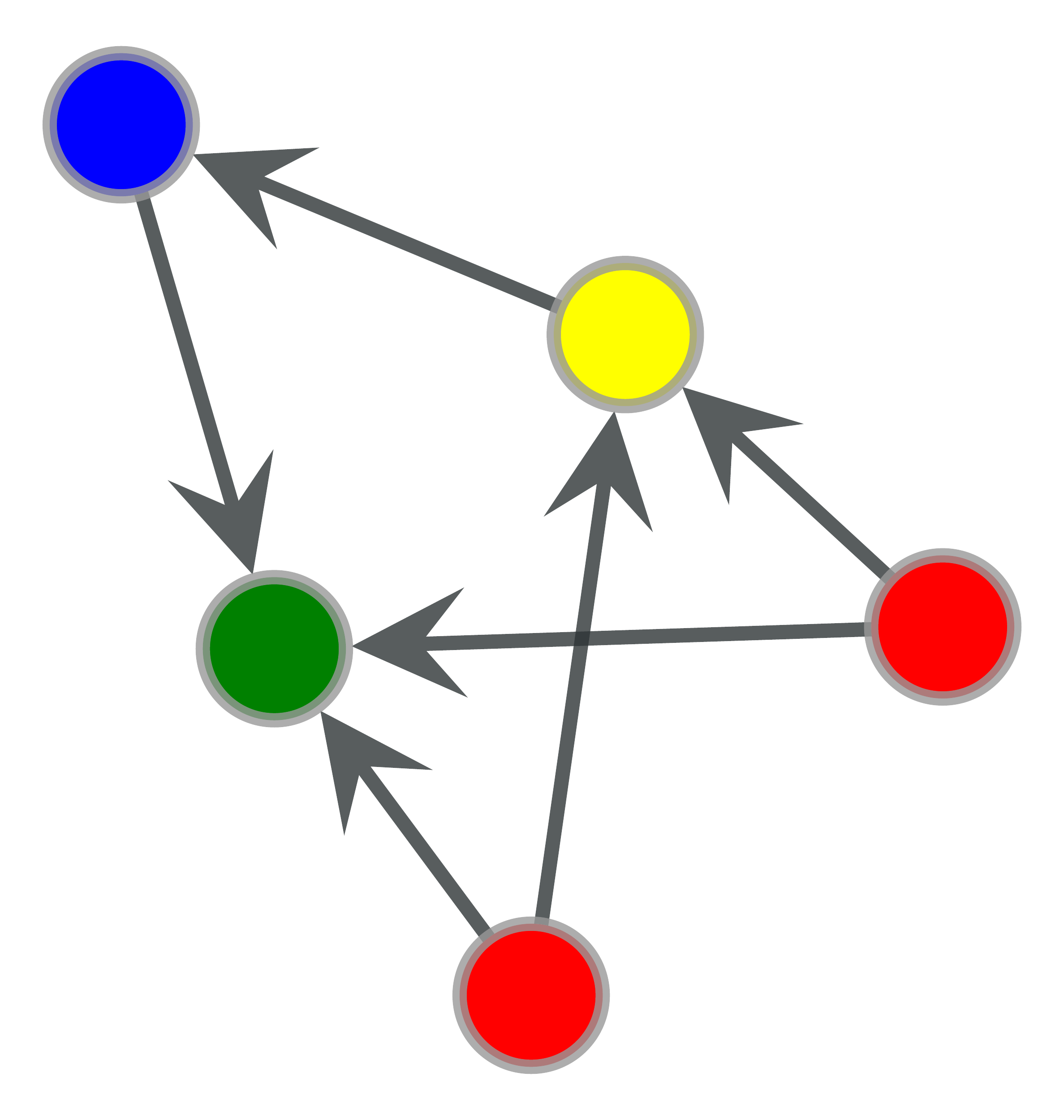}& \includegraphics[height=0.09\textwidth]{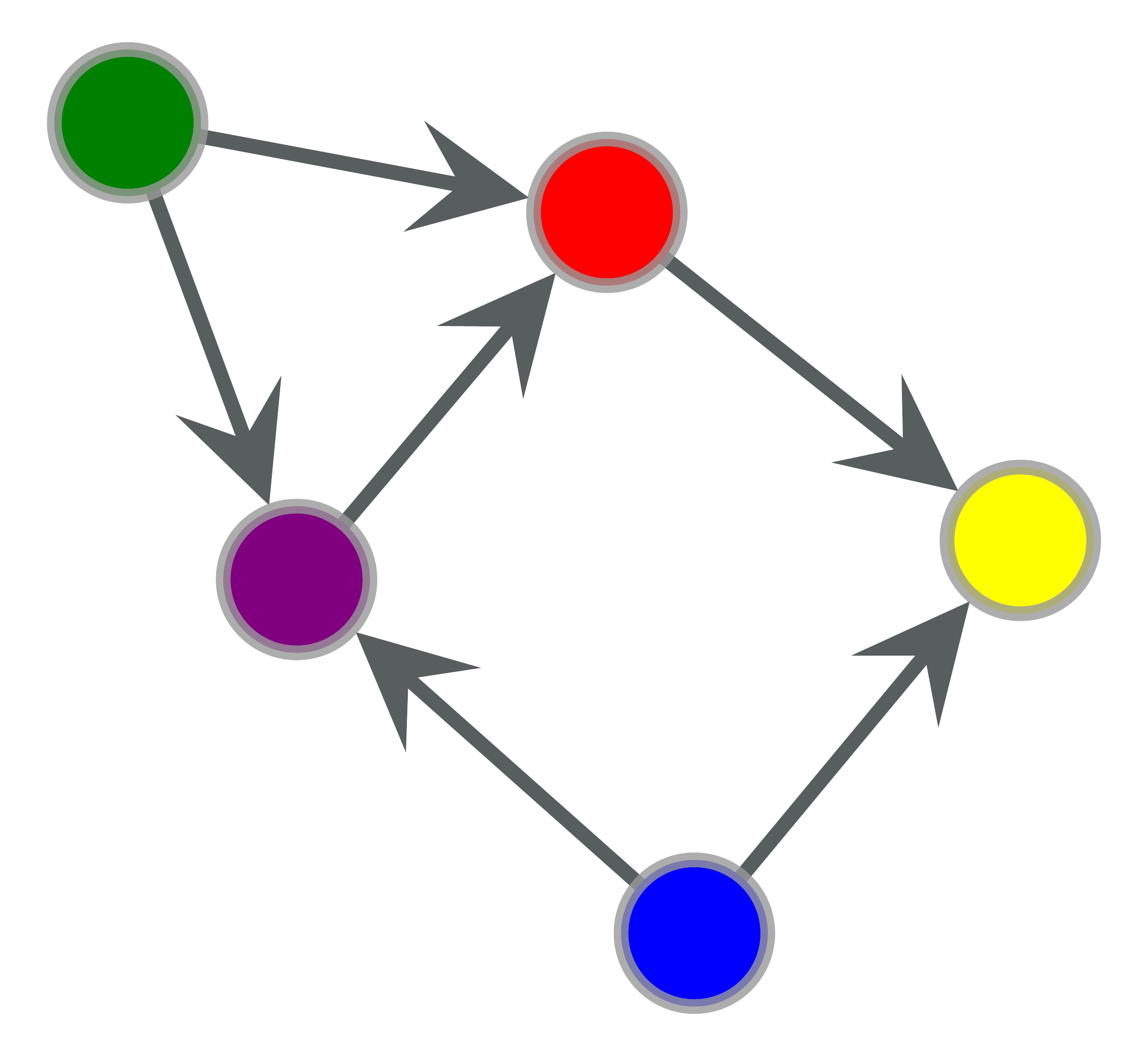}& \includegraphics[height=0.09\textwidth]{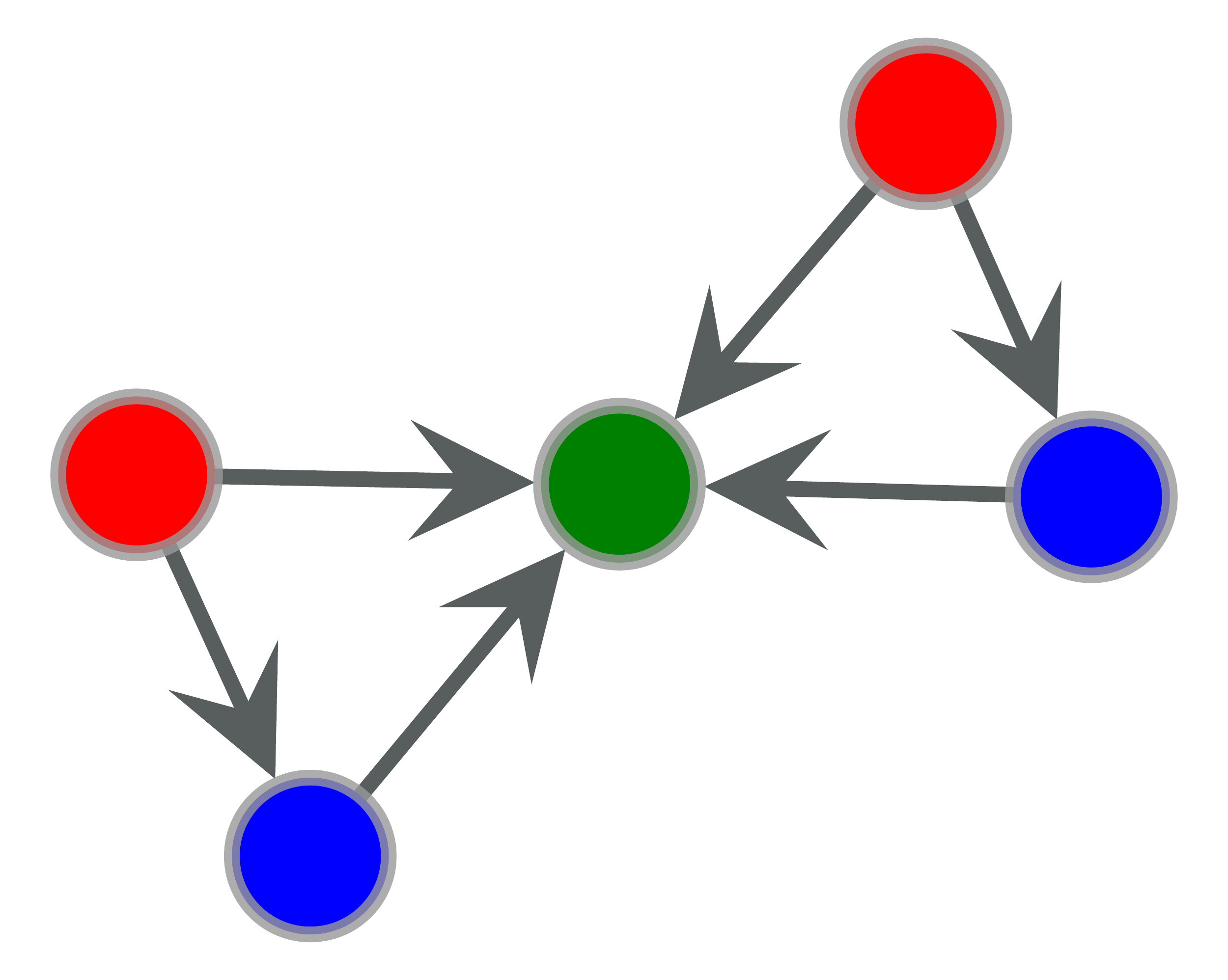}& \includegraphics[height=0.09\textwidth]{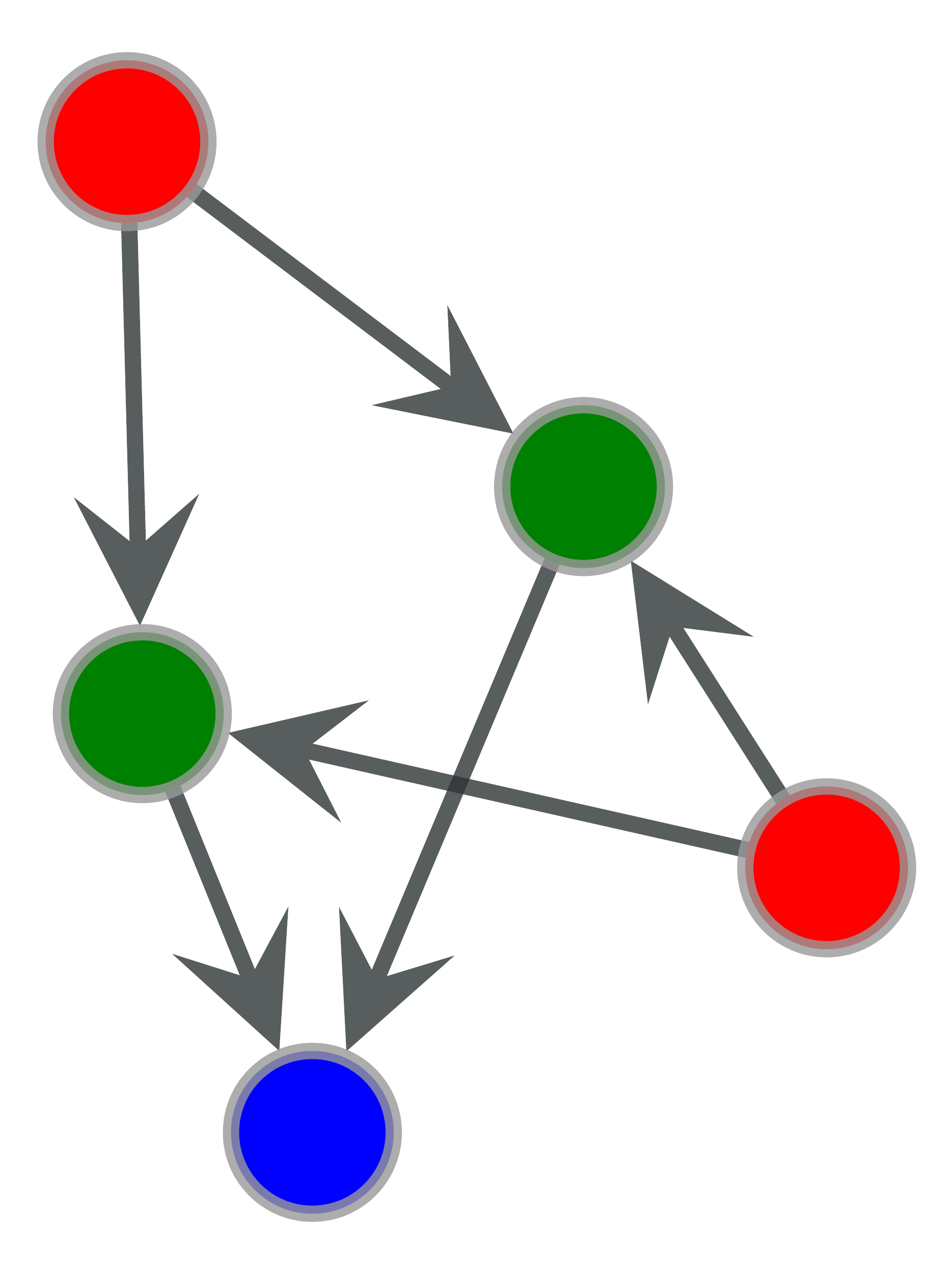}& \includegraphics[height=0.09\textwidth]{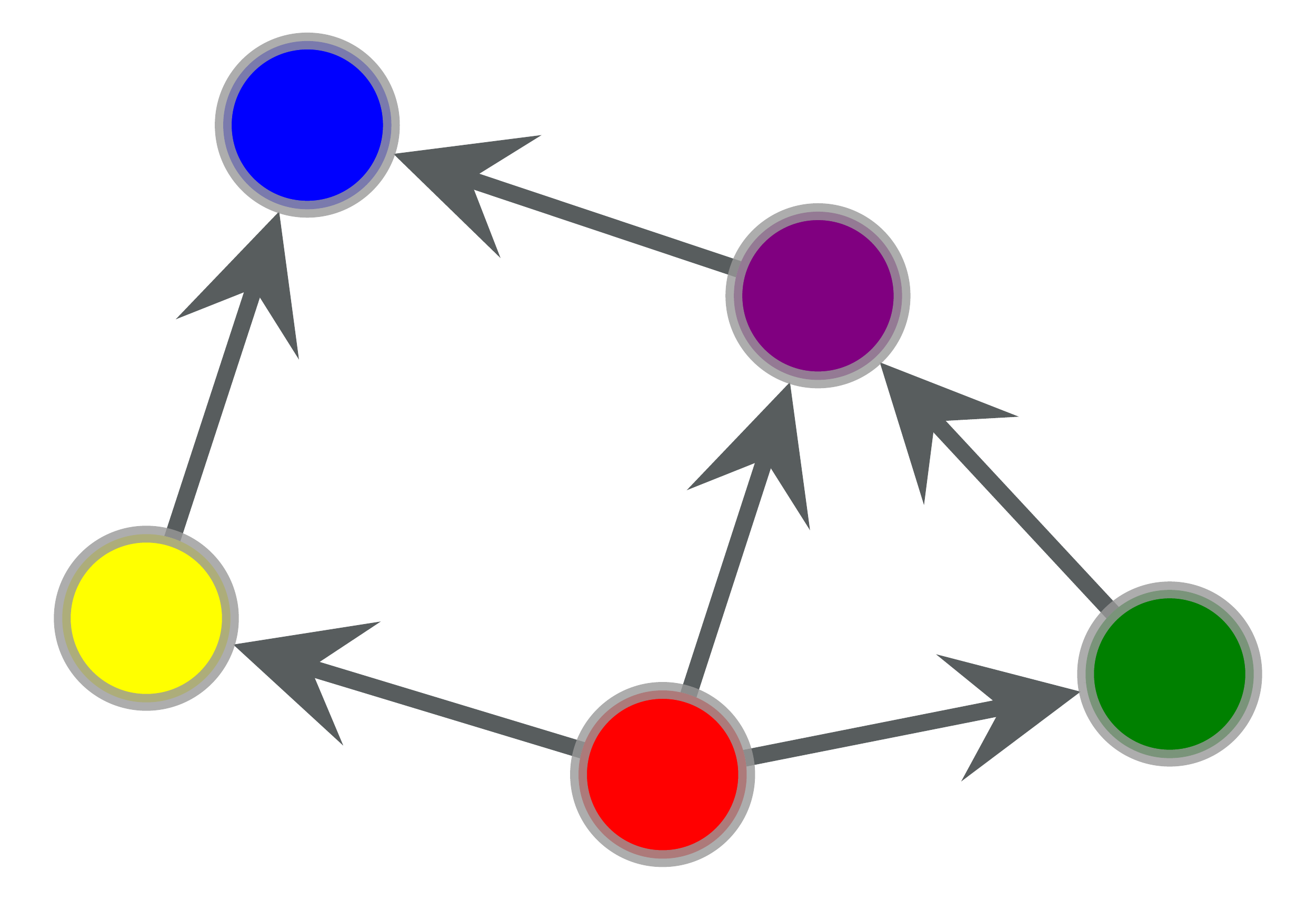}&
\includegraphics[height=0.09\textwidth]{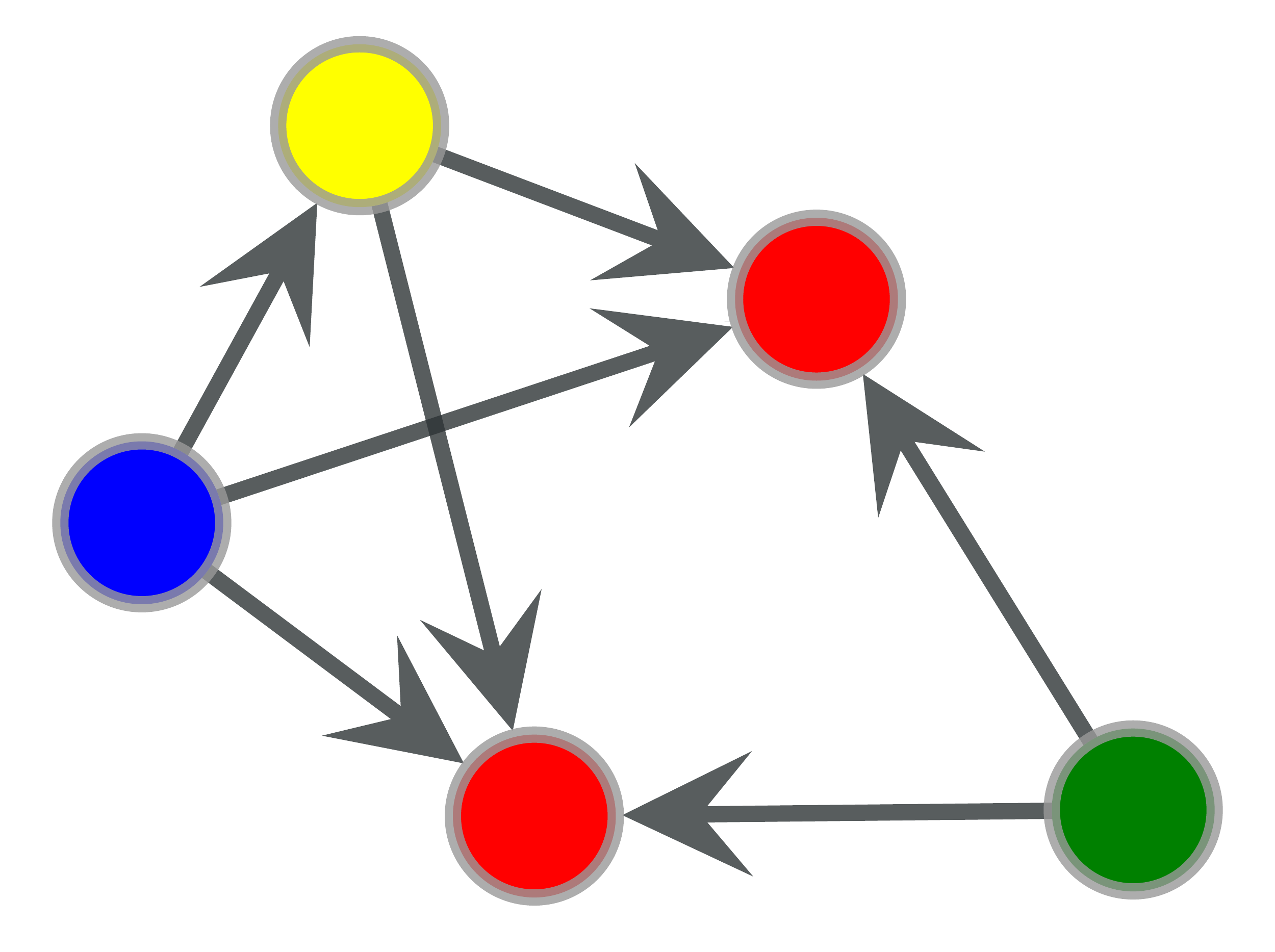}& \includegraphics[height=0.09\textwidth]{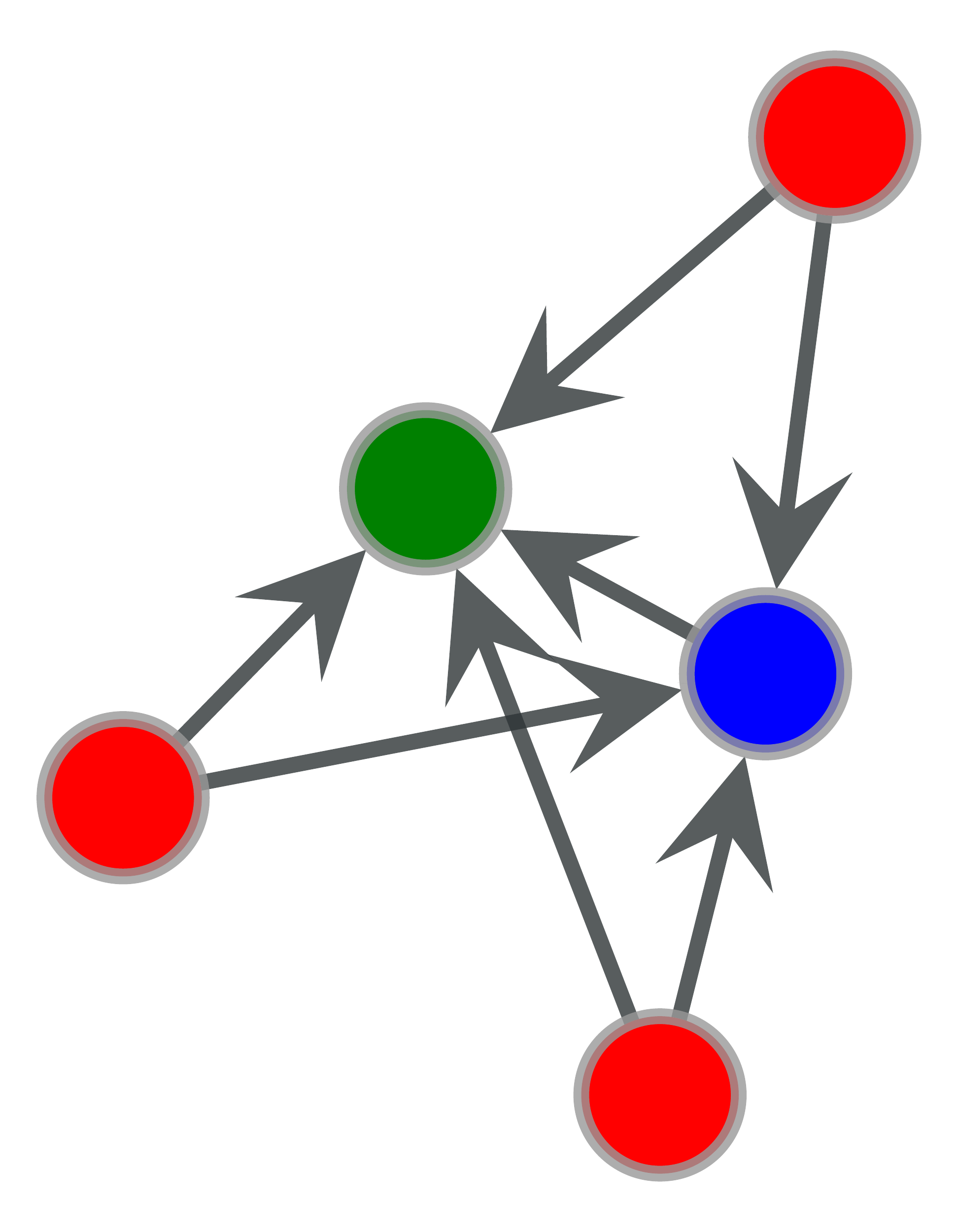}\\

\hline
Id&7& 8& 9& 10& 11& 12& 13\\
\hline
$n_m$&13& 14& 11& 6& 12& 39& 11\\
\hline
$m$& \includegraphics[height=0.09\textwidth]{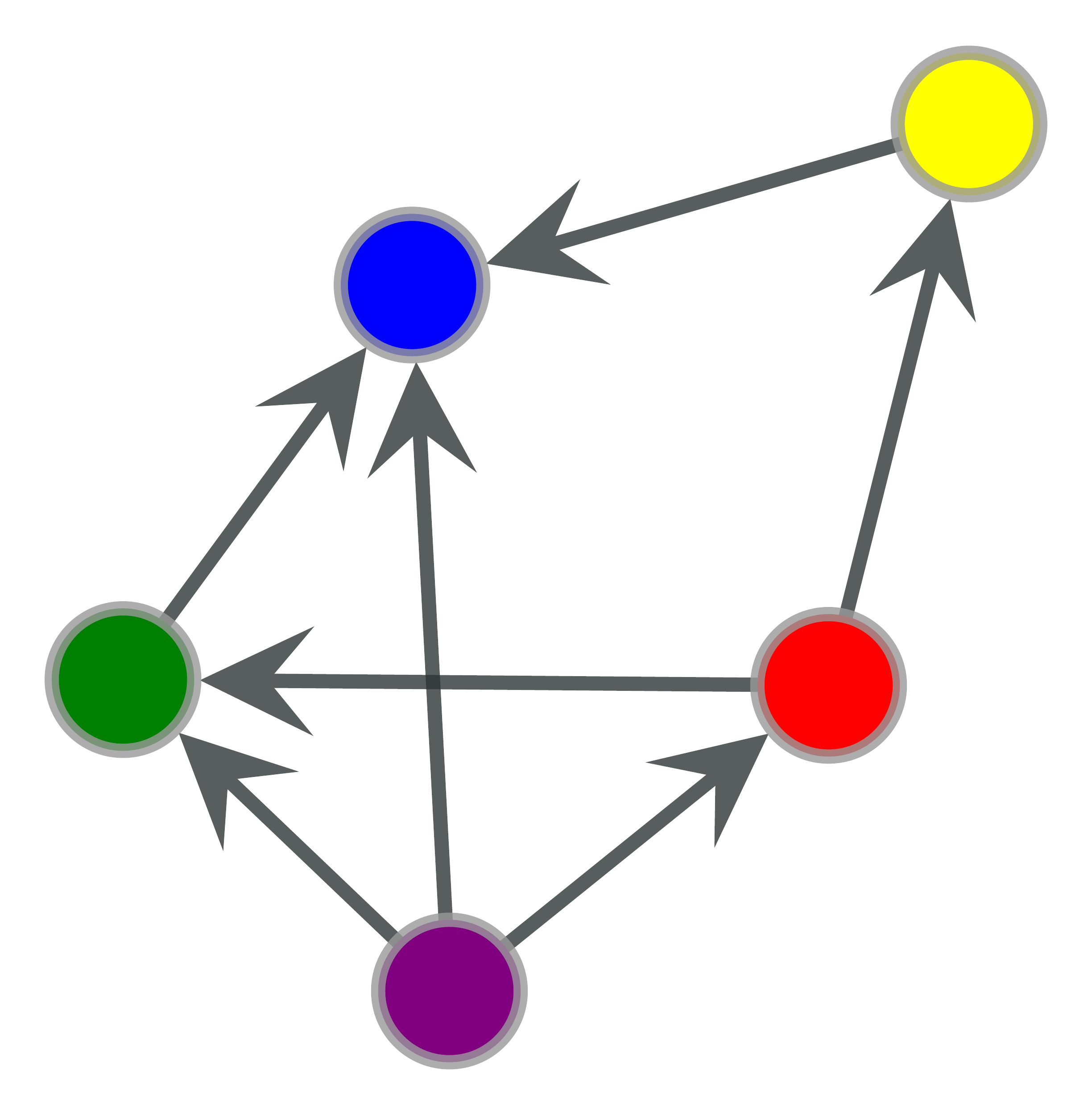}& \includegraphics[height=0.09\textwidth]{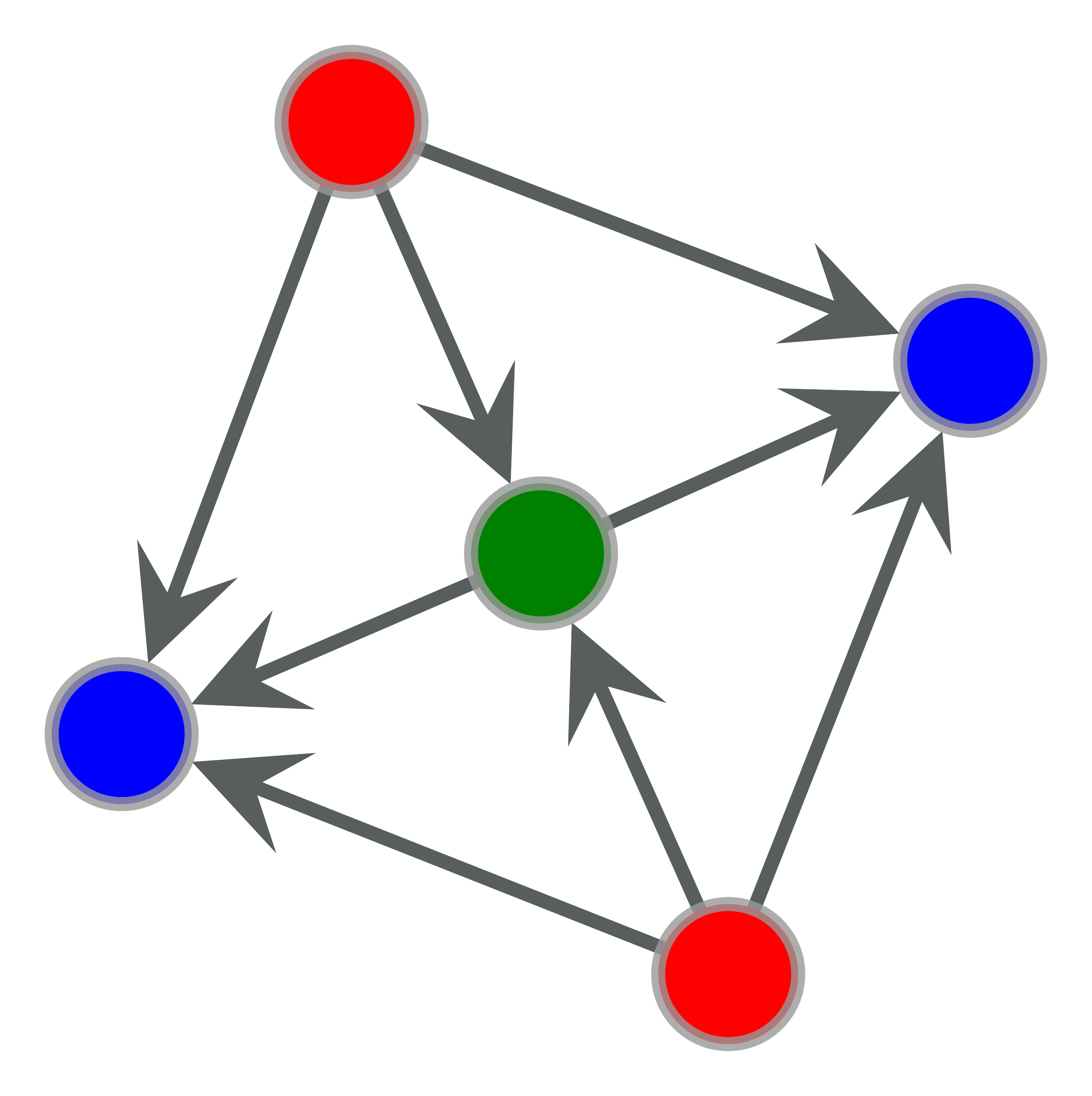}& \includegraphics[height=0.09\textwidth]{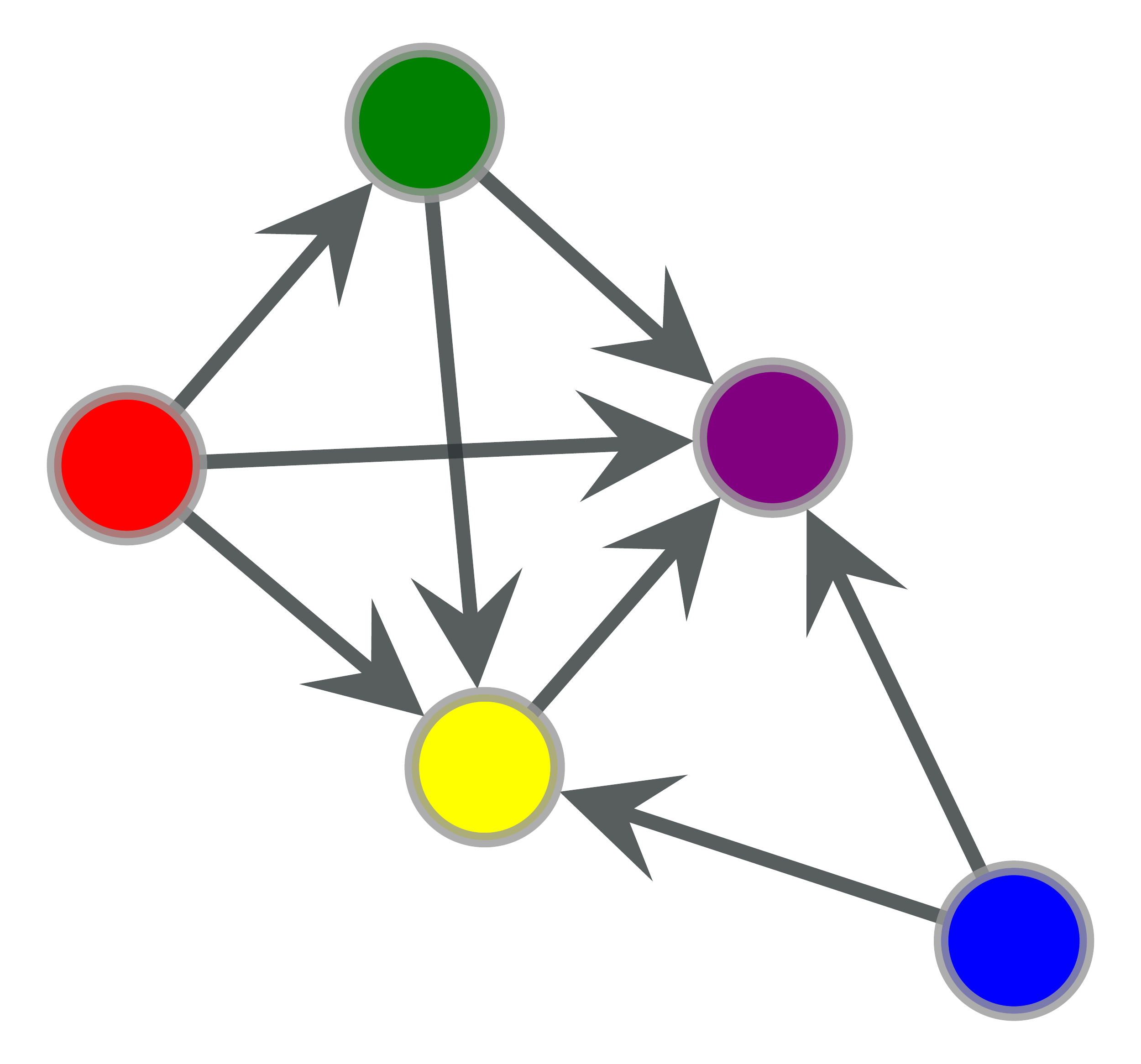}& \includegraphics[height=0.09\textwidth]{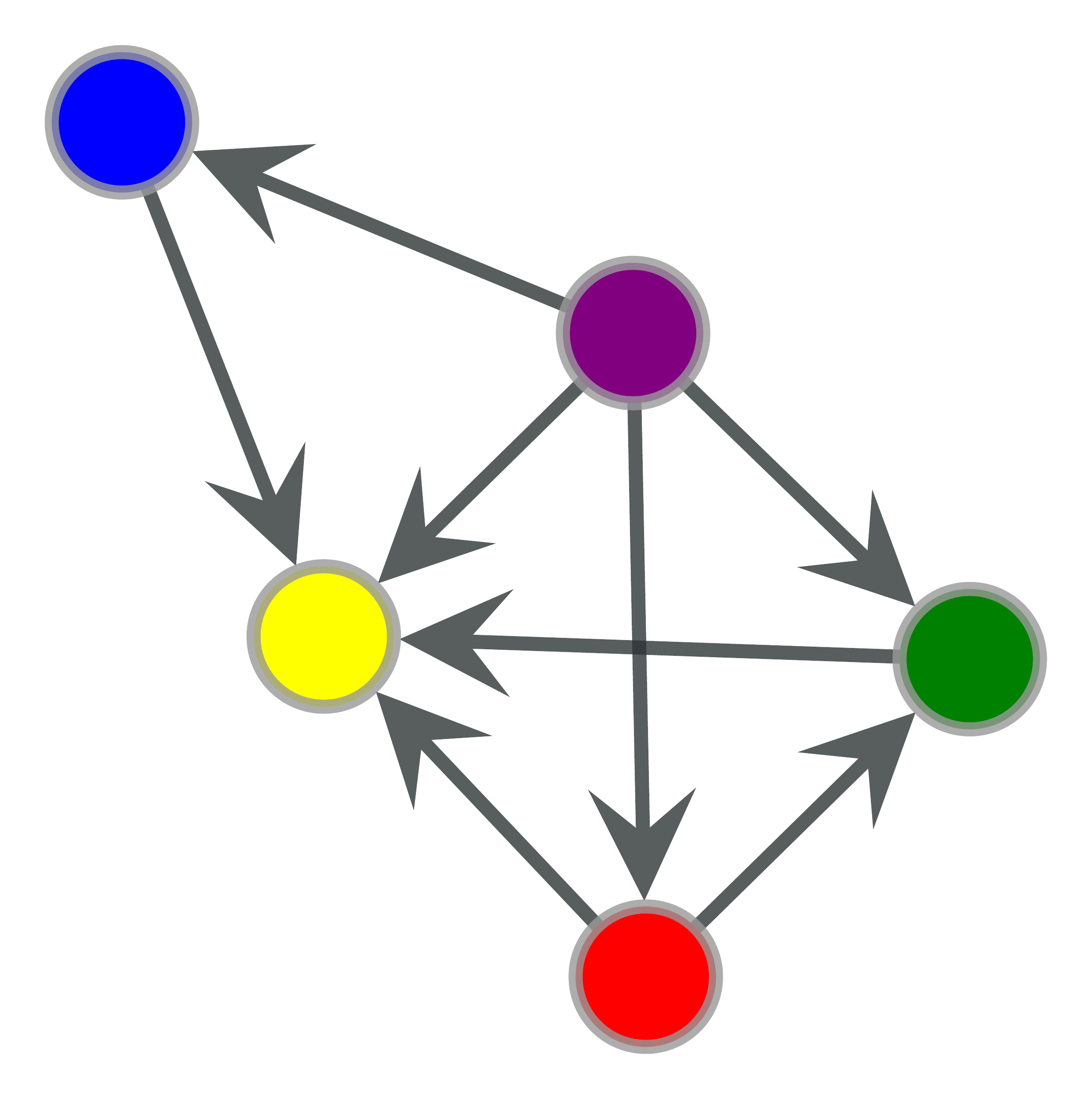}& \includegraphics[height=0.09\textwidth]{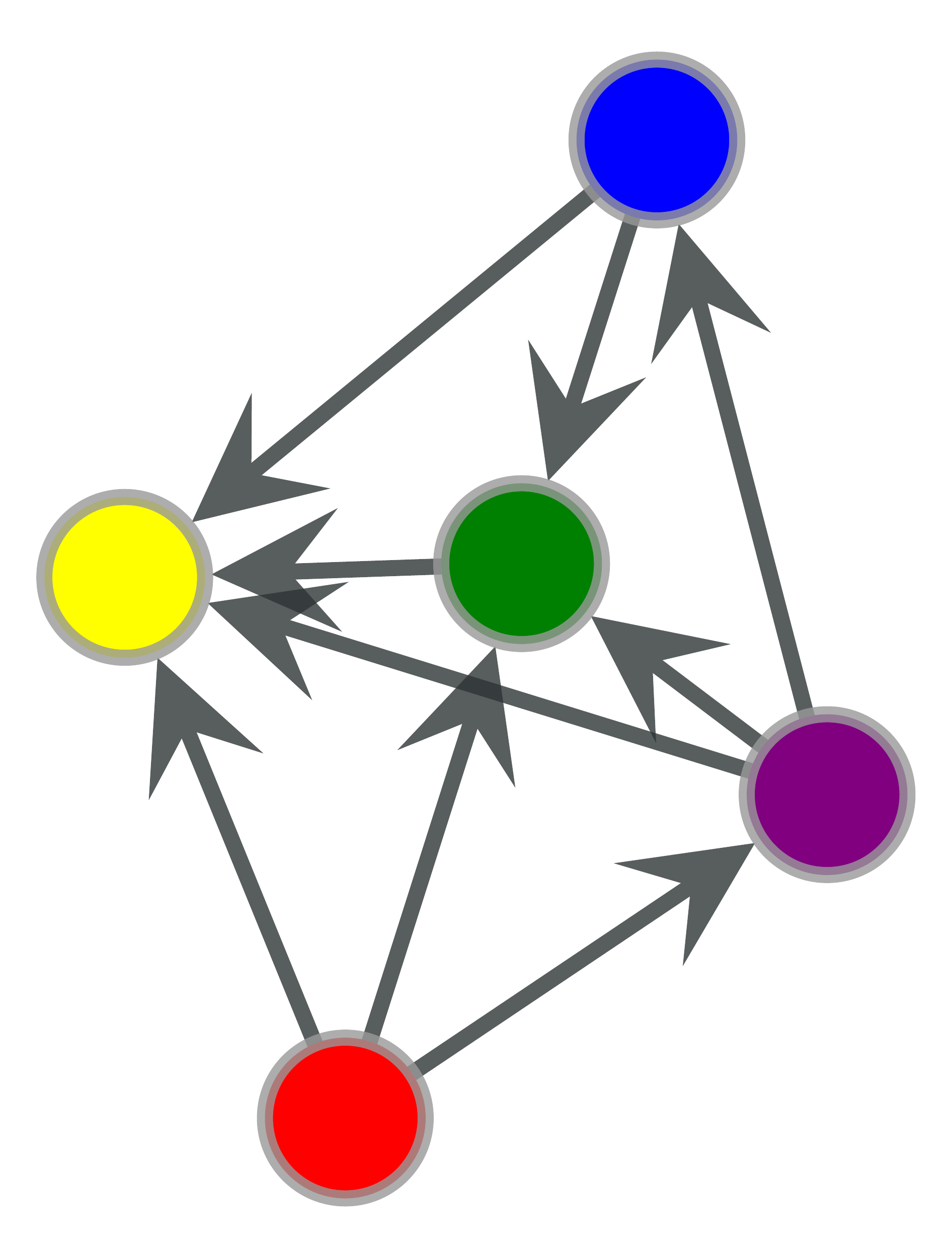}& \includegraphics[height=0.09\textwidth]{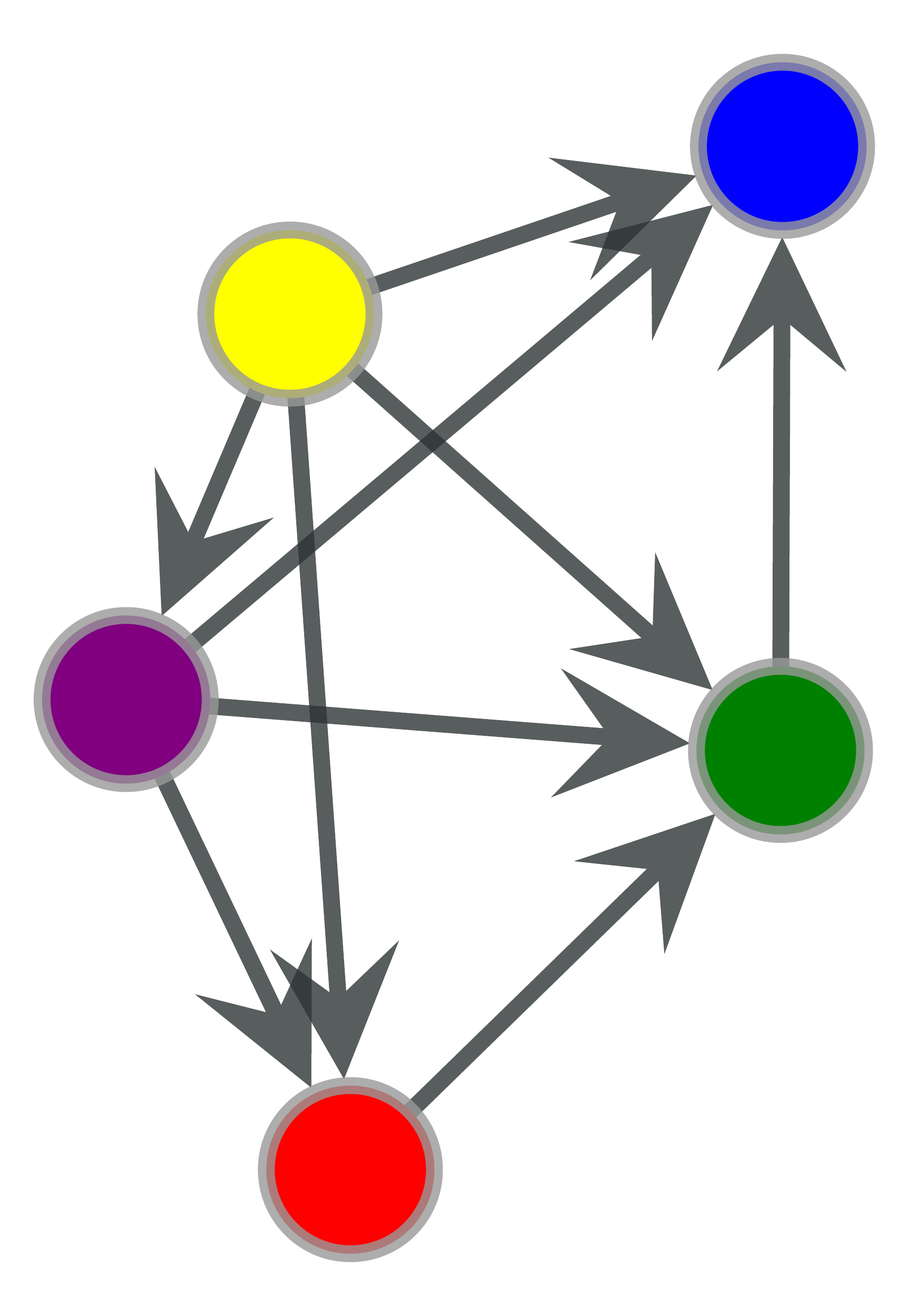}& \includegraphics[height=0.09\textwidth]{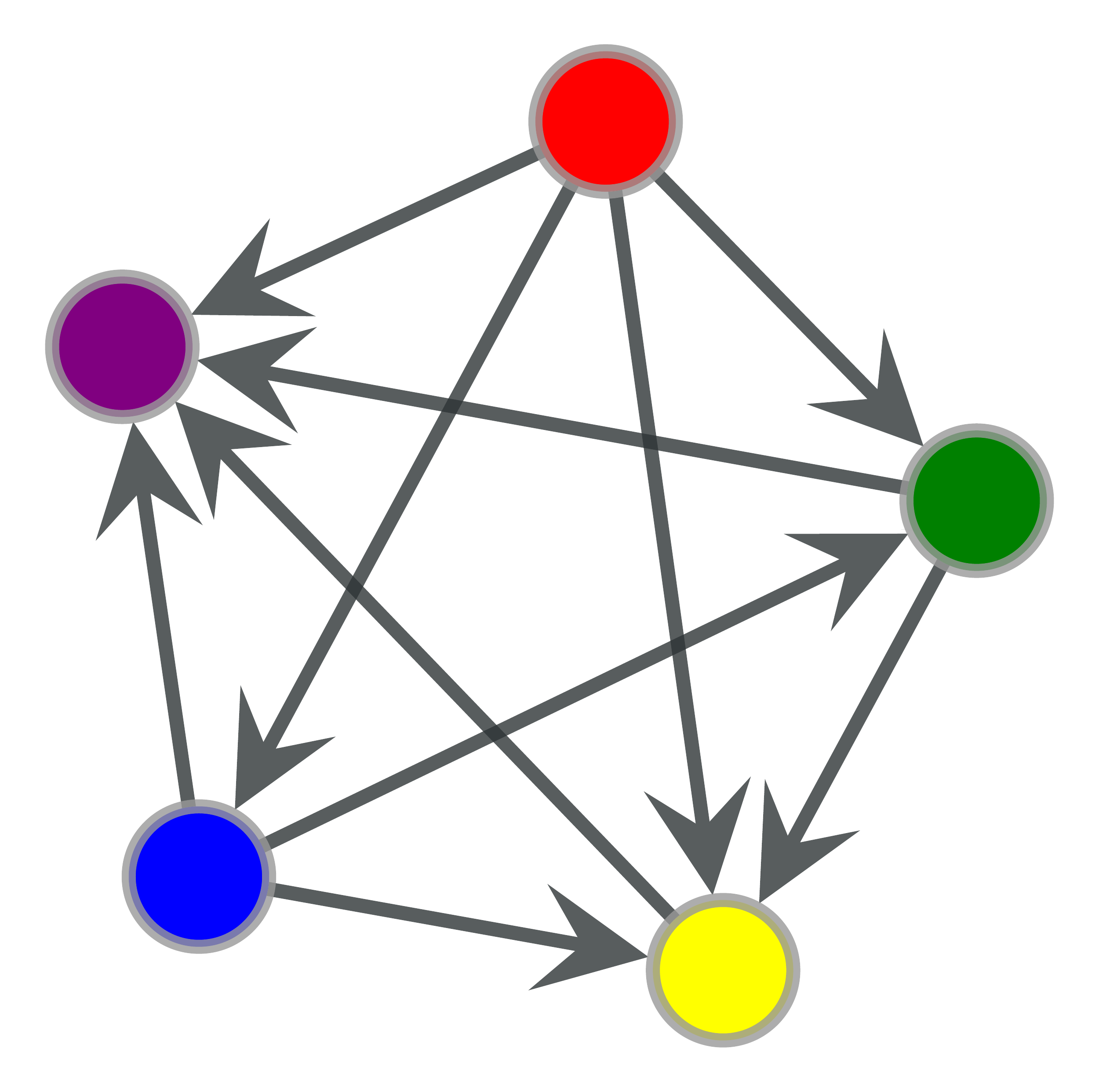}\\

\hline
Id&14& 15& 16& 17& 18& 19& 20\\
\hline
$n_m$&18& 22& 19& 13& 26& 6& 54\\
\hline
\end{tabular}
\end{adjustbox}
\end{center}

\caption{Directed motifs in the subgraph MAP-configuration for the directed connectome of the human brain and their respective counts ($n_m$).}
\label{dibrain}
\end{table}

\subsubsection*{{\em E.coli} Metabolic network}
Next we consider the metabolic network of {\em E.coli} which is a part of a collection of 43 metabolic networks from \cite{meta}. For the {\em E.coli} metabolic network the MAP-configuration contains 11 non-trivial atoms (Table \ref{metaM}) which cover about 97\% of the edges. The MAP configuration  for the \emph{E.coli} metabolic network has strong resemblance to a decomposition of the network into metabolic pathways. We find that the atoms in the MAP configuration of the metabolic network capture important features of metabolism and include motifs commonly used in biochemistry to describe metabolic pathways. For instance cycles (Atoms 3 and 9 in Table \ref{metaM}) and long bidirectional paths (Atom 10 in Table \ref{metaM}) are known to play an crucial role in regulating core metabolic pathways such as Gluconeogenesis \cite{Hanson2013Gluconeogenesis} and substrate cycles \cite{Curi2016RegulatoryNow}. We also find atoms that can be interpreted as irreversible pathways (Atoms 2 and 6 in Table \ref{metaM}) where one metabolite is transformed into another in a controlled step-way fashion and atoms that can be interpreted as breaking down complex molecules into simpler ones which are then in turn combined to construct other larger molecules (Atoms 4 and 8 in Table \ref{metaM}). We also analysed the remaining 42 metabolic networks in the data set \cite{meta} and found that the method finds very similar atoms across the networks in the data set (See SI).
\begin{table}[h!]\label{metamm}
\begin{center}
\begin{adjustbox}{max width=\textwidth}

\begin{tabular}{|c|c|c|c|c|c|c|}
\hline

$m$&\includegraphics[height=0.09\textwidth]{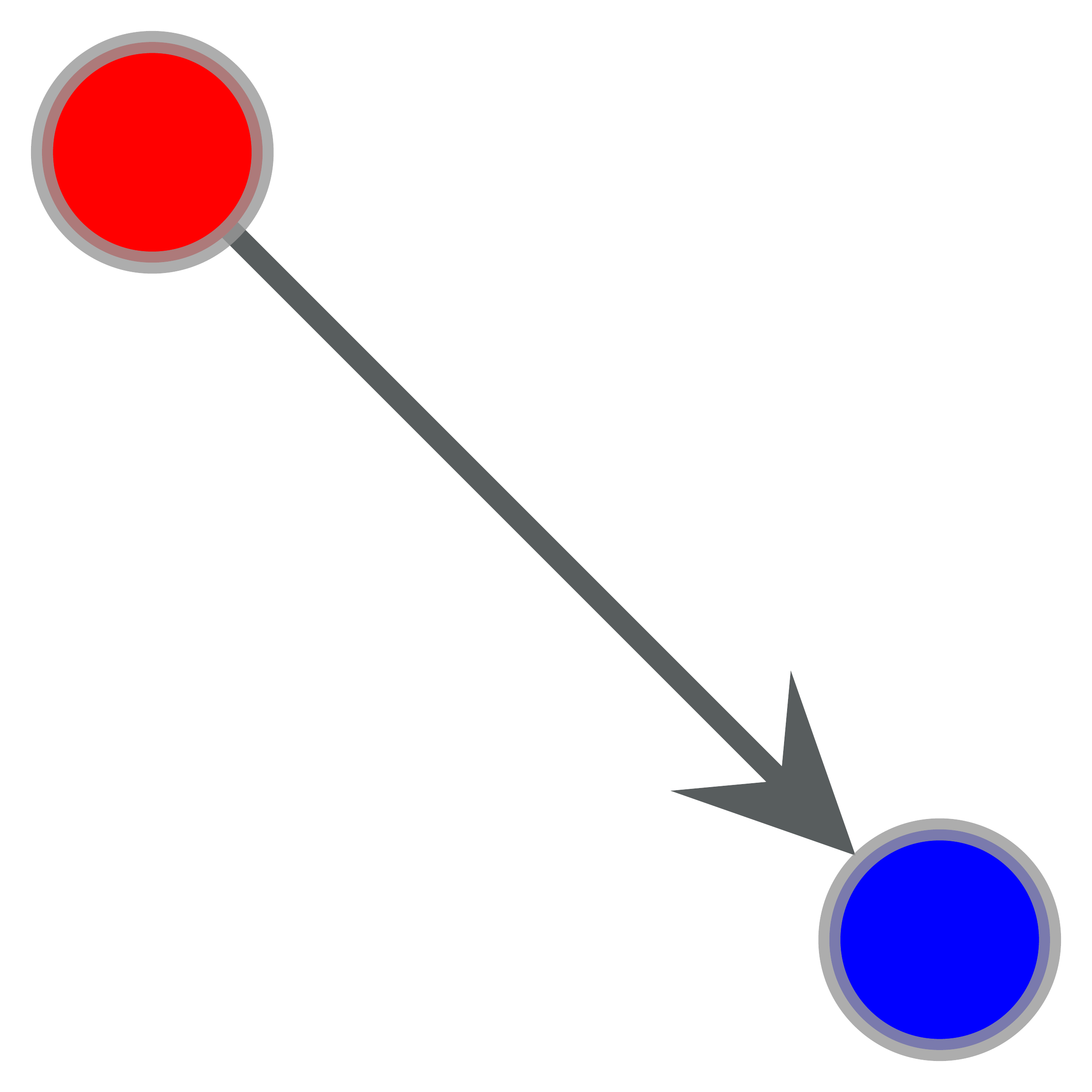}& \includegraphics[height=0.09\textwidth]{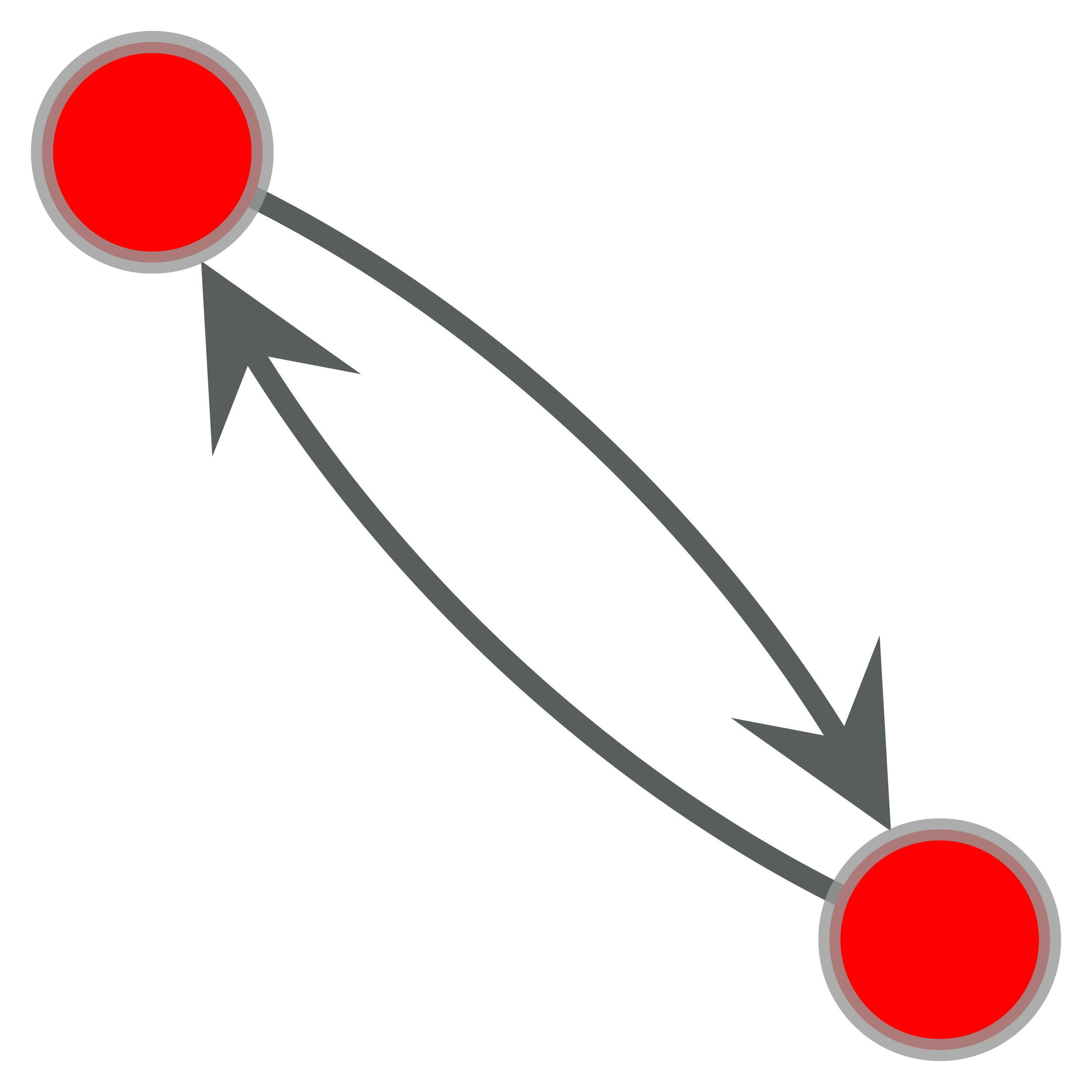}& \includegraphics[height=0.09\textwidth,angle=45]{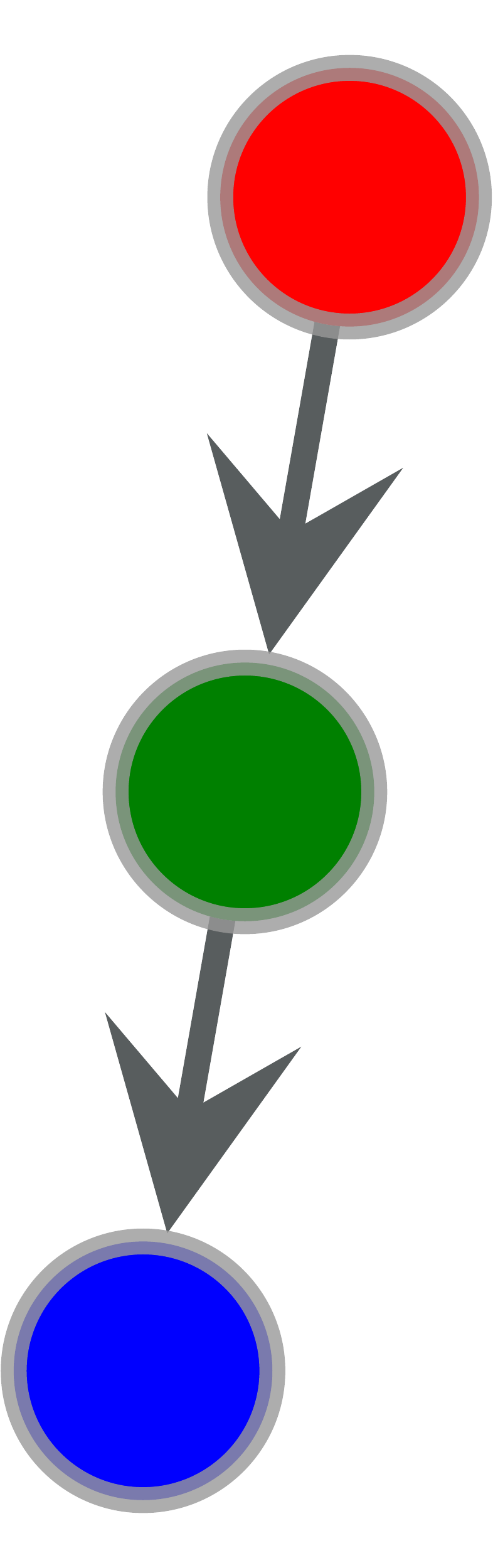}& \includegraphics[height=0.09\textwidth]{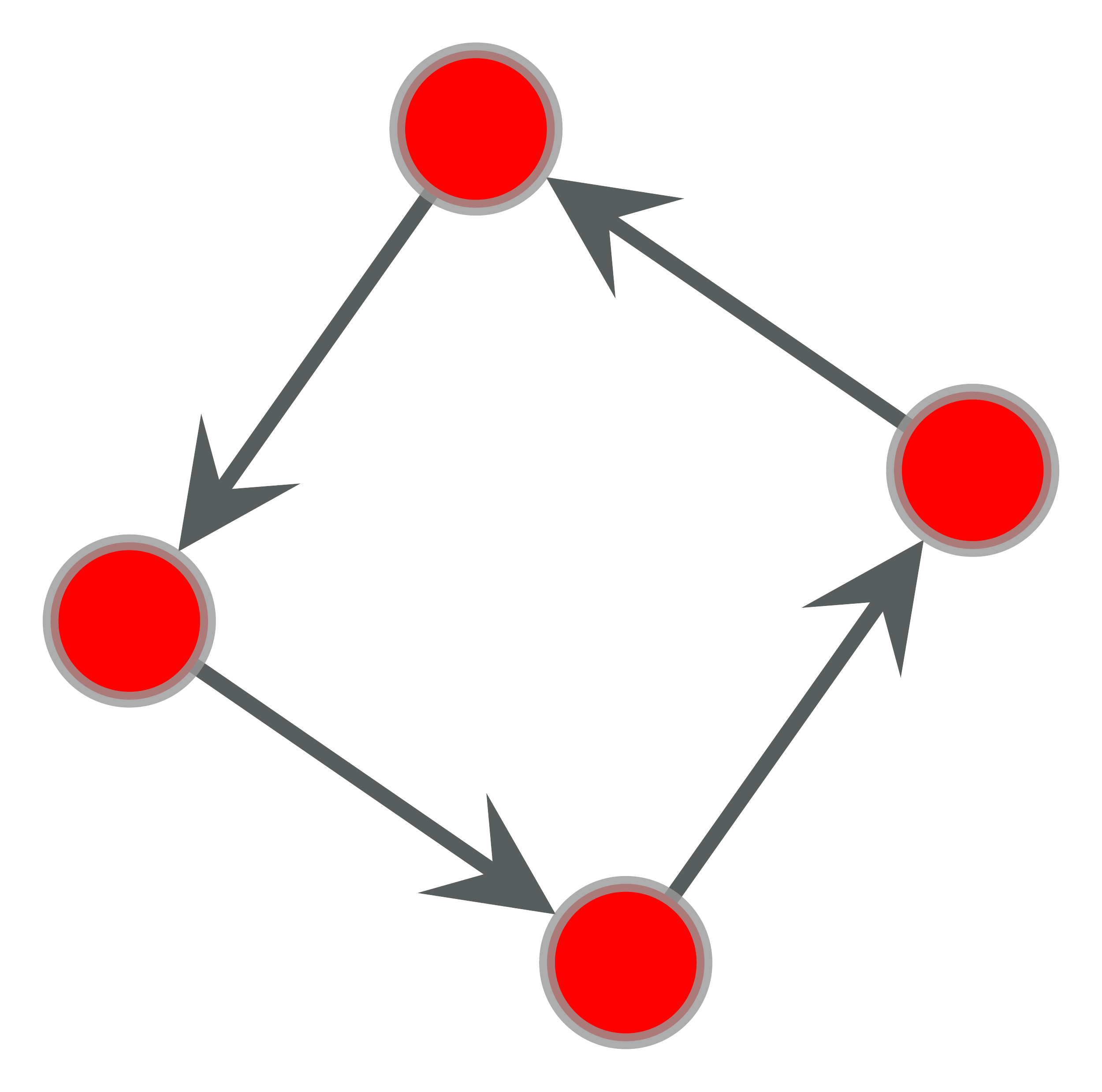}& \includegraphics[height=0.09\textwidth]{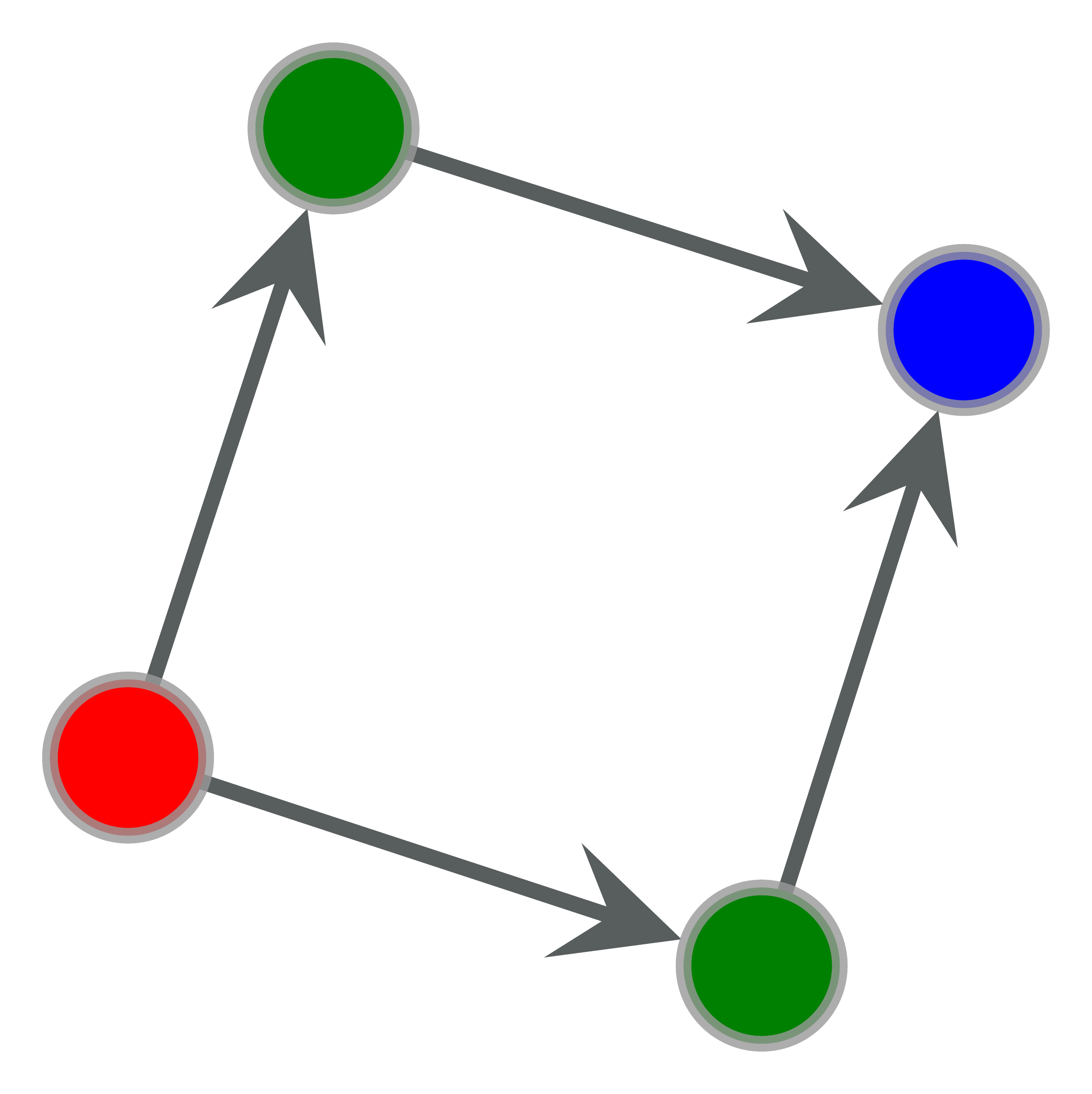}& \includegraphics[height=0.09\textwidth]{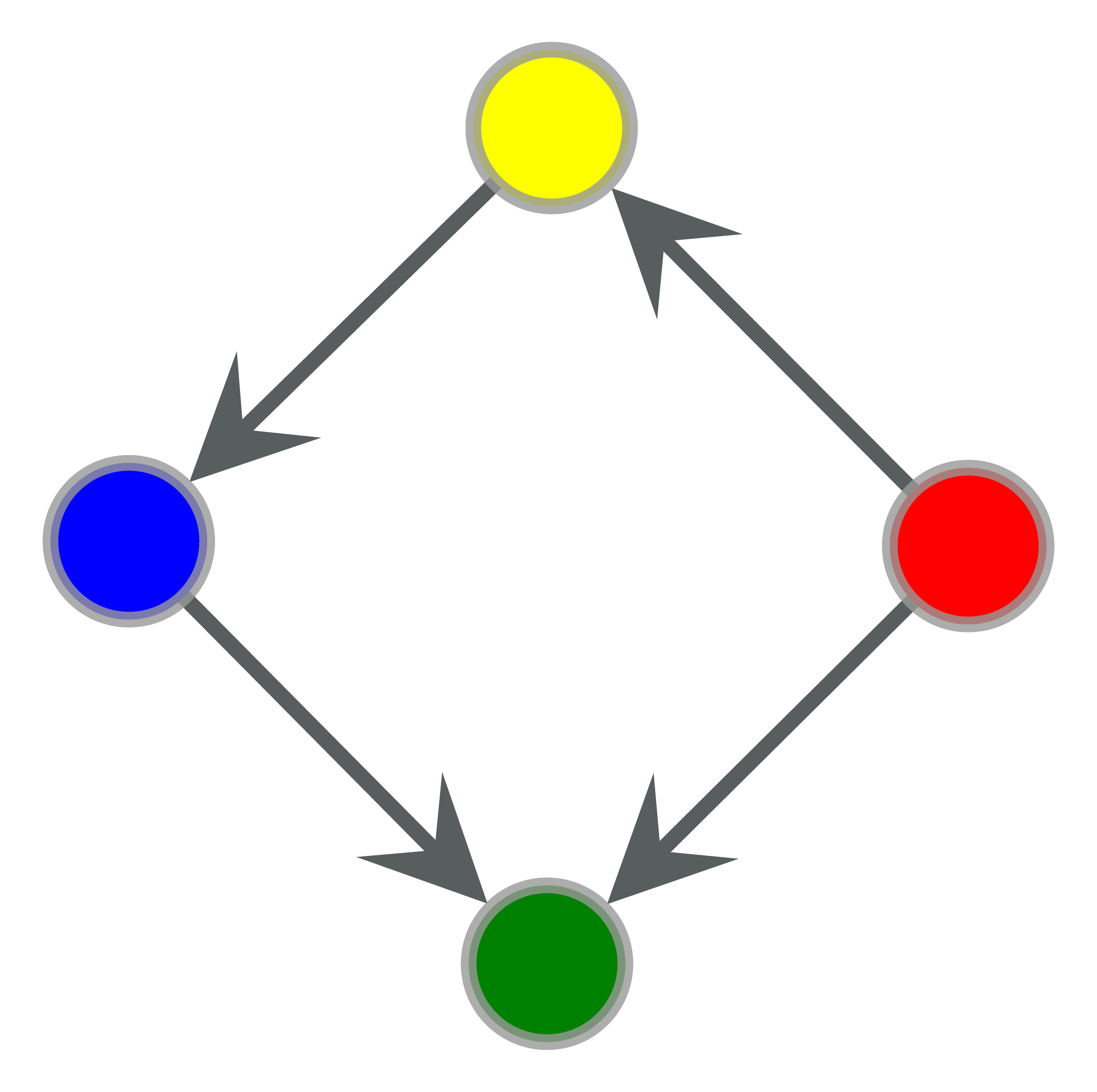}\\
\hline
Id&0&1&2&3&4&5\\
\hline
$n_m$&135& 51& 145& 262& 82& 7\\

\hline
$m$& \includegraphics[height=0.09\textwidth]{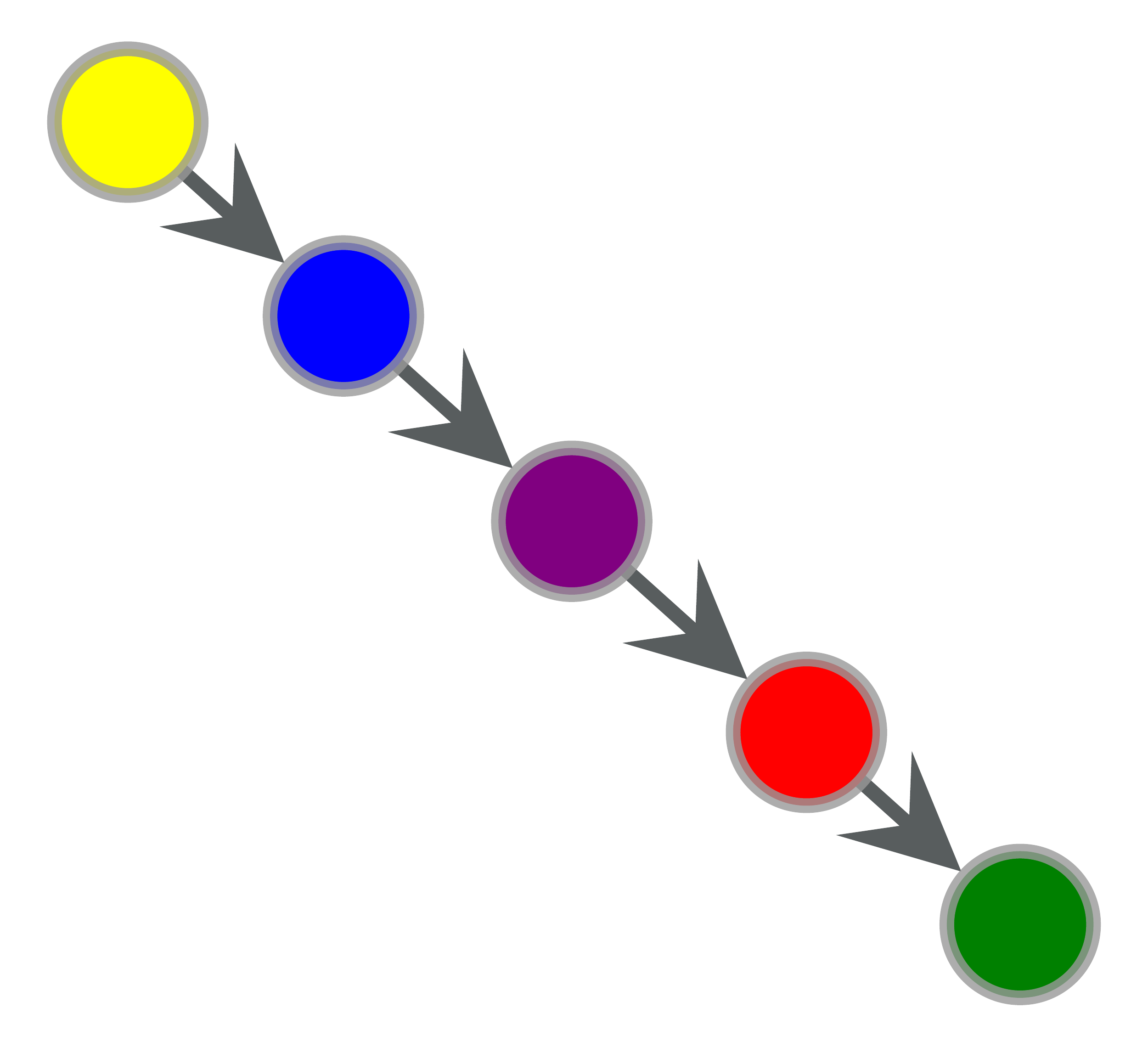}&\includegraphics[height=0.09\textwidth]{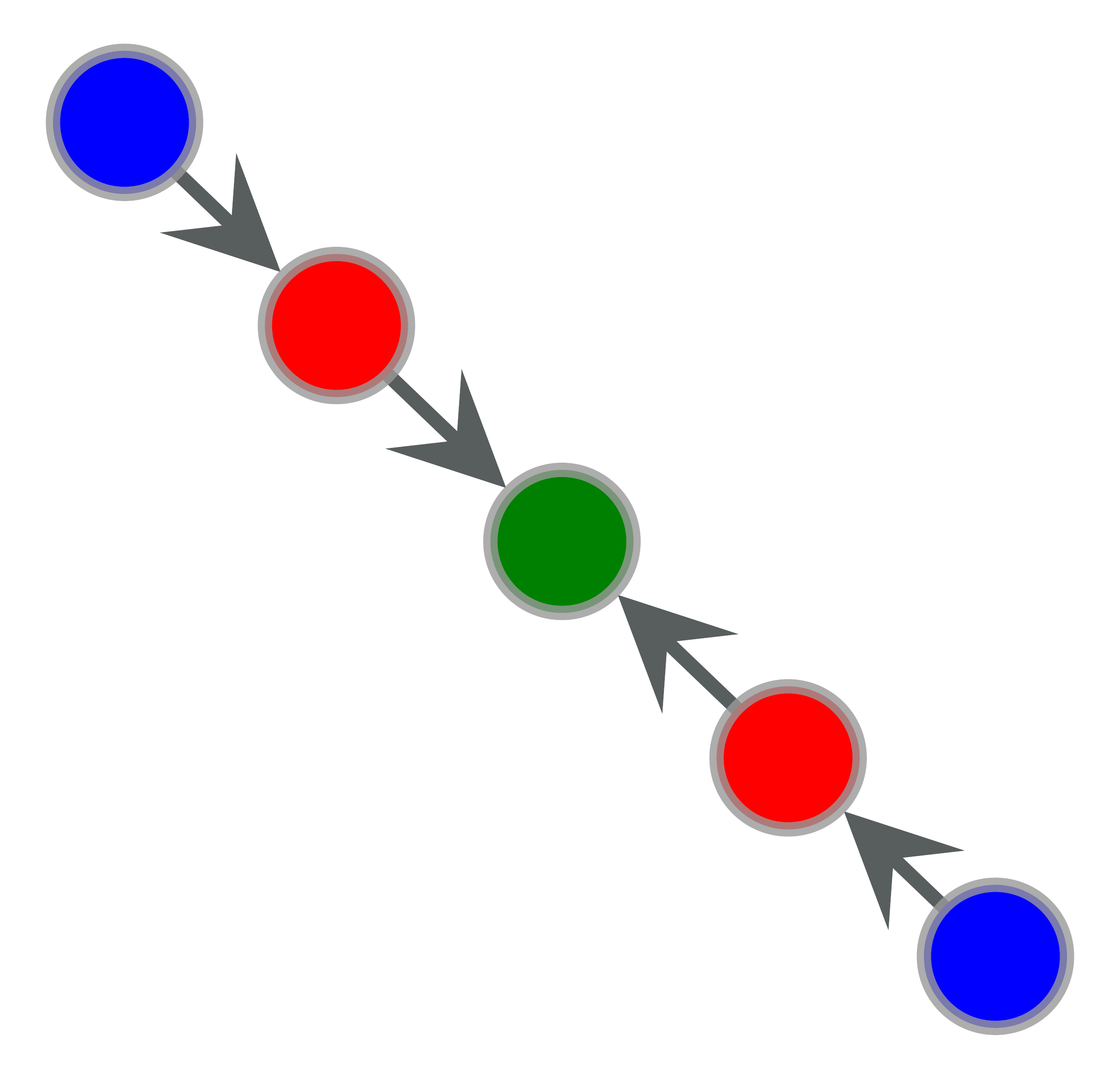}&\includegraphics[height=0.09\textwidth,]{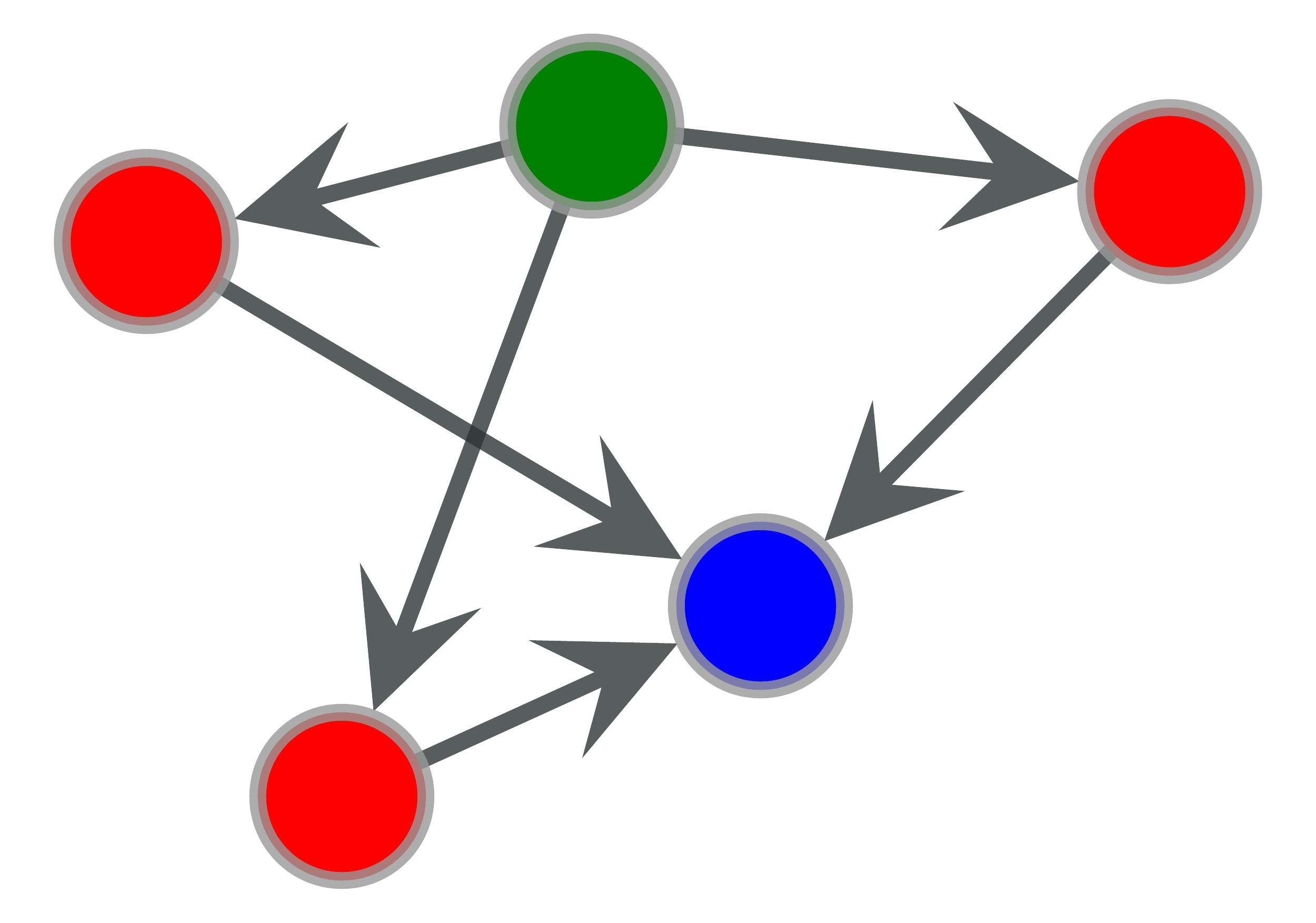}& \includegraphics[height=0.09\textwidth]{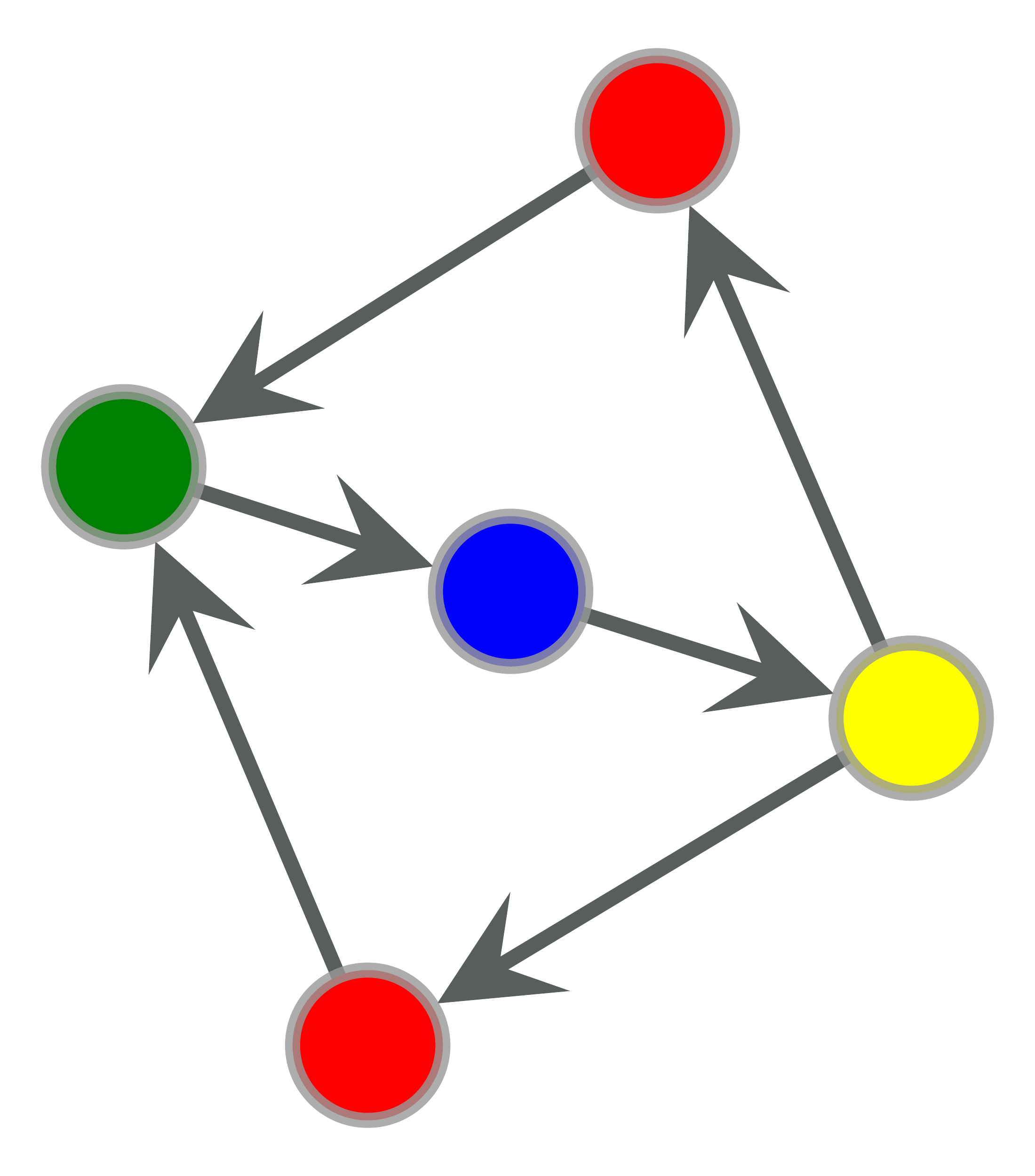}& \includegraphics[height=0.09\textwidth]{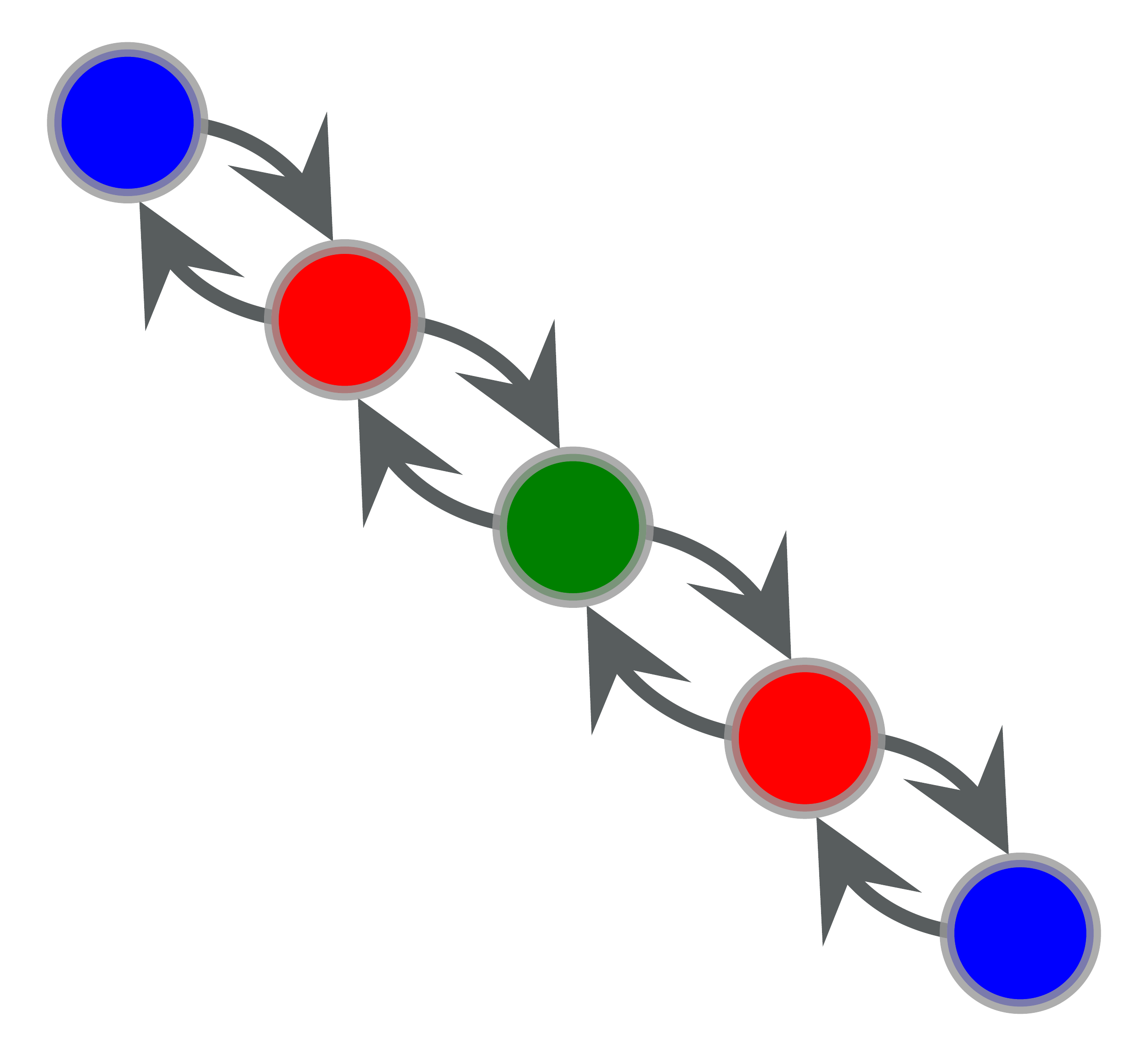}& \includegraphics[height=0.09\textwidth]{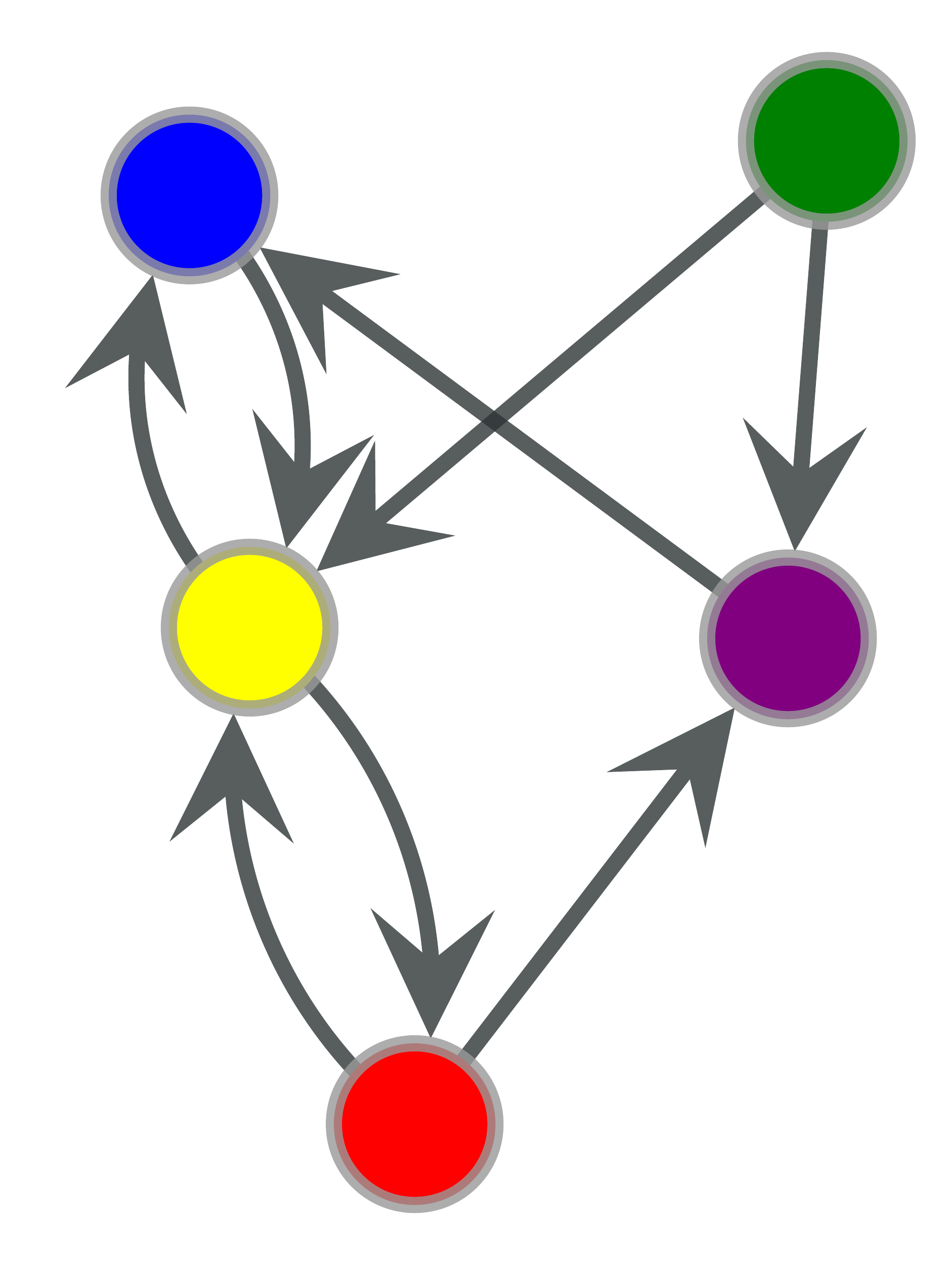}\\
\hline
Id&6&7&8&9&10&11\\
\hline
$n_m$&91& 20& 368& 170& 17& 3\\

\hline
\end{tabular}
\end{adjustbox}
\end{center}
\caption{Atoms found in the MAP configurations of the metabolic network of {\em E.coli} and their respective counts ($n_m$).}\label{metaM}
\end{table}

\subsubsection*{C.elegans neuronal network of synaptic connections}
 Finally, we consider the network of synaptic connections in the adult hermaphrodite worms {\em C. elegans} \cite{cook2019whole}. This network has 454 nodes and 4,841 directed edges. The network corresponding to male adults and the undirected networks representing gap junctions are discussed in the SI.   
The MAP configuration for the synaptic network 24 nontrivial atoms that cover approximately 70\% of edges. We find that the MAP configuration contains a large number of bi-fan motifs (Atom 4 in Table \ref{CEH}) and multiple motifs that correspond to various combinations of feed--forward--loops (Atoms 6,7,8 and 10 in Table \ref{CEH}) indicating that feed--forward--loops usually associated to neuronal networks are organized into larger atoms in network. In addition to these the MAP configuration also contains multiple atoms resulting from symmetric combinations of the triangular connection pattern ($A \longleftrightarrow B$, $C\rightarrow A$, $C\rightarrow B$) i.e.  atoms 12, 14, 18 and 19 in Table \ref{CEH}. The combinations we find in general contain three or more smaller motifs in specific combinations suggesting the presence of motifs than go beyond pairwise combinations of lower order motifs as studied in \cite{adler2022emergence}. Finally, we also observe atoms containing chains and cycles of bidirectional edges such as the atoms 1,2,15,17,20,21 in Table \ref{CEH} and directed cliques. We also find the same general classes of atoms in the MAP configuration of the network of the male {\em C.elegans} for which the results can be found in the SI.

\begin{table}[h!]
\begin{center}
\begin{adjustbox}{max width=\textwidth}

\begin{tabular}{|c|c|c|c|c|c|c|c|c|}
\hline

$m$& \includegraphics[height=0.09\textwidth]{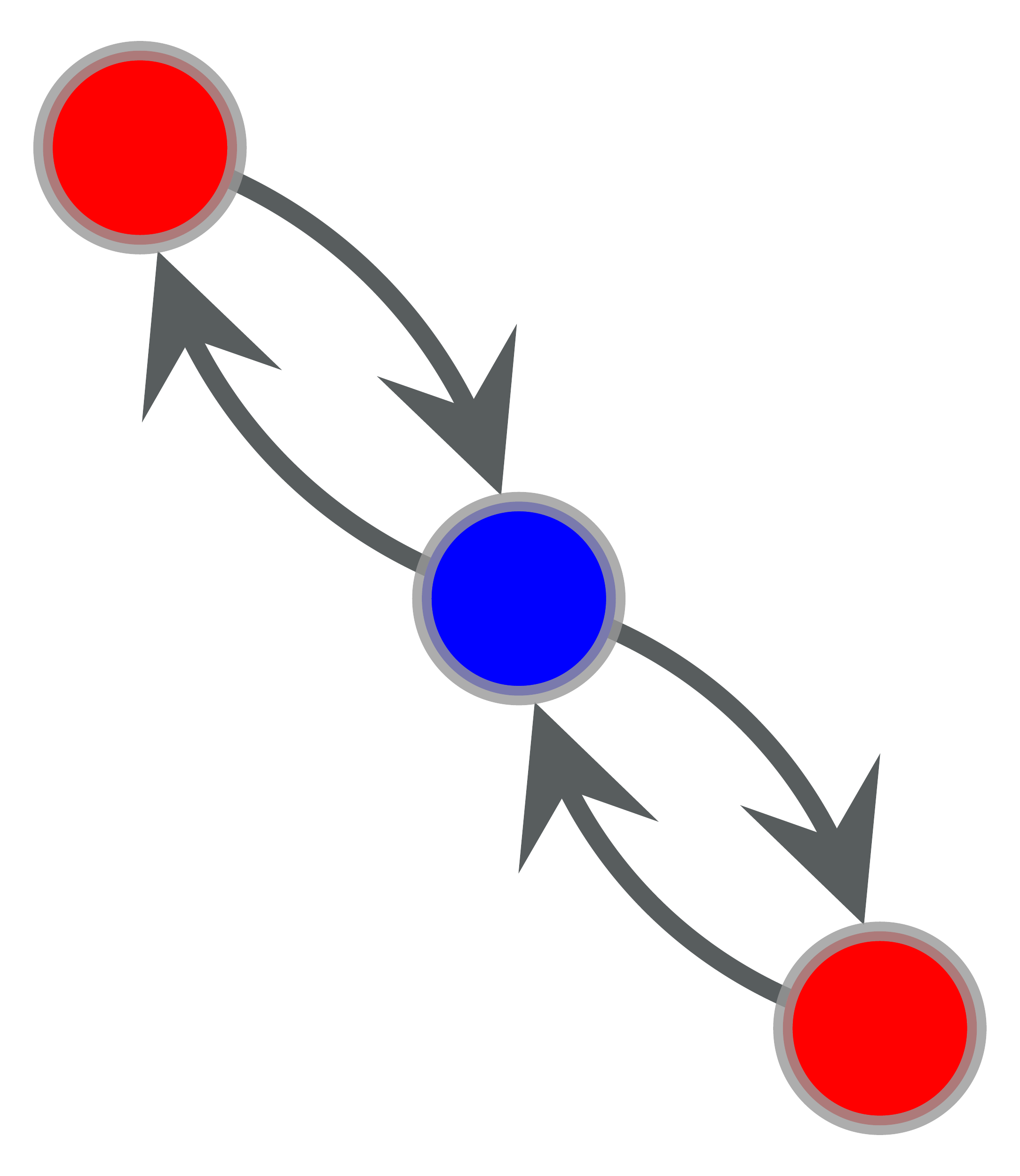}& \includegraphics[height=0.09\textwidth]{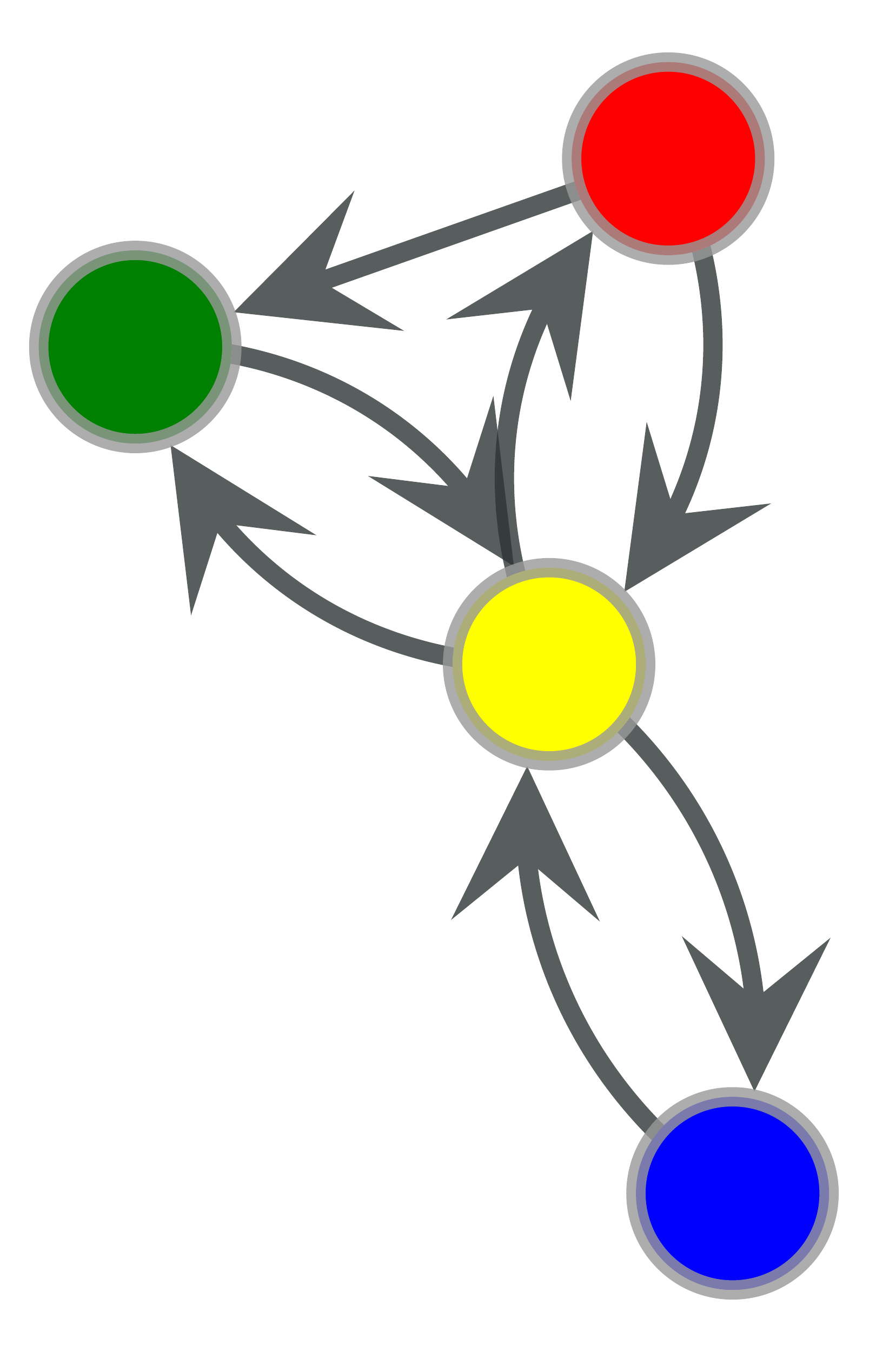}& \includegraphics[height=0.09\textwidth]{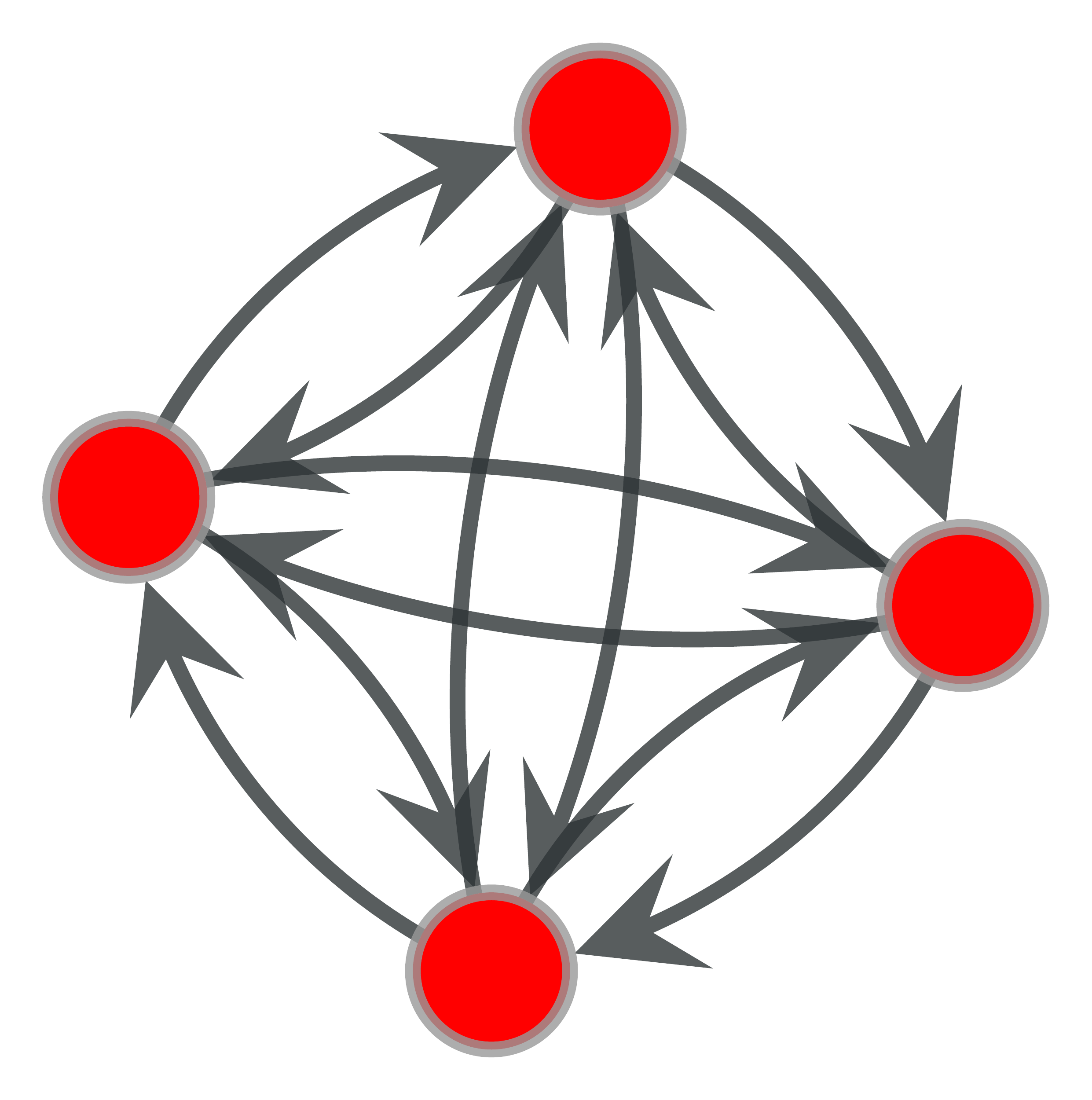}& \includegraphics[height=0.09\textwidth]{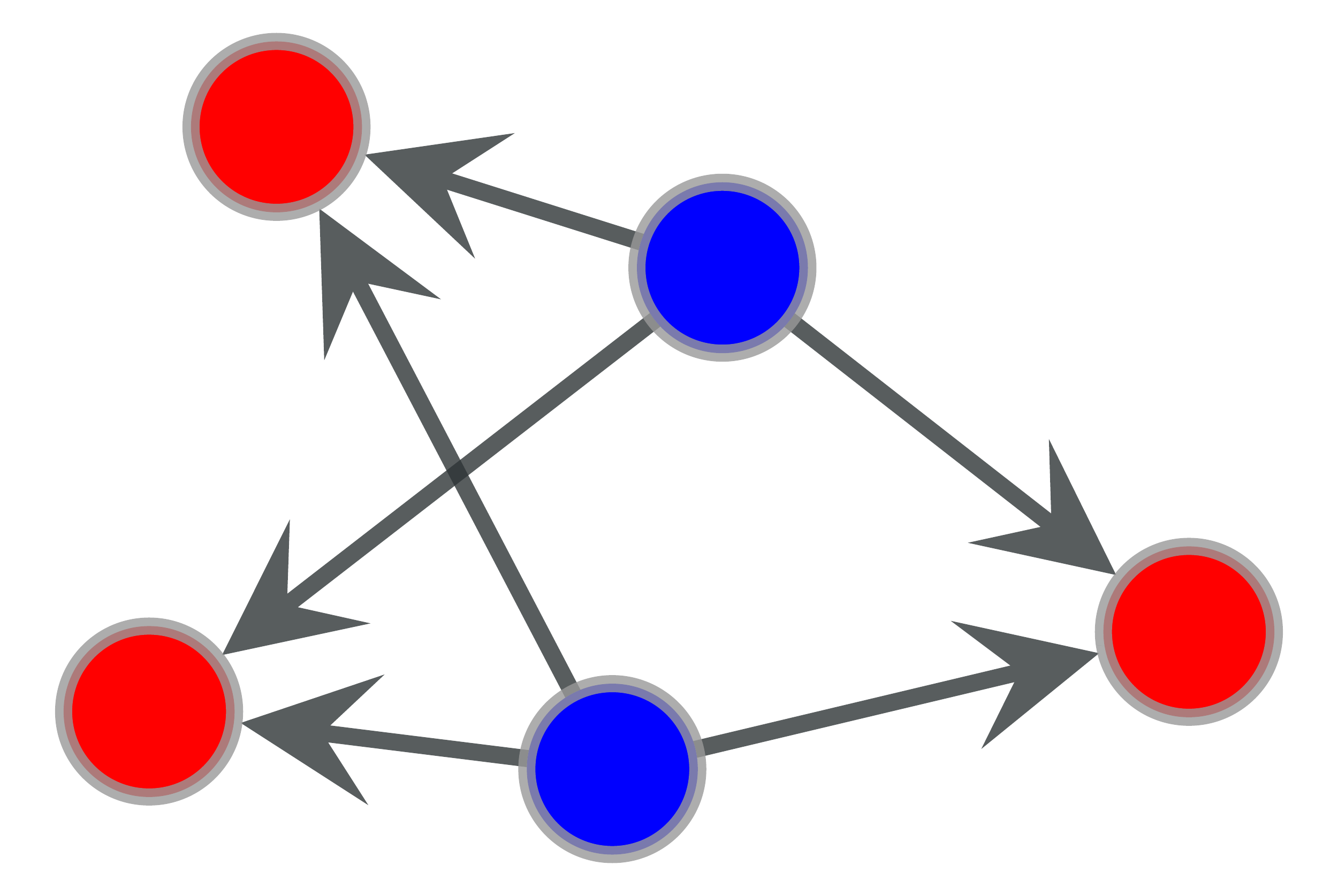} &\includegraphics[height=0.09\textwidth]{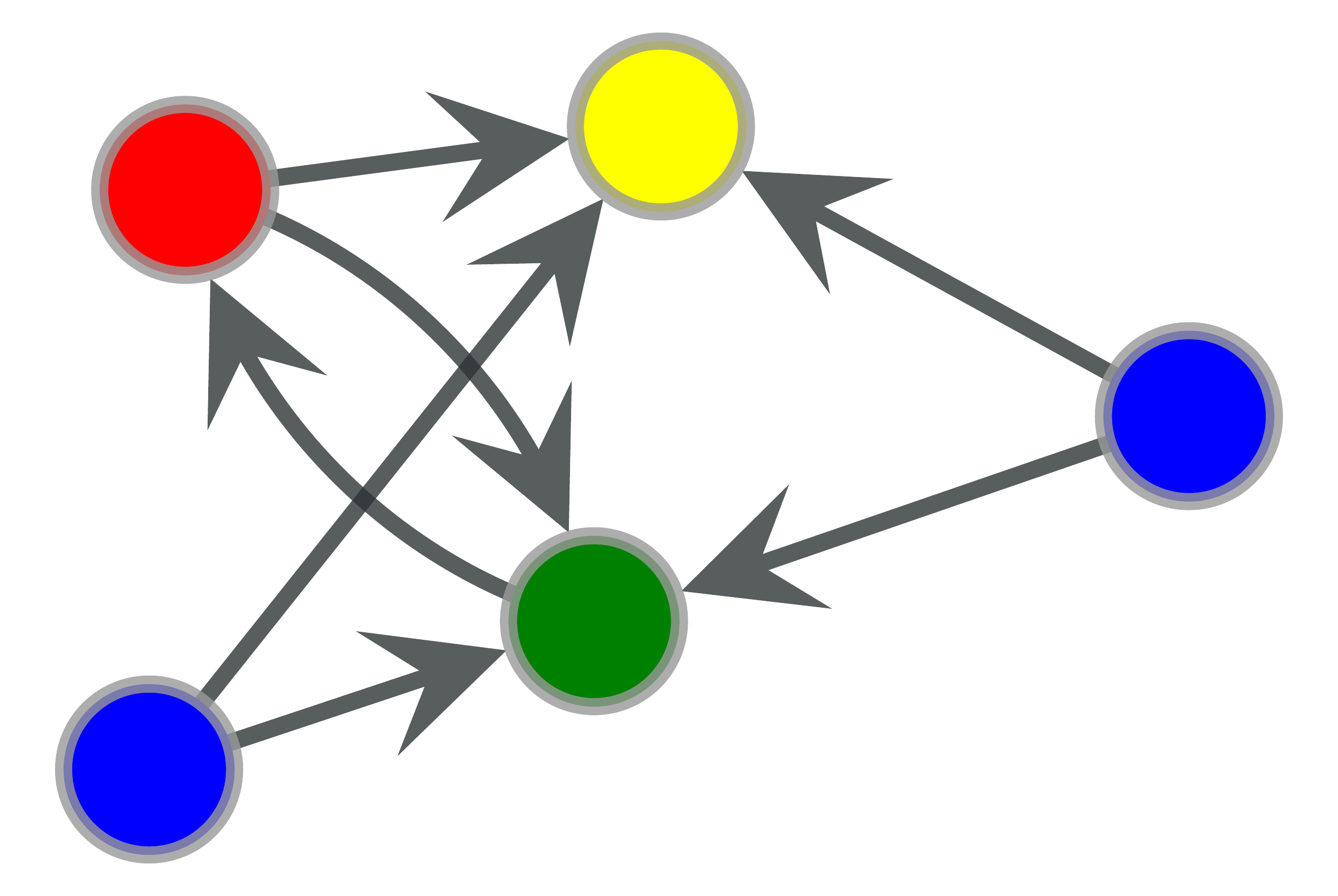}& \includegraphics[height=0.09\textwidth]{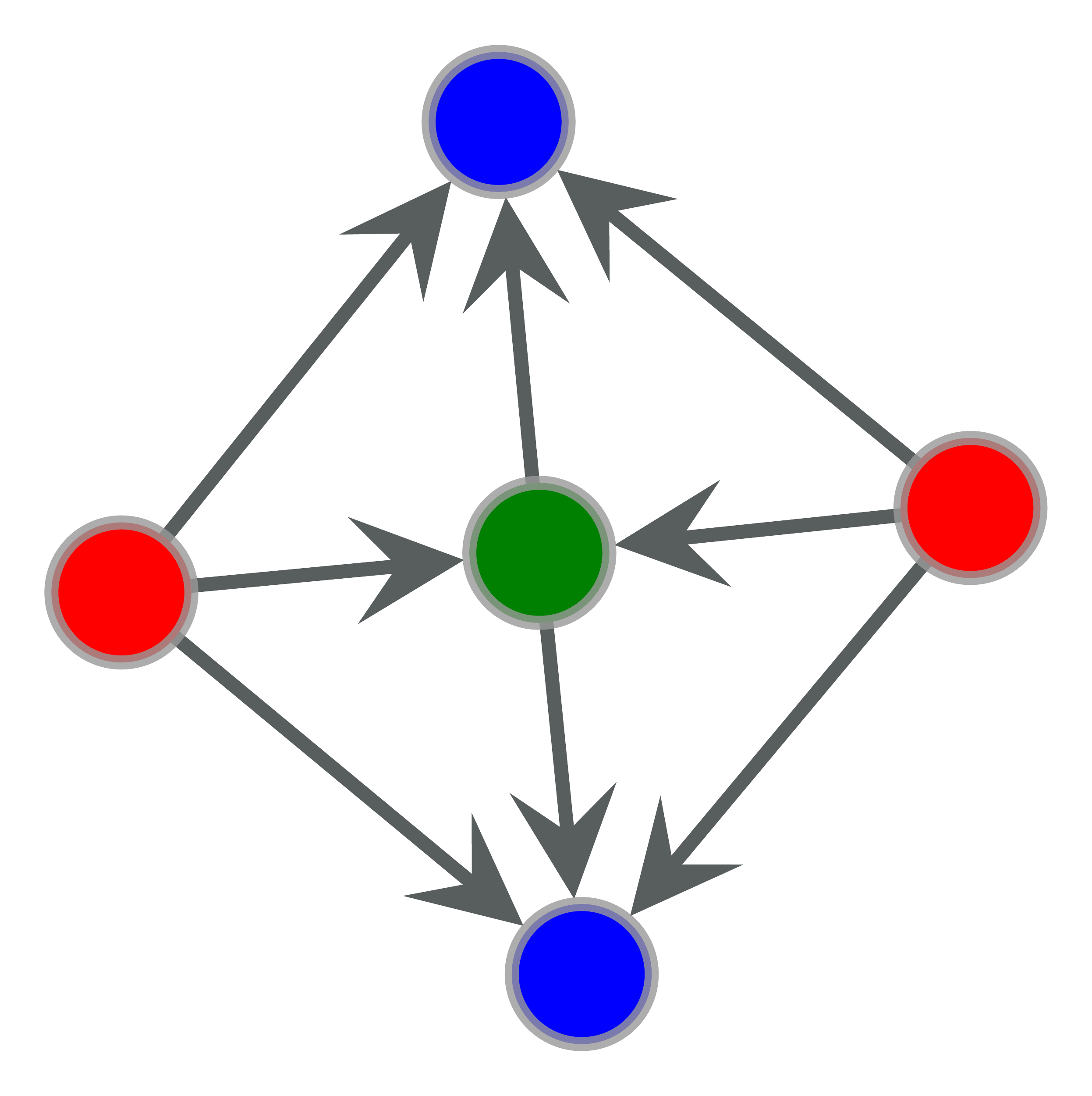}&\includegraphics[height=0.09\textwidth]{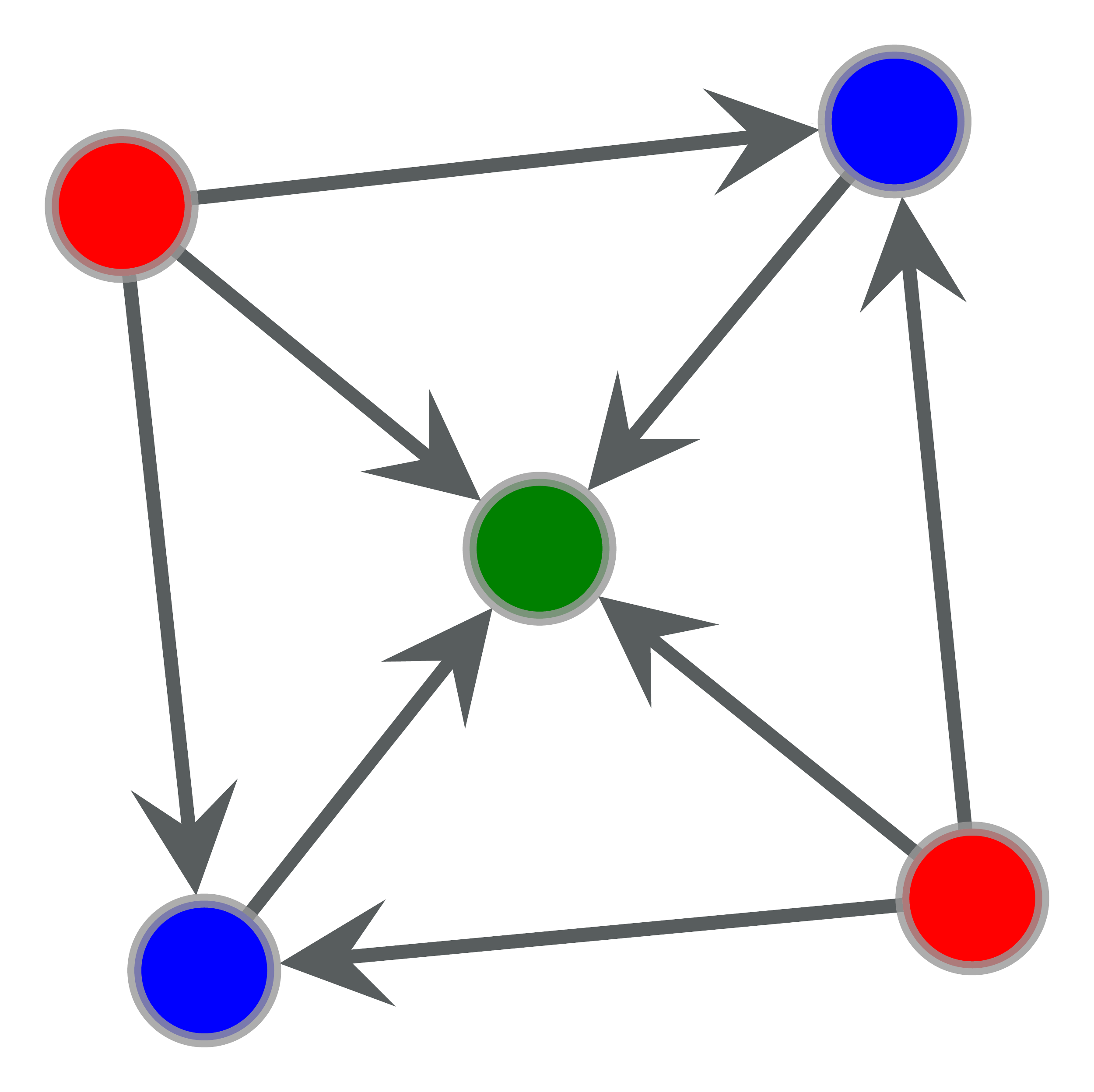}&\includegraphics[height=0.09\textwidth]{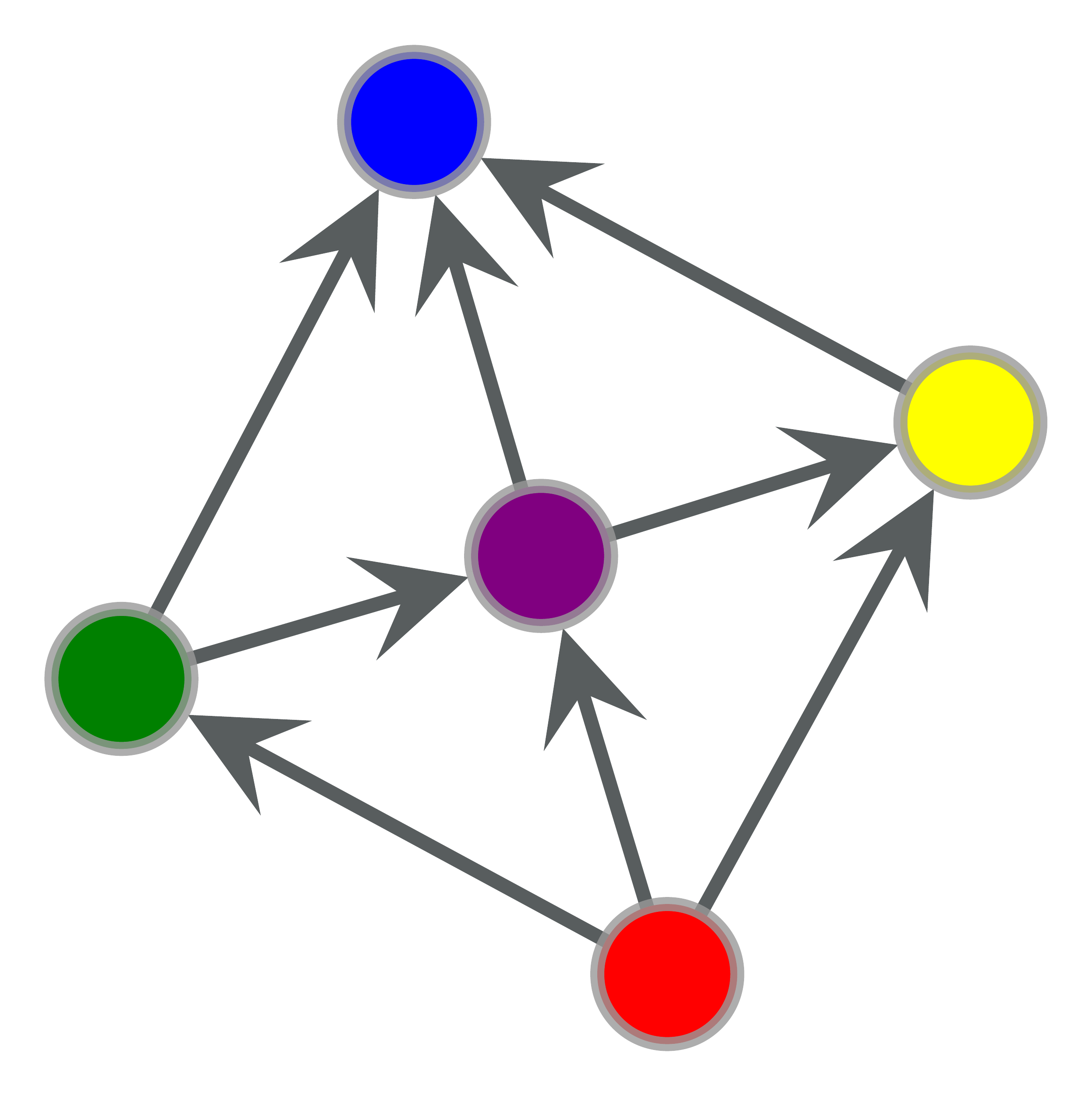}\\
\hline
Id&1&2&3&4&5&6&7&8\\
\hline
$n_m$&9& 9& 12& 138& 3& 25& 16& 13\\

\hline
$m$&  \includegraphics[height=0.09\textwidth]{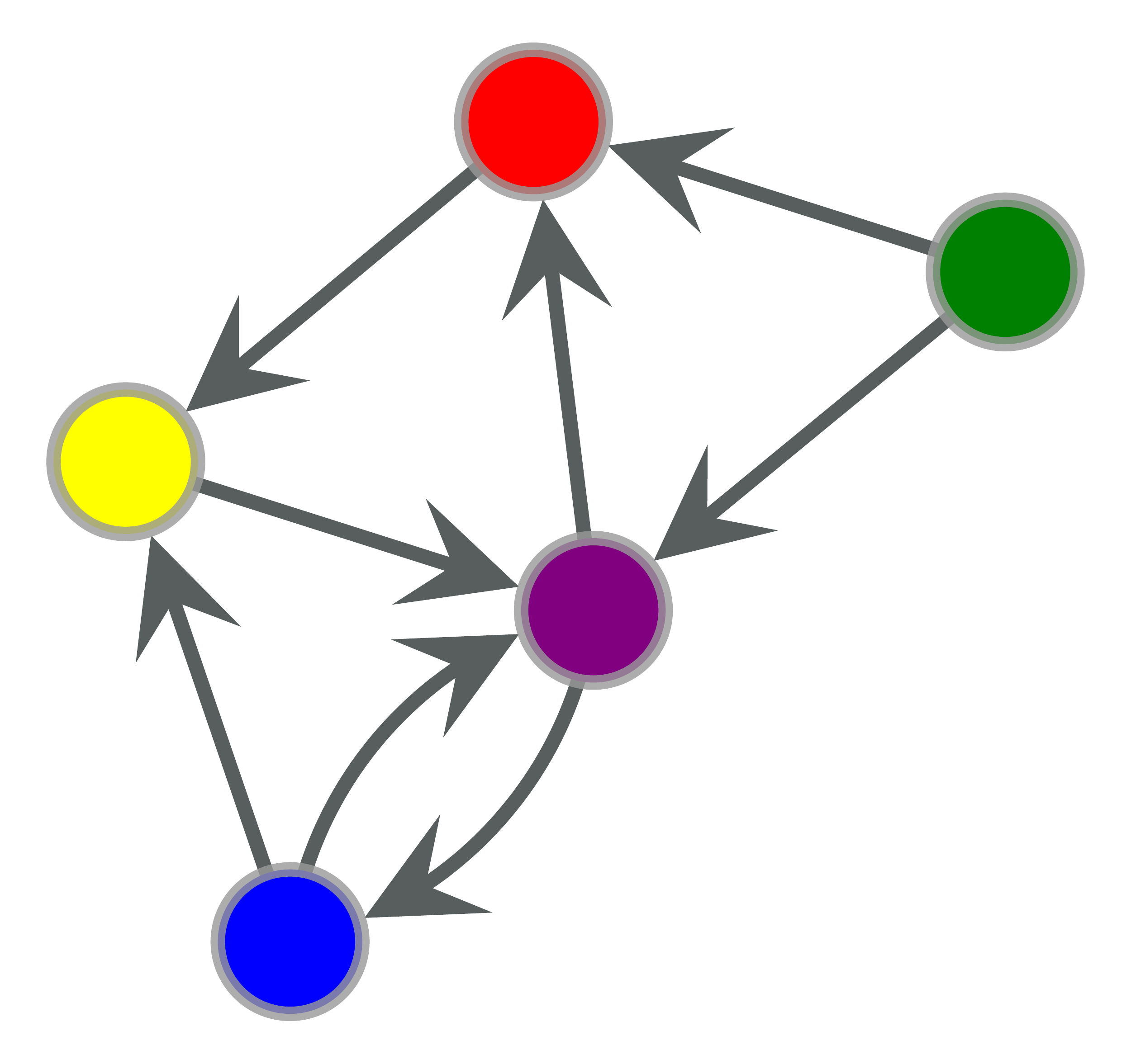}&\includegraphics[height=0.09\textwidth]{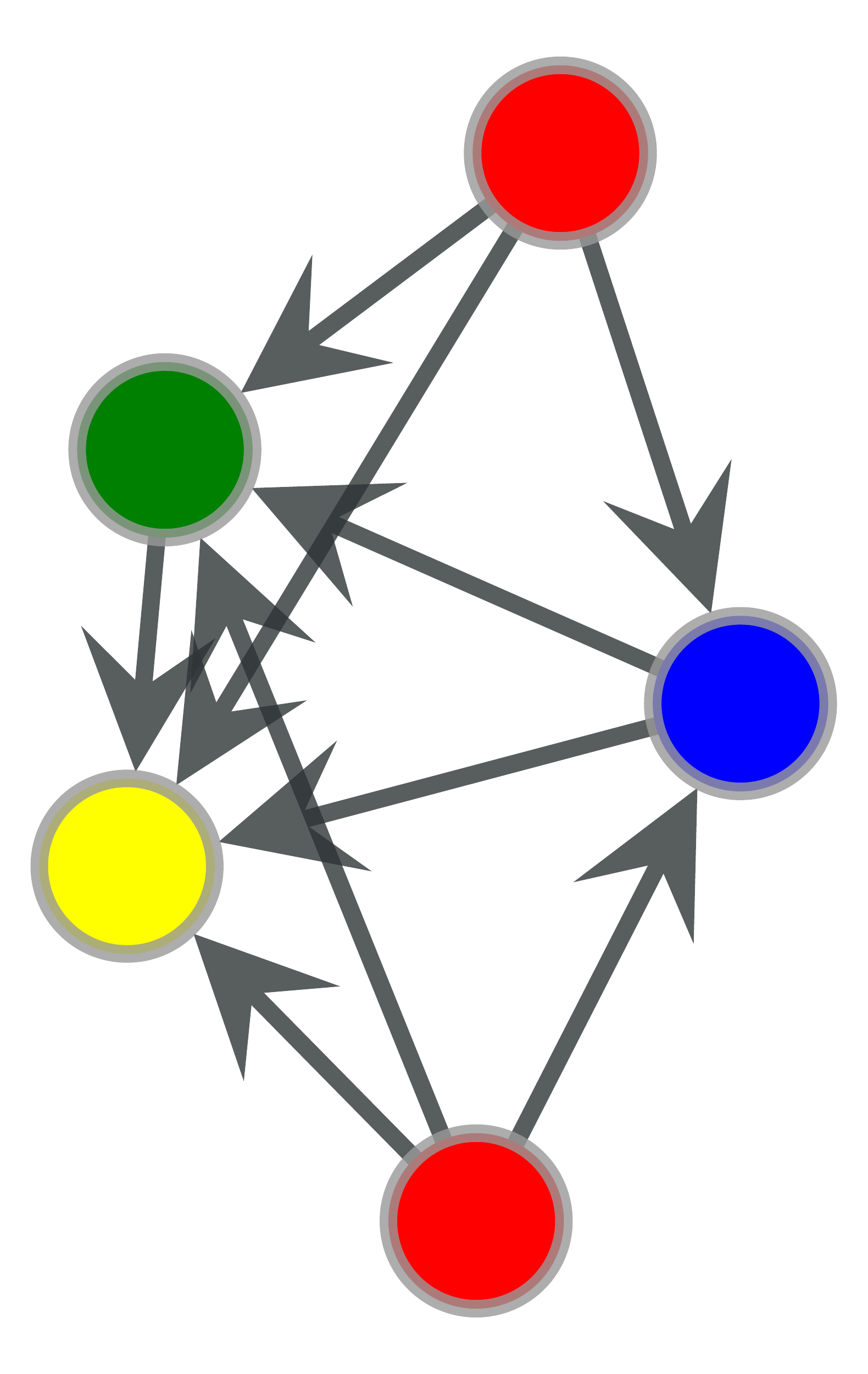}& \includegraphics[height=0.09\textwidth]{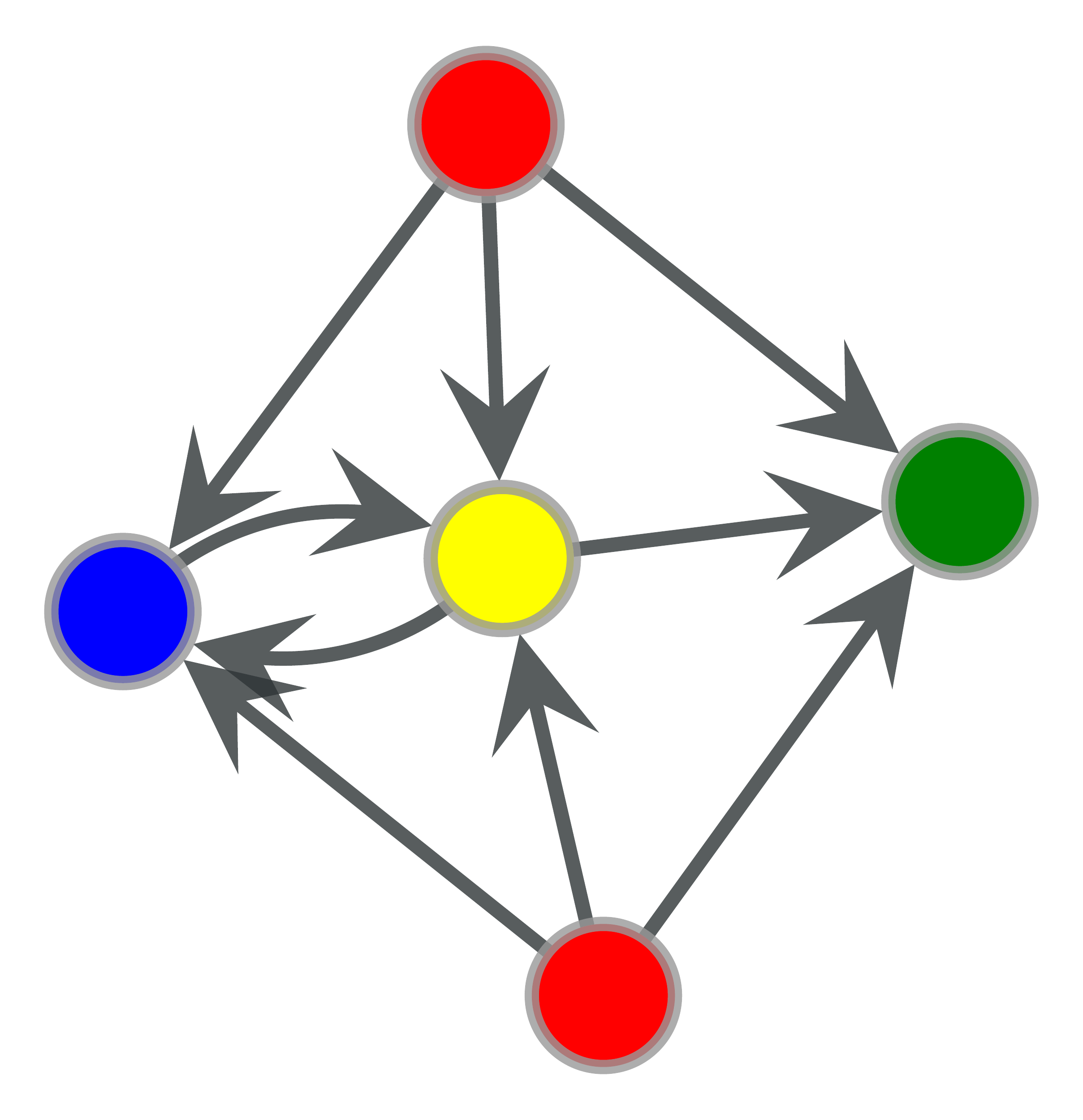}&\includegraphics[height=0.09\textwidth]{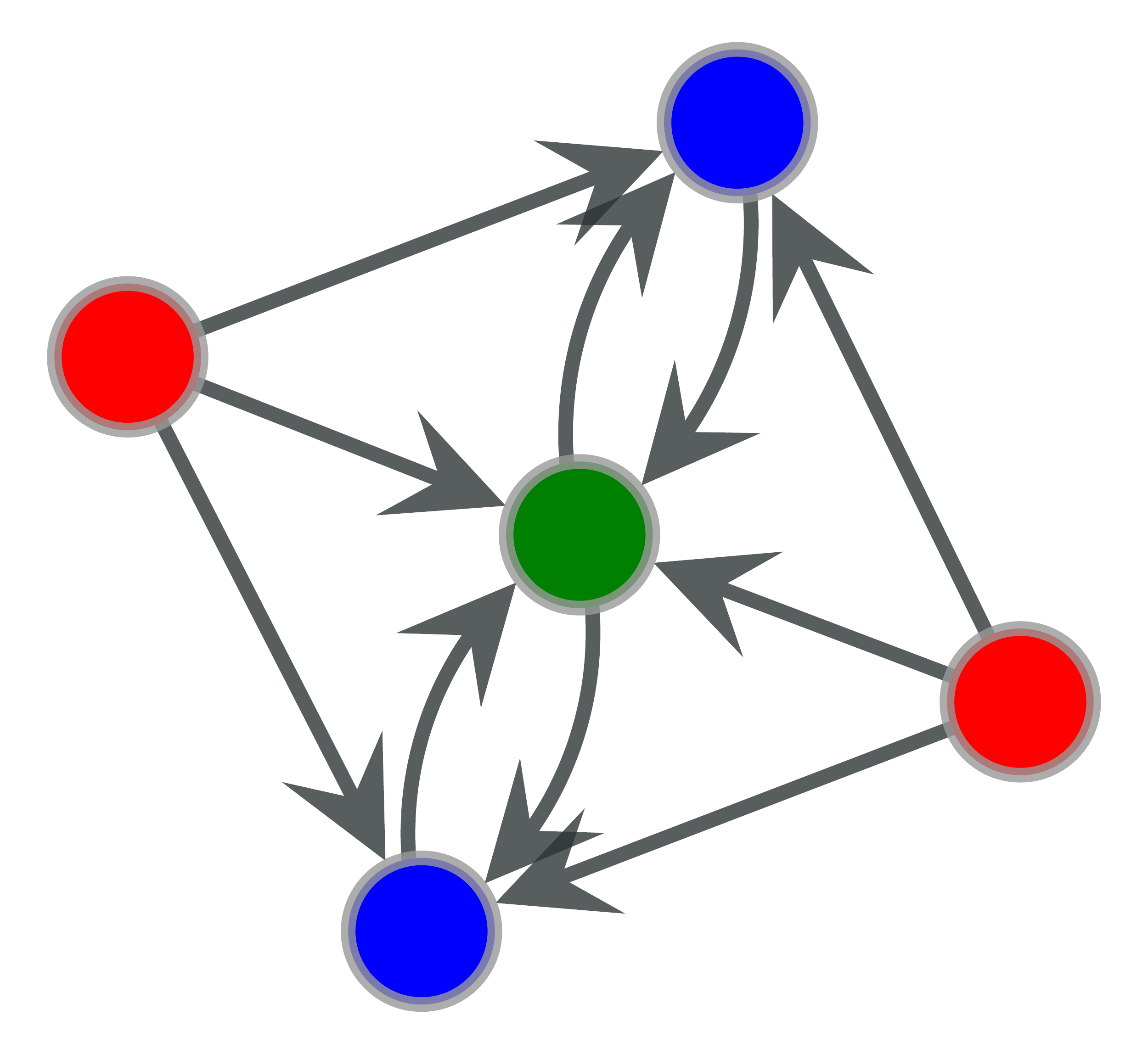}& \includegraphics[height=0.09\textwidth]{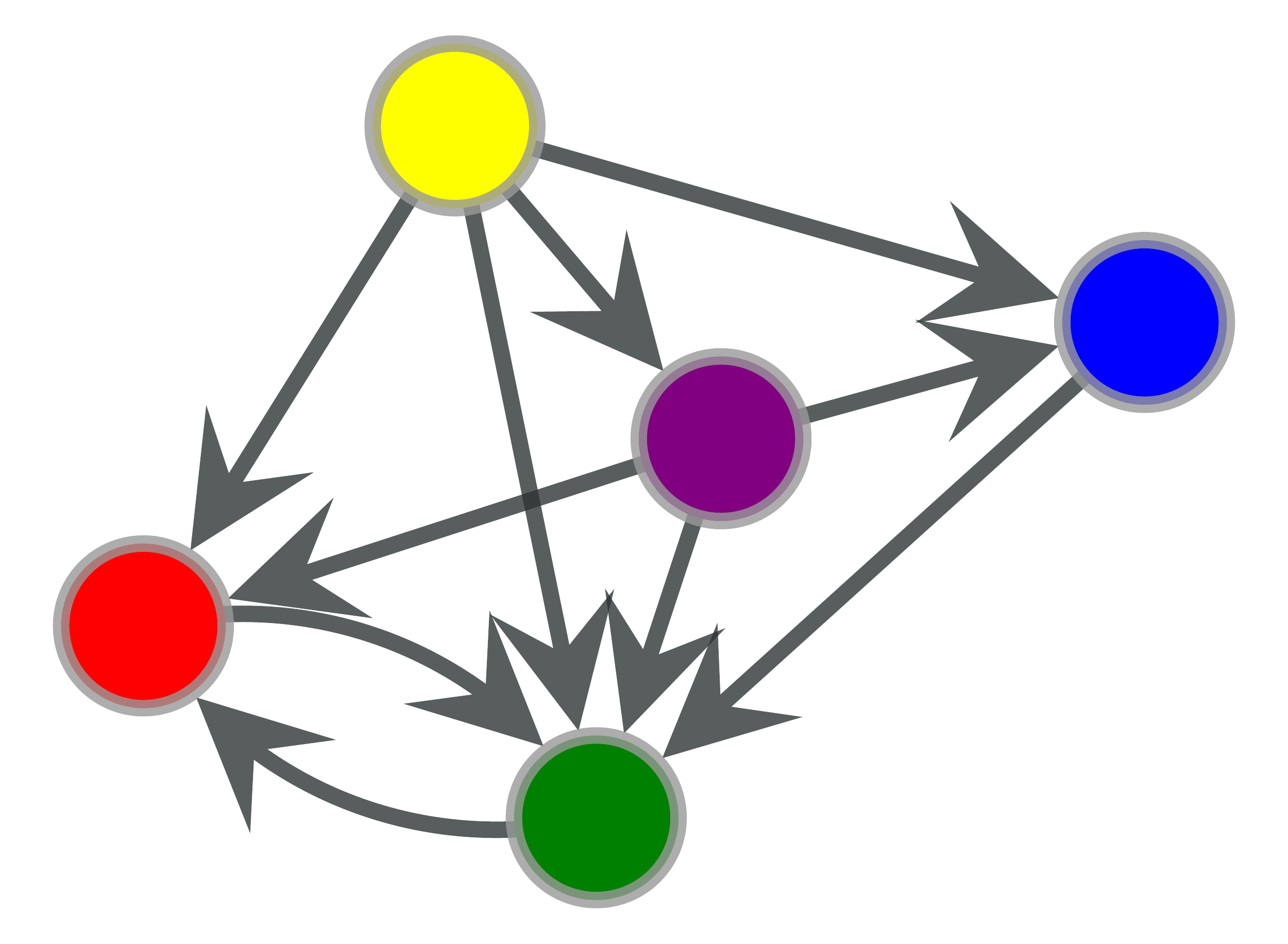}&\includegraphics[height=0.09\textwidth]{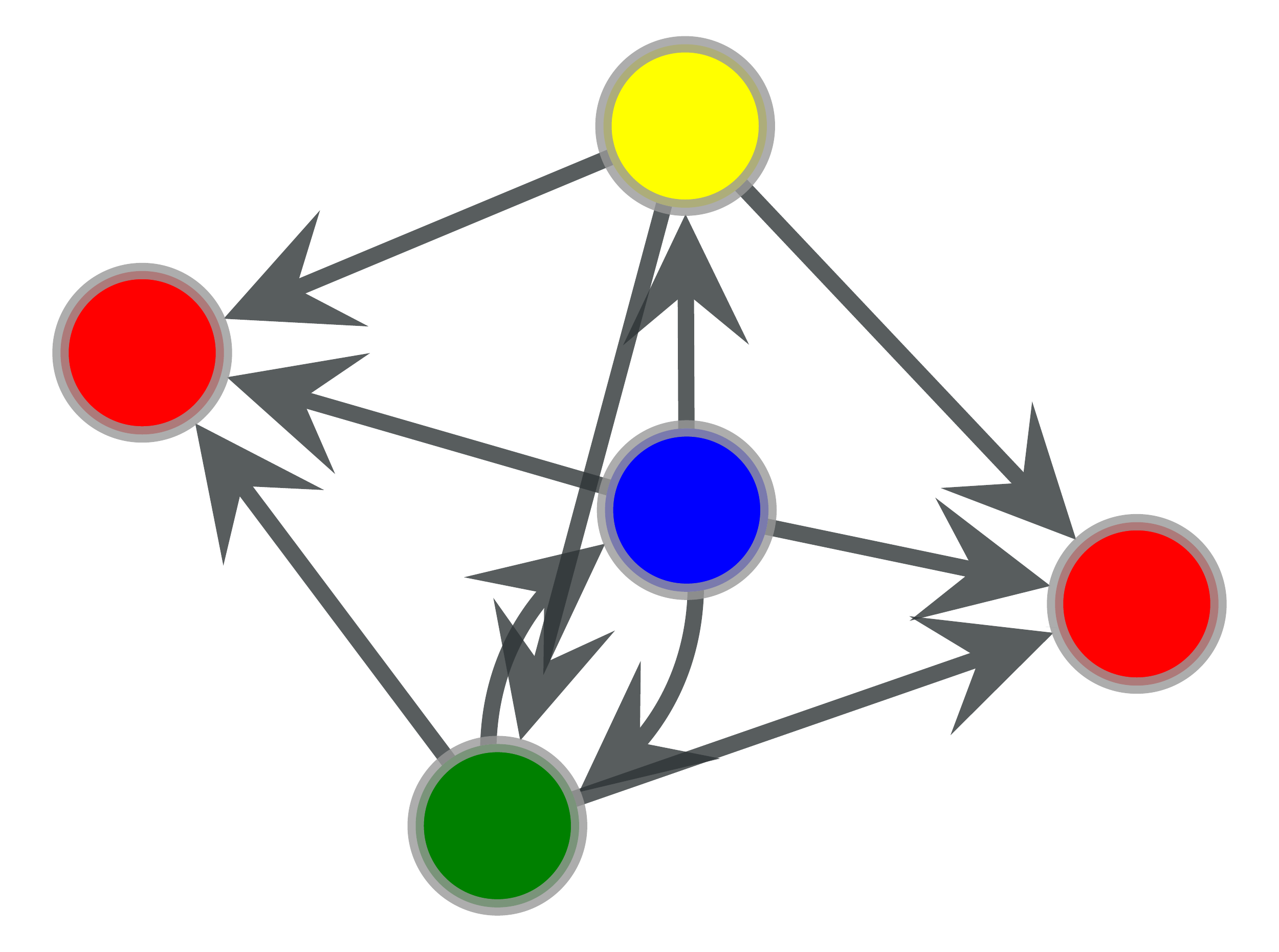}&\includegraphics[height=0.09\textwidth]{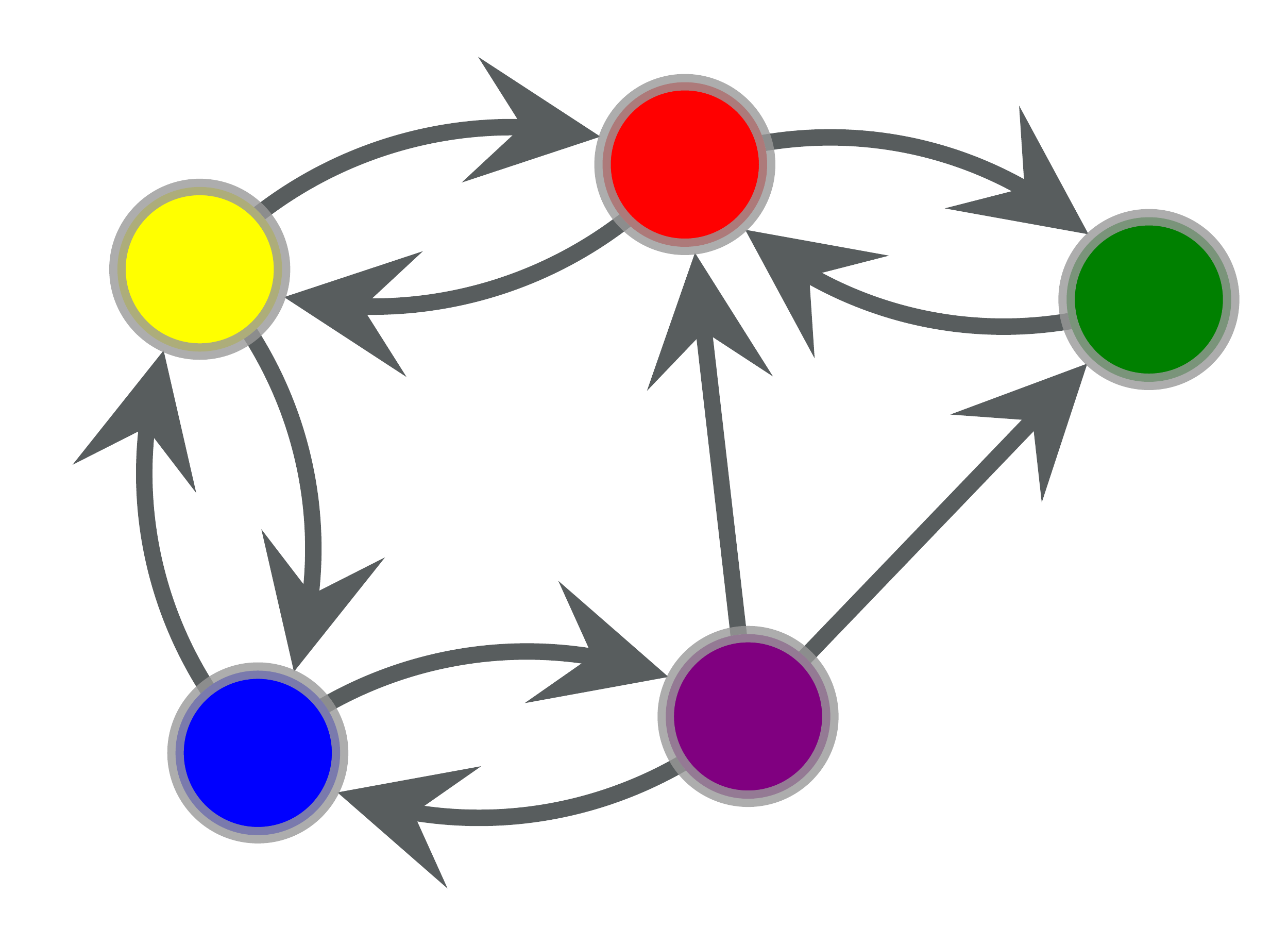}&\includegraphics[height=0.09\textwidth]{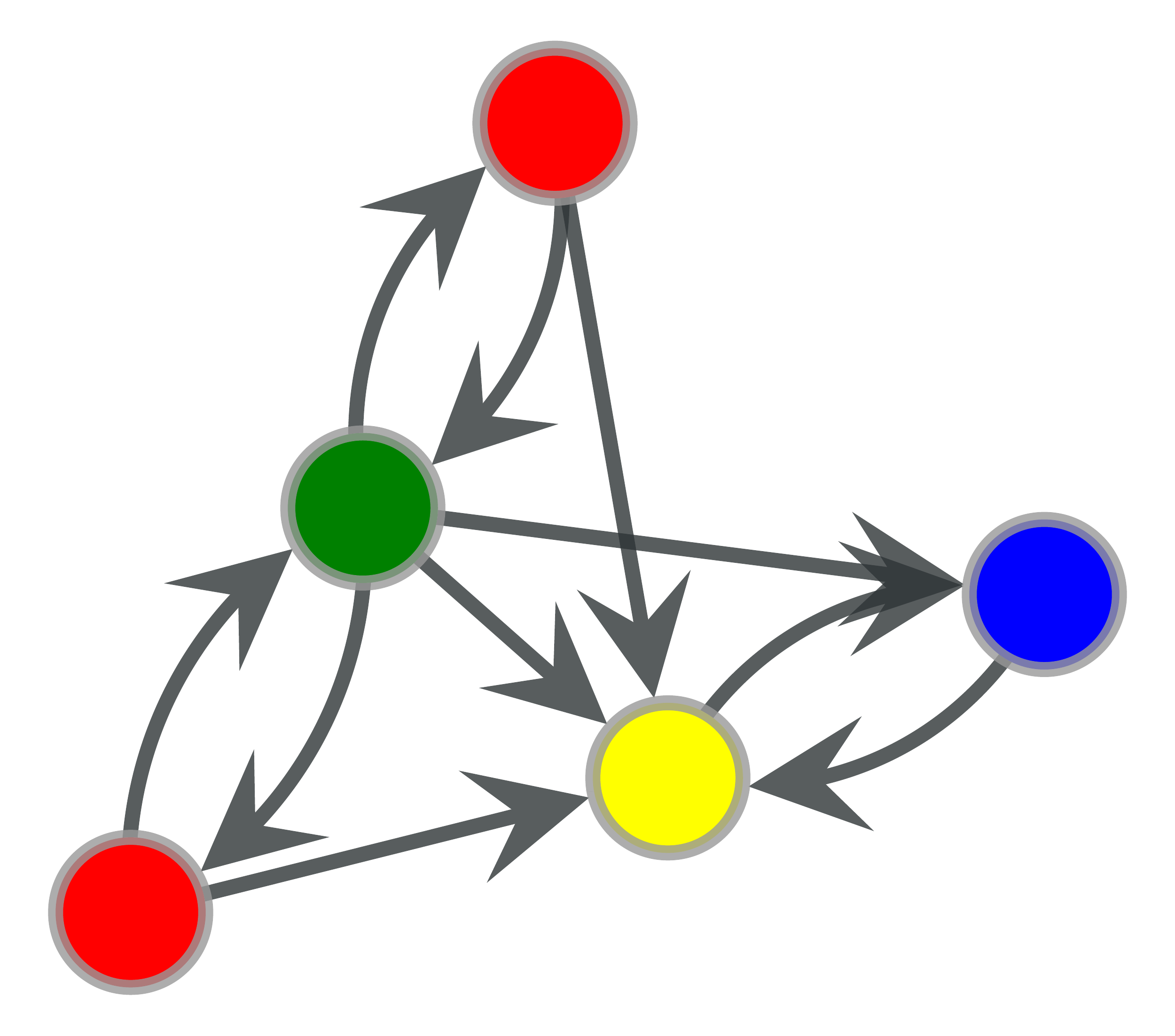} \\
\hline
Id&9&10&11&12&13&14&15&16\\
\hline
$n_m$&4& 20& 10& 26& 13& 6& 5& 2\\

\hline
$m$& \includegraphics[height=0.09\textwidth]{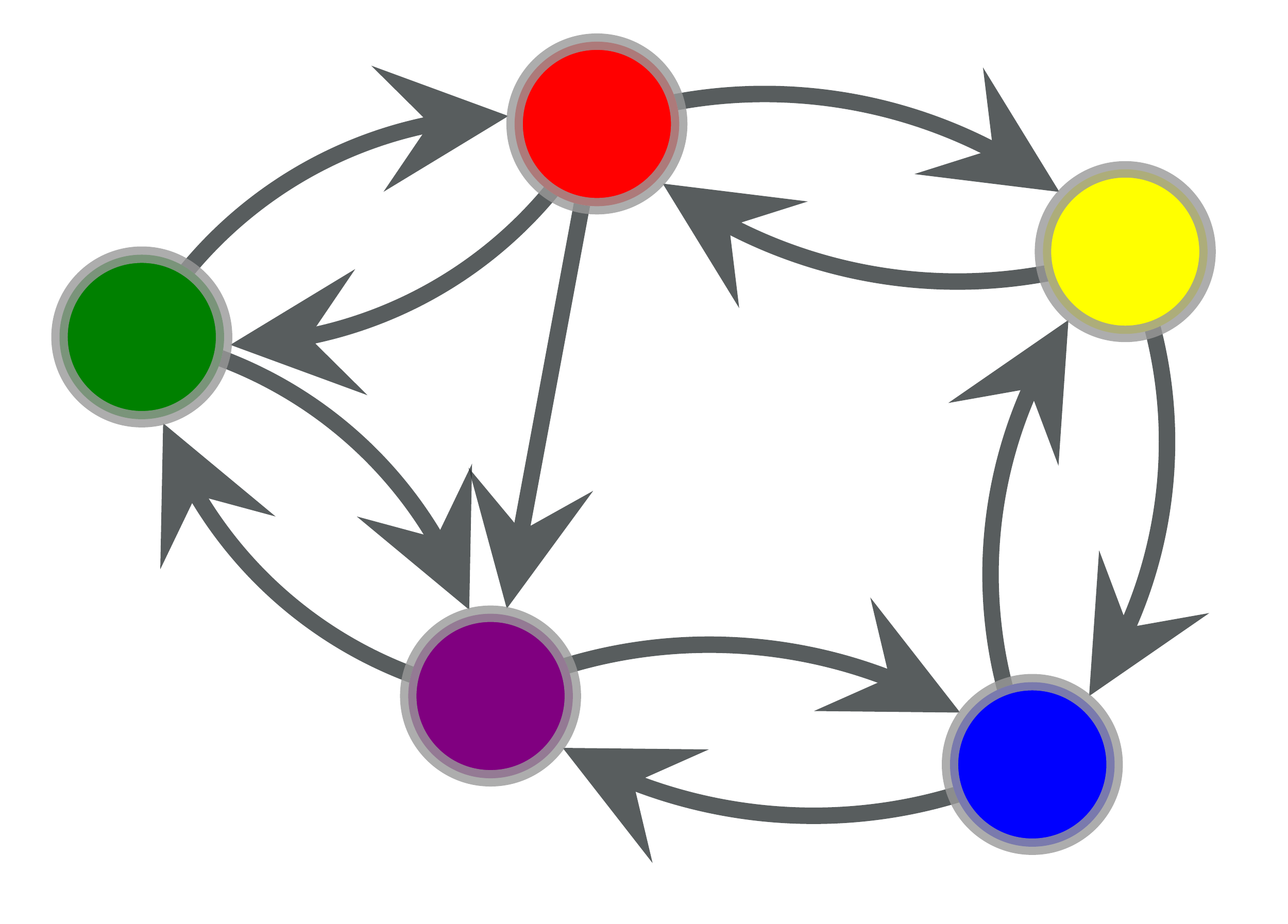}&\includegraphics[height=0.09\textwidth]{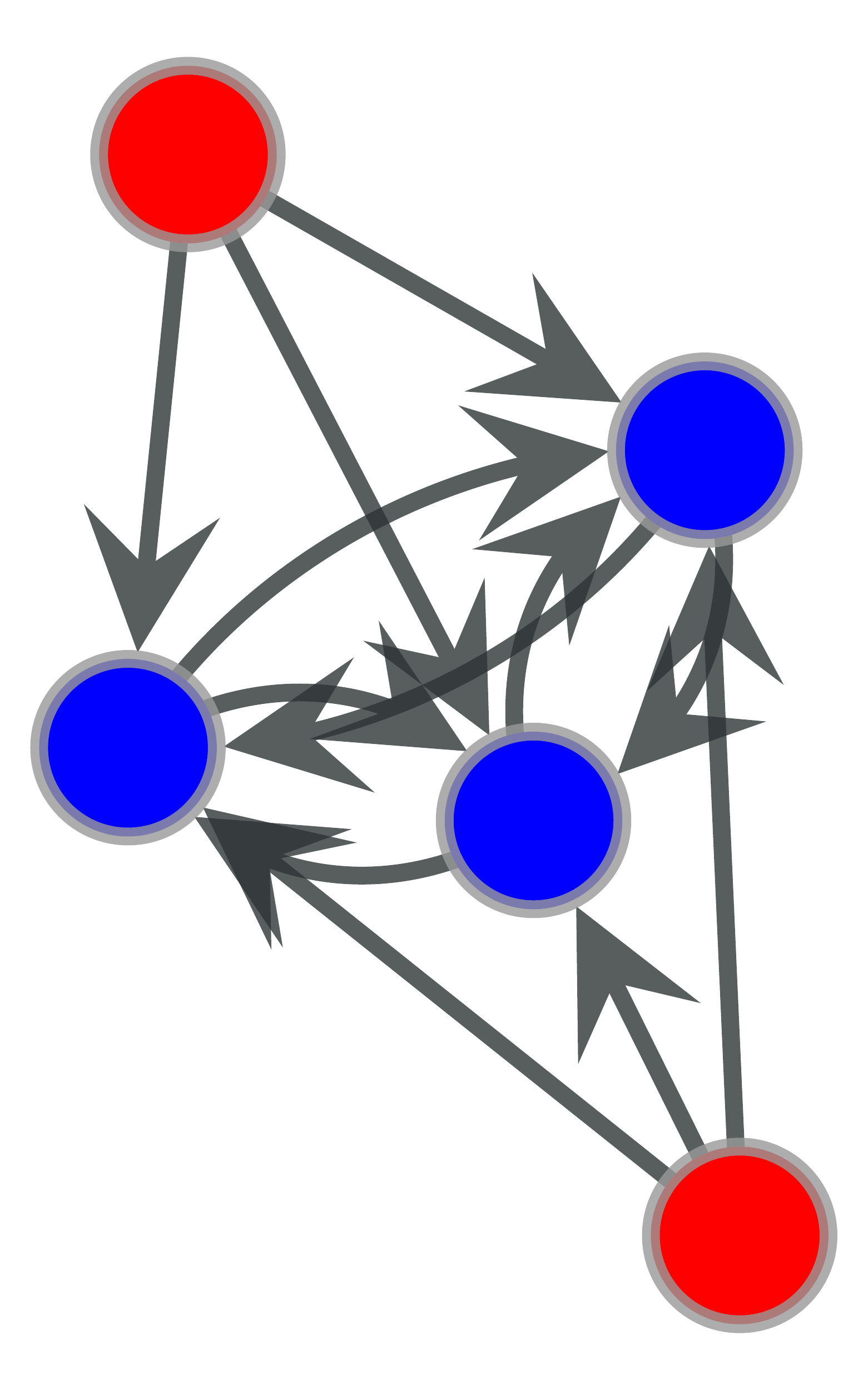}& \includegraphics[height=0.09\textwidth]{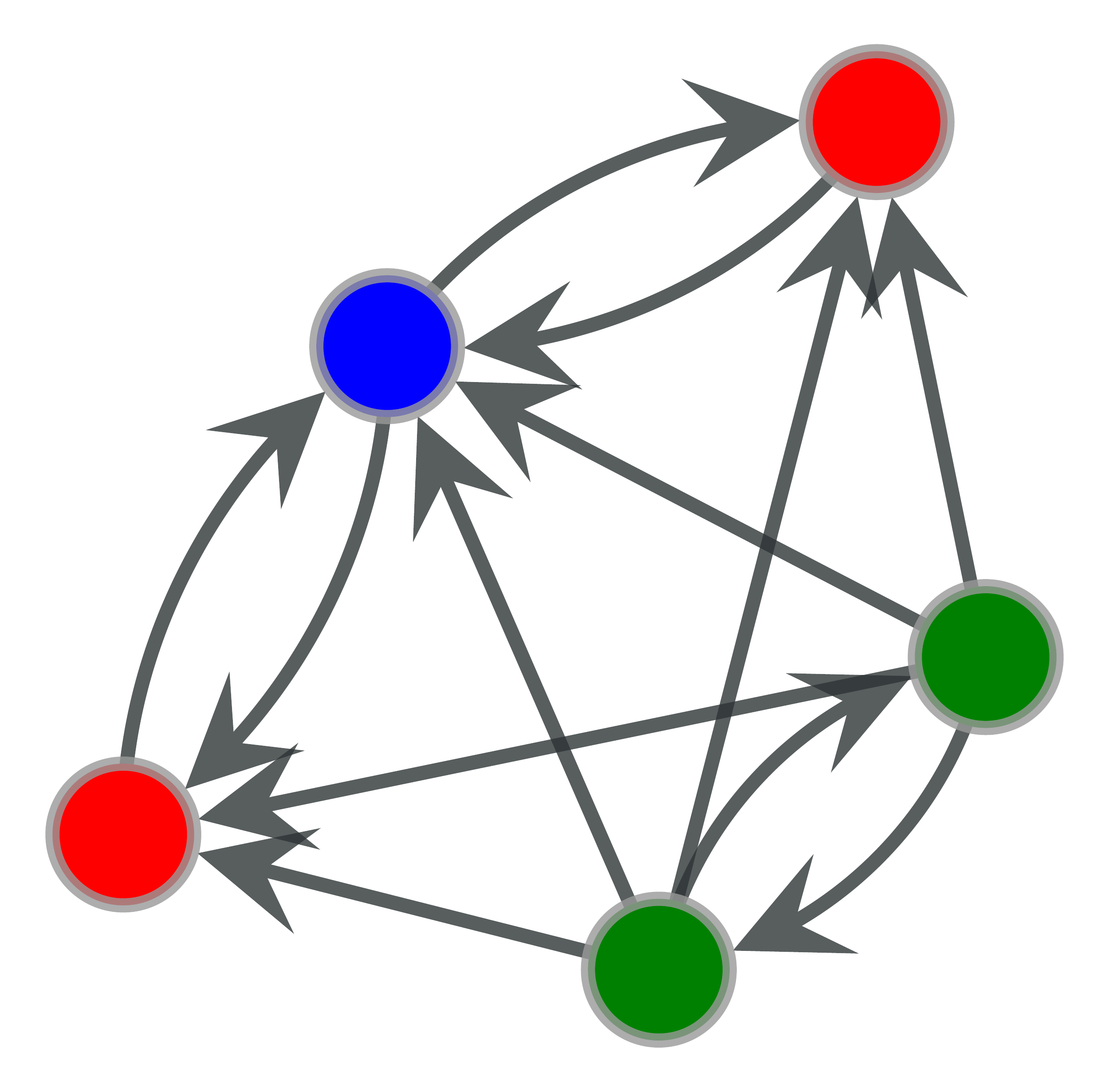}&\includegraphics[height=0.09\textwidth]{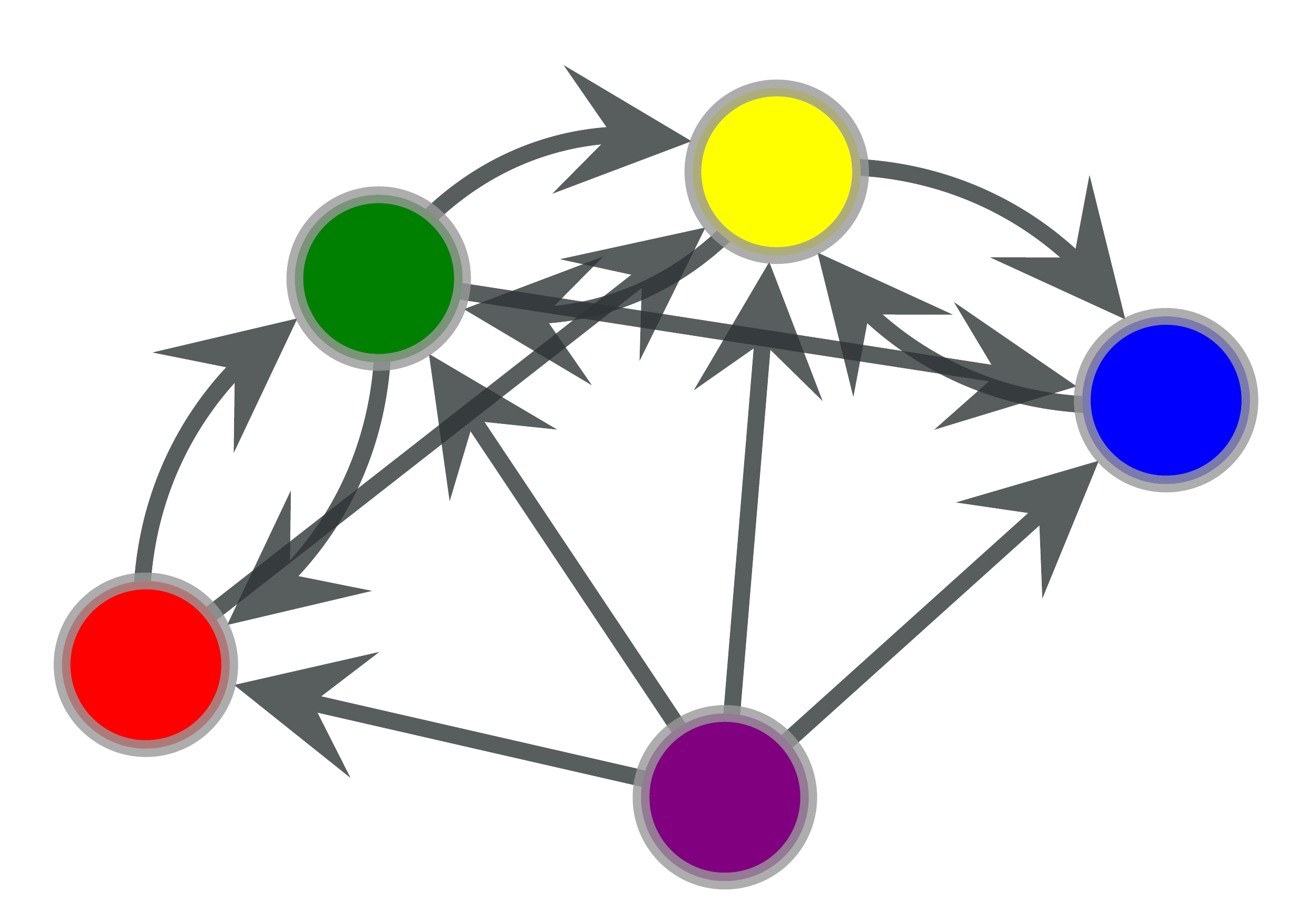}& \includegraphics[height=0.09\textwidth]{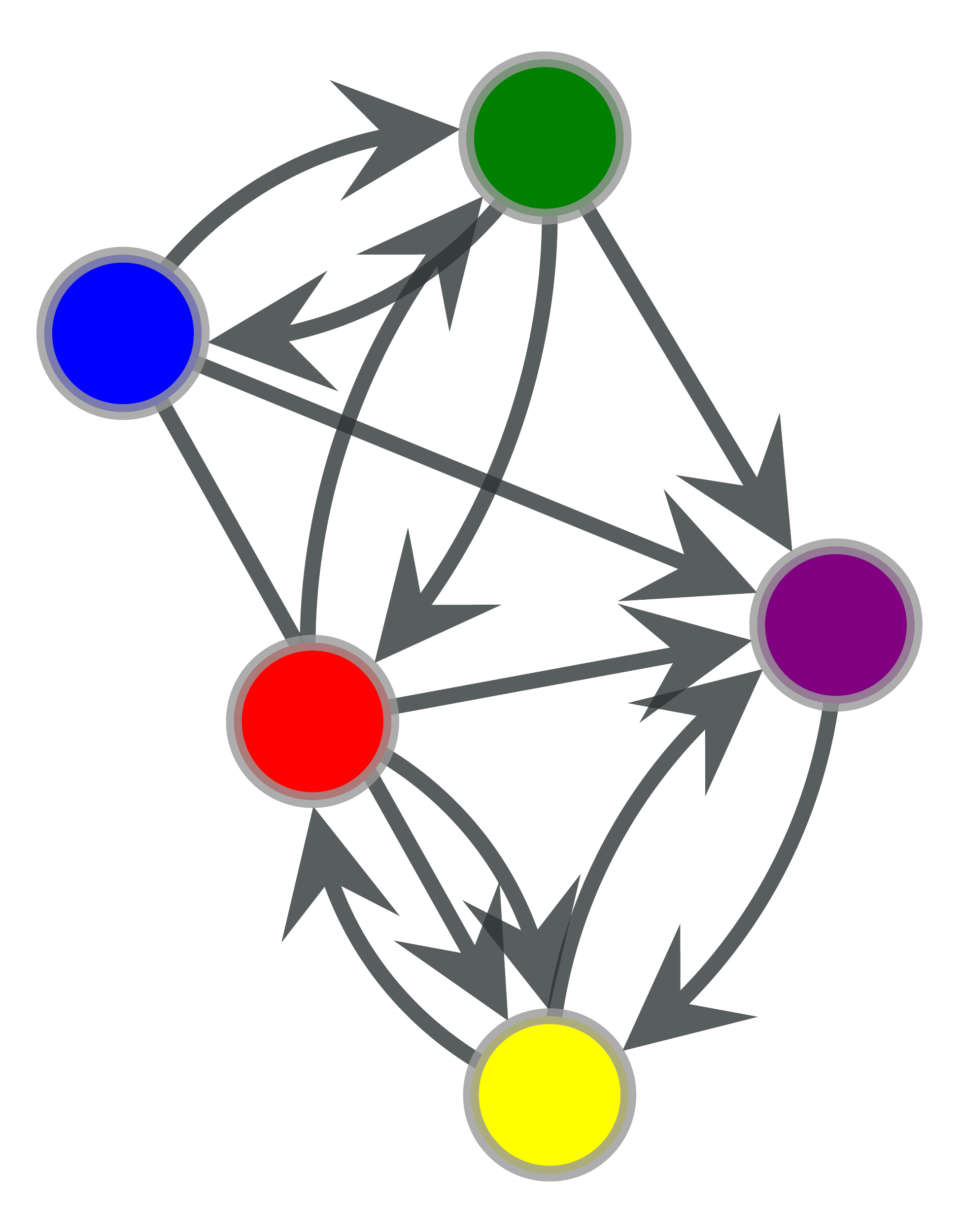}&\includegraphics[height=0.09\textwidth]{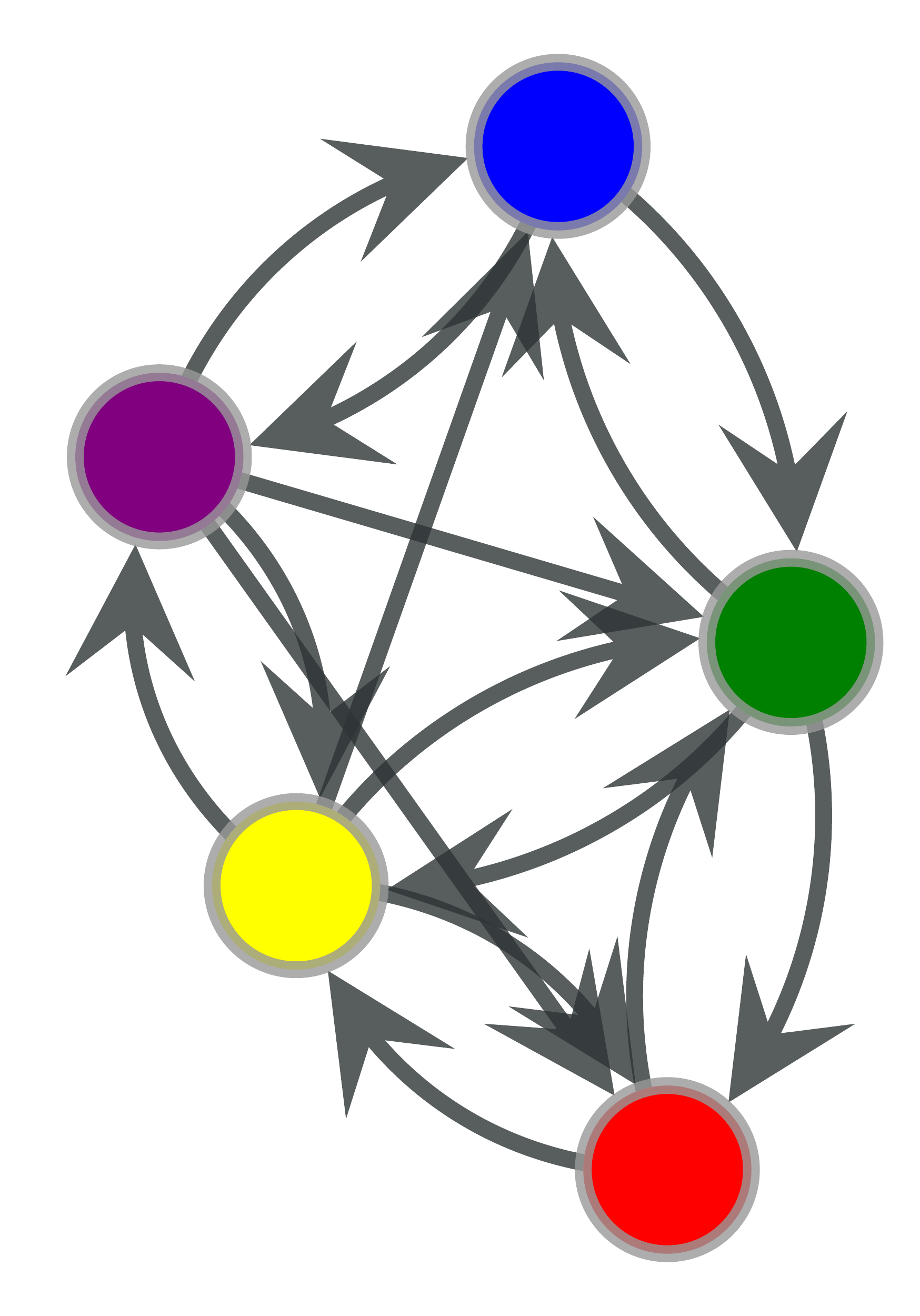}&\includegraphics[height=0.09\textwidth]{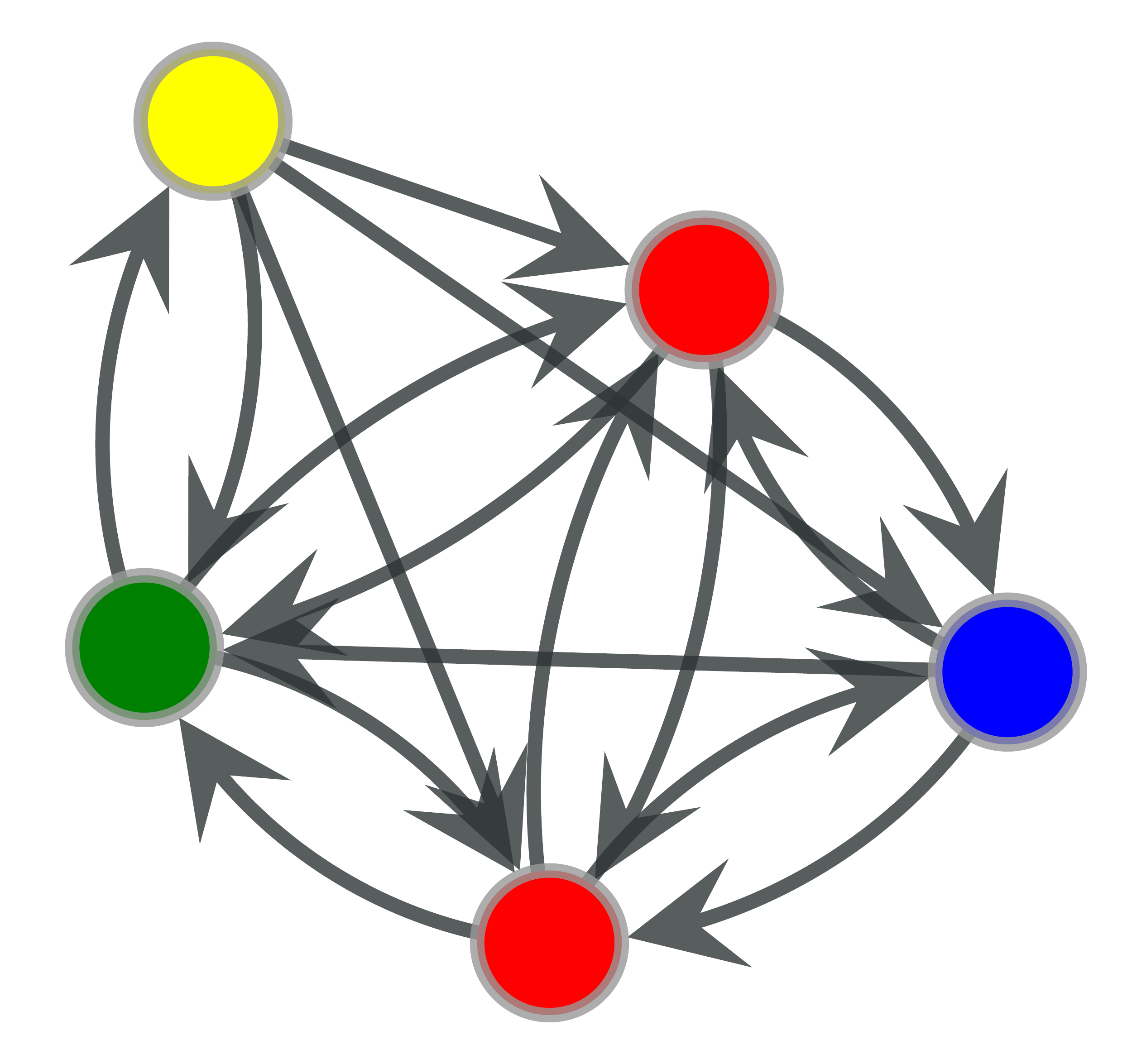}& \includegraphics[height=0.09\textwidth]{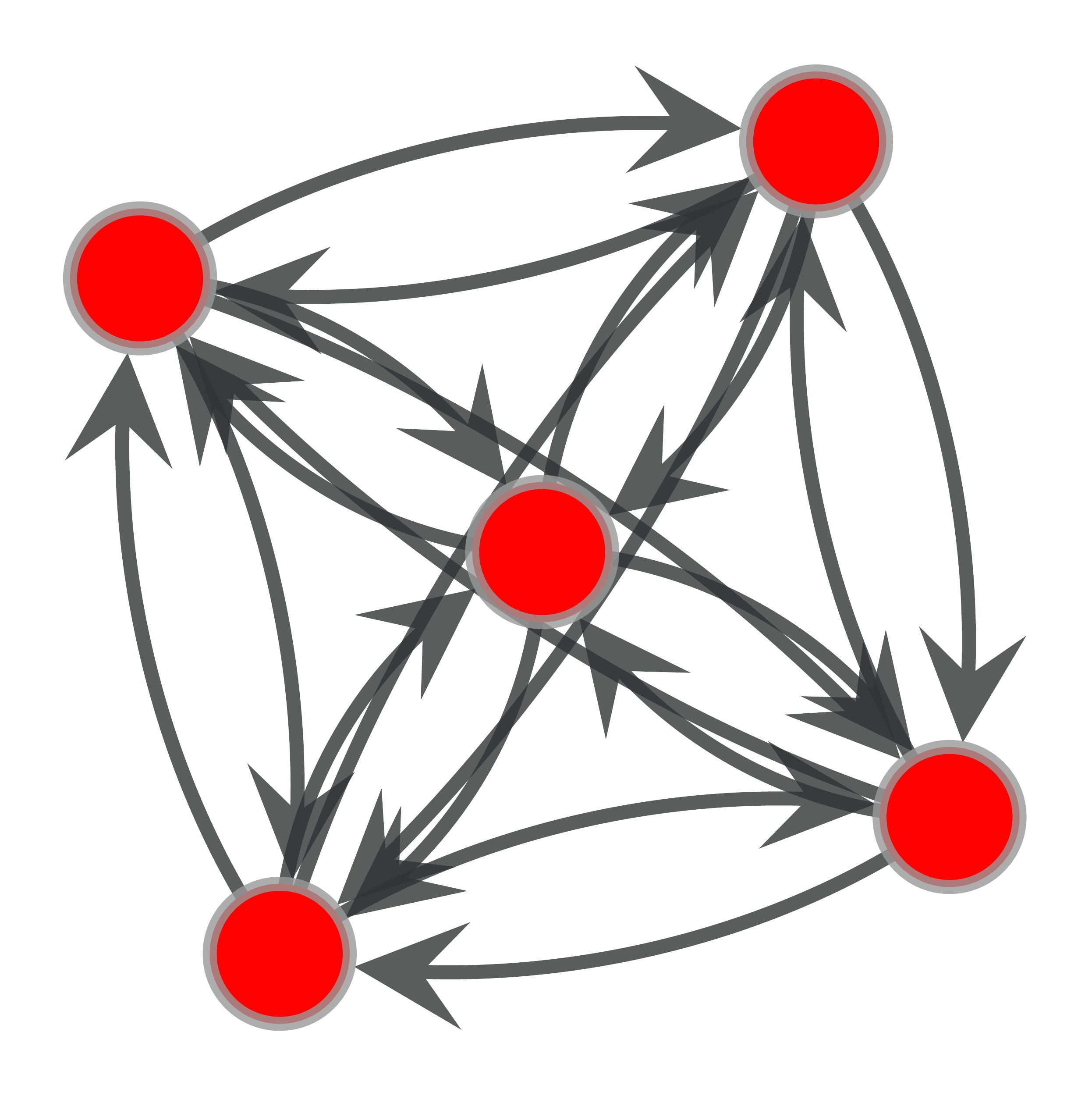} \\
\hline
Id&17&18&19&20&21&22&23&24\\
\hline
$n_m$&6& 23& 14& 12& 6& 10& 7& 2\\

\hline
\end{tabular}
\end{adjustbox}
\end{center}
\caption{Non trivial atoms found in the MAP configurations of the network of synaptic connections of the hermaphrodite {\em C. elegans} together with their respective counts ($n_m$). }
\label{CEH}

\end{table}

\section*{Discussion}\label{conc}
Despite the ubiquity and importance of network data, many problems are still outstanding in the modelling and analysis of networks, especially in terms of local or microscopic features. The quantitative analysis of local network structures poses theoretical and computational challenges, not the least due to the large variety of potential local structures and their highly interdependent nature. In this paper, we addressed some of these problems by introducing a nonparametric method that is based on the assumption that networks are made of not only edges but also higher order building blocks. When considering real data this assumption naturally leads to the questions whether the network under consideration can be effectively decomposed into higher order building blocks and what shape or types of these building blocks occurring in such an optimal decomposition are? We address this problem by introducing an objective function in the form of a prior that allows evaluation/comparison of decompositions of networks into small subgraphs on the basis of the statistical evidence provided by the data. By considering large sets of potential higher order interactions simultaneously in a holistic fashion the method is able to infer concise sets of patterns from within thousands of candidates that for many empirical networks include higher order interactions of known structural and functional significance. 

Due to its non-parametric nature the method does not require any prior assumptions on the types or frequencies of higher order interactions allowing these to be inferred from the data. Furthermore the method produces not merely a set of statistically significant patterns but an explicit decomposition of the network into higher order subgraphs that can be further used to examine the higher order structure of the network and its topological and dynamical implications. For instance the use of these representations in detection methods for higher order communities such as \cite{Benson2016Higher-orderNetworks.} could provide insights into the interplay between local and large scale network structures.  

The method also addresses the long standing problem of modelling the prevalence of triangles and other highly connected subgraphs observed in many real world networks by providing a fit of the network to analytically tractable models that better reflect its local structure than models based on conditionally independent dyadic interactions. Our empirical results show that inclusion of higher order interactions leads to more parsimonious representations for many real world networks providing empirical support for the inclusion of higher order interactions in models and representations of complex networks even when the  network is initially given in terms of dyadic interactions only. 

Our analysis still leaves some important questions open. For instance the generative models we use implicitly assume that pairs of atomic subgraphs very rarely intersect at more than one vertex. Consequently, models that incorporate more general intersections and coupling between higher order subgraphs and their use in inference based methods could lead to further refinement of our approach. Another possibility would be to incorporate community structures into generative models which could potentially lead to models and representations that can capture structures over multiple scales. 

Finally, while in this article we opted for a greedy heuristic to find an optimal representation, developing alternative inference algorithms is an area for interest for future research. Especially MCMC algorithms that could potentially provide a more refined view of the posterior distribution. Ideally, such algorithms would combine optimization with subgraph discovery which is especially relevant for large networks where exhaustively finding subgraphs poses computational challenges.

\section*{Methods}
\subsection*{Likelihood for degree-corrected SGCMs}
In the degree corrected microcanonical ensemble \cite{Wegner2021AtomicNetworks} the likelihood is given by: $\log(P(C|\mathbf{d}_{m,i}))=-\log(\Omega(\mathbf{d}_{m,i}))$ where $\Omega(\mathbf{d}_{m,i})$ is the number of all configurations with atomic degree sequence $\mathbf{d}_{m,i}$. In order to compute the likelihood we consider the model by Karrer and Newman \cite{Karrer2010RandomSubgraphs} which generalizes the edge configuration model to the case of general sets of atoms. In this model atomic stubs or partial subgraphs are attached to each vertex $v$ reflecting its orbit degree $d_{m,i}(v)$. These are then matched at random in appropriate combinations to generate a subgraph configuration with desired orbit degree sequence. The matching processes for different atoms are independent and the number of possible matchings for a given $m$ with atomic degree sequence $\mathbf{d}_{m,i}$ is: 
\begin{equation}\label{stubS}
\Omega_m(\mathbf{d}_{m,i})=\frac{\prod_i (|O_{m,i}|n_m)!}{(\prod_i |O_{m,i}|!)^{n_m}n_m!\prod_{i,v}d_{m,i}(v)!}\mu_m^{n_m},
\end{equation}
where $n_m$ is the number of $m$ subgraphs and $$\mu_m=\frac{\prod_i|O_{i,m}|!}{|Aut(m)|}$$ is the number of distinct  $m$-subgraphs that can be formed given the orbit memberships of its vertices. 

However, Eq.~\ref{stubS} does include cases where two or more stubs corresponding to the same vertex are matched together resulting in a subgraph where two or more vertices are contracted into a single vertex. Moreover the expression also includes cases where a given subgraph is created more than once. In order to discount for such cases one can consider the probability ($P_v(\mathbf{d}_{m,i})$) that no stubs attached to the same vertex are matched together and the probability $P_1(\mathbf{d}_{m,i})$ that no subgraph is created more than once during the generation process. Resulting in the following expression for $P(C|\mathbf{d}_{m,i},M)=\prod_m(\Omega_m(\mathbf{d}_{m,i})P_v(\mathbf{d}_{m,i})P_1(\mathbf{d}_{m,i}))^{-1}$:
\begin{equation}\label{Cent}
\begin{split}
\log(P(C|\mathbf{d}_{m,i},M))
&=-\sum_m\Bigg[-\log(n_m!)-n_m\log(|\mathrm{Aut(m)}|)\\&+\sum_i\big[\log((|O_{m,i}|n_m)!)-\sum_v \log(d_{m,i}(v)!)\big]\\
&-\frac{|\mathrm{Aut(m)}|n_m^2}{2}\prod_i\frac{1}{(n_m|O_{m,i}|)^{|O_{m,i}|}}\Big(\frac{\langle d_{m,i}^2\rangle}{\langle d_{m,i}\rangle}-1\Big)^{|O_{m,i}|}\\
&-\frac{1}{2}\Big[|m|(\frac{\langle (\sum_i d_{m,i})^2\rangle}{\langle \sum_i d_{m,i}\rangle}-1) -\sum_i \big(\frac{\langle d_{m,i}^2\rangle}{\langle d_{m,i}\rangle}-1\big)  \Big]\Bigg].\end{split}
\end{equation}

When the only atom is the single edge the above calculations are equivalent to counting number of graphs with a given degree sequence and the expressions above reduces to known results given in ~\cite{bianconi2009entropy,Bender1978AsymptoticFunctions} when $M$ only contains the single edge in both the undirected and directed case. Similar closed form expressions can be calculated for the other variants of the model following the same techniques~\cite{Wegner2021AtomicNetworks}. A brief derivation of these likelihoods is given in the SI.

\subsubsection*{Priors on atoms}
Maybe the most challenging aspect of finding an optimal subgraph configuration is to identify a suitable set of atoms. For this, one requires priors over subsets of atoms/motifs. Although ideally the set of potential atoms should be kept as general as possible, in practice one is faced with the fact that the number of motifs increases super-exponentially with size as well as the computational complexity of finding subgraphs. Therefore, in most practical settings one is forced to restrict the set of potential atoms included in the analysis. For such finite sets of candidate motifs $\mathcal{M}$ one could in principle use an uninformative prior that assigns the same probability to every non empty subset $M$ of $\mathcal{M}$ i.e.  $P(M)=(2^{|\mathcal{M}|}-1)^{-1}$. However, such a prior would assign each atom the same prior probability regardless of its size or structure whereas intuitively we would expect larger and more complex atoms to be less likely than simpler ones such as the single edge.  Therefore we consider a prior which assumes that motifs occur independently with probabilities $0<p_m<1$. Resulting in a prior of the form:
\begin{equation}\label{Mpr}
P(M)=\prod_{m\in M} p_m \times \prod_{m \not\in M} (1-p_m). 
\end{equation}
In order to assign the probabilities  to motifs we consider a prior that is inspired by Rissanen's universal prior for integers ~\cite{rissanen}. For this we initially assume that the set of candidate motifs $\mathcal{M}$ is infinite. To ensure that the prior is proper we require that $\sum_{m\in\mathcal{M}} p_m =\alpha<\infty$ for which the prior can be written as:
\begin{equation}\label{Mpz}
P(M)=\frac{1}{Z} \times \prod_{m\in M} \frac{p_m}{1-p_m},
\end{equation}
where $Z=\prod_{m\in\mathcal{M}} (1-p_m)^{-1}$ and $Z$ is well defined as a result of $\sum_{m\in\mathcal{M}}p_m =\alpha<\infty$. Following Rissanen's construction for the integers we want the prior to be maximally non--informative which in this translates to the condition that the entropy diverges to infinity. The entropy is given by:
\begin{eqnarray}
H(P_M)=\sum h(p_m),
\label{MP}
\end{eqnarray}
where $h(p_m) = -p_m \log(p_m) -(1-p_m)\log(1-p_m)$ is the binary entropy.  Making use of the expansion $h(p)=-p\log(\frac{p}{e})+O(p^2)$ for $p_m<<1$ the divergence condition becomes equivalent to $\sum_{m\in\mathcal{M}} -p_m\log(p_m)$ being divergent. Without loss of generality we assume that motifs in $\mathcal{M}$ are partially ordered according to $p_m$ (i.e. $m'\leq m \iff p_{m'}\leq p_m$). Consequently, putting aside the normalization this reduces to the definition of Rissanen \cite{rissanen} of a universal prior. The above equations have no unique solution but such priors can be shown to agree with respect to their leading order term. Consequently, one way of constructing a prior that satisfies the above conditions is to order motifs (i.e. map them on to integers) in a way that reflects prior expectations regarding the likelihoods of motifs and set $p_m$ to be equal to the probability of the integer index $n(m)$ of $m$ under a universal prior for the integers for instance $p_m=2^{ -\log_2^*(n(m))+c}$. Here $\log_2^*(\cdot)$ is the iterated logarithm which is given by the sum  $\log^*_2(x)=\log_2(x)+\log_2(\log_2(x))+...$ over positive terms and $c$ a normalizing constant. Approaches to ordering motifs are discussed in the SI.

\subsubsection*{Prior for count of atoms}
For the number $\mathbf{n}_m$ of atomic subgraphs  we consider the maximum entropy prior subject to the condition 
\begin{equation}\label{expE}
\big<\sum_m n_m e_m \big > =E,
\end{equation}
where $E$ is the number of edges in $G$ and $e_m$ is the number of edges of $m$. This results in a prior that has the following form:
\begin{equation}
P(\mathbf{n_m}|E,M)= Z^{-1} e^{-\lambda \sum_m n_m e_m }. 
\end{equation}
Where $Z$ is the normalizing constant:
\begin{eqnarray*}
Z=\sum_{\mathbf{n} \in \mathbb{Z}^{+|M|}} e^{-\lambda\sum_m n_m e_m }=\prod_m \frac{1}{e^{\lambda e_m}-1}.
\end{eqnarray*}
Imposing the constrains given by Eq.\ref{expE} we get:
\begin{eqnarray*}
E&=&-\frac{\partial \log(Z)}{\partial \lambda}\\
&=& \sum_m \frac{e_m}{1-e^{-\lambda e_m}},
\end{eqnarray*}
which can be solved numerically for $\lambda$. 

\subsubsection*{Priors for atomic degree sequences}
The number of components in the atomic degree sequence depends on the set of atoms and the type of model under consideration. Moreover, various components of the atomic degree sequence can in principle differ significantly with respect to their distribution. In order to ensure that the prior effectively captures potential regularities in different types of atomic degree sequences we consider two potential priors. Then given a component of an atomic degree sequence we evaluate both priors and choose the one which assigns the degree sequence under consideration the highest probability. 

We first discuss the priors for one dimensional degree sequences which form the basis for our prior for multi dimensional degree sequences. Given the sum over degrees $\sum_v d(v)=D$ of the degree sequence $d(v)$ on $N$ vertices one can simply assume that every degree sequence that is compatible with this condition is equally likely with probability:

\begin{equation}\label{dp1}
    P_1(d|D)=\multiset{N}{D}^{-1}, 
\end{equation}
where $\multiset{m}{n}={n+m-1 \choose m}$ is the number of $m$ combinations with repetitions from a set of size $n$. 

However, this prior favors Poisson type degree distributions in contrast to highly heterogeneous degree distributions found in many real world networks. Hence we also consider the hyper-prior introduced in Ref.\cite{peixoto2017nonparametric} that is compatible with more general degree distributions. In this prior the degree sequence is conditioned on a degree distribution $\mathbf{\eta}=\{n_k\}$ where $n_k$ is the number of vertices having degree $k$, i.e:

\begin{equation}\label{dp2}
P_2(d|D)=P(d|\mathbf{\eta})P(\mathbf{\eta}|D).
\end{equation}
Assuming uniform priors for both factors:
$$P(d|\eta)= \prod_{m,i}\frac{\prod_k n_k!}{N!},$$
and
$$P(\eta|D)=\prod_{m,i}q(D,N)^{-1}, $$
where $q(m,n)$ is the number of restricted partitions of the integer $m$ into at most $n$ non-zero parts. Although no exact closed form expression is known for $q(m,n)$ in practice it can be efficiently approximated to a high degree of accuracy, as described in \cite{peixoto2017nonparametric}.

\subsubsection*{Multi dimensional atomic degree sequences}
The simplest way of generating multi dimensional degree sequences is to assume that each component of the sequence is generated independently by a one dimensional prior $P(d_k|D_k)$ where $P$ is chosen to be whichever one of $P_1$ (Eq.\ref{dp1}) and $P_2$ (Eq.\ref{dp2}) assigns $d_k$ a higher probability:
$$P(\mathbf{d}_k|\mathbf{D}_k)=\prod_k max[P_1(d_k|D_k),P_2(d_k|D_k)]. $$

Here the range of $k$ depends on the model under consideration. For instance, if consider, the orbit degree model $k$ runs over the orbits of  the atoms of the model whereas if we consider the atomic degree model $k$ runs over the atoms of the model.

\subsection*{Inference algorithm}\label{algorithms}
Finding a subgraph configuration that maximizes the posterior is a set covering problem \cite{Caprara2000AlgorithmsProblem} where we seek to cover the edges of $G$ using subgraphs of $G$ while maximizing the posterior. Covering problems are known to be NP-complete and hence it is unlikely that the problem can be solved exactly in polynomial time. 
Therefore we consider a greedy heuristic that, starting from an empty configuration, at each step identifies the atom that is most effective in covering edges where the effectiveness of an atom is measured in terms of the description length per edge. The effectiveness of atom $m$ is determined by finding a set of $m$-subgraphs $C_m$ that minimizes:
\begin{equation}
\sigma_{m,t}=\frac{\Delta\Sigma_t(C_m)}{|(E(G)-E(C_t)) \cap E(C_m)|},
\end{equation}
where $E(C_t)$ is the set of edges covered by the configuration $C_t$ at iteration $t$, $E(C_m)$ is the set of edges contained in the subgraphs in $C_m$ and $\Delta\Sigma_t(C)$ is the change in $\Sigma$ when $C_m$ is added to the current state i.e. $\Sigma(C_t\cup C_m)-\Sigma(C_t)$.
Finding a set of $m$-subgraphs that minimizes $\sigma$ is in itself a non-trivial problem which we approximate using a heuristic that seeks to identify a set of non-intersecting $m$-subgraphs that has maximal cardinality and contains no edges  covered by $C_t$. For this, the algorithm iteratively finds $m$-subgraphs, on the set of edges not yet covered, where the search for such subgraphs first visits nodes having lower degrees in order to increase the number of non-intersecting subgraphs in $C_m$. At each step of the algorithm the $C_m$ that minimizes $\sigma_{m,t}$ is selected and added to the configuration until all edges are covered which in general occurs when the most effective atom is the single edge. Our implementation uses the LAD subgraph isomorphism algorithm \cite{solnon2010alldifferent}. A more detailed description of the inference algorithm is given in the SI.

\subsubsection*{Synthetic networks}
We tested the algorithm on synthetic networks generated from various types of configuration models covering a wide range of atoms and atomic degree distributions. We found that the method in general correctly recovers the set atoms in synthetic examples and that the inferred and true (ground truth) subgraph configurations agree to a high degree of accuracy with respect to the subgraphs they contain. In all cases the model selection procedure was able to correctly identify the model that was used to generate the data. We also tested the method on networks generated by the edge only configuration model and found that the method correctly identifies that these do not contain any atoms other than the single edge. Moreover, we found that if based on non-degree corrected models the method sometimes identifies spurious atomic subgraphs in networks with heavy tailed degree distributions generated using the edge configuration model highlighting the importance of using degree--corrected models. Details of results and procedures used to generate synthetic networks can be found in the SI.

Here we should also note that as with almost any type of inference procedure there exist cases where our method would fail to identify all atomic substructures either due to sub-optimal solutions found by the inference algorithm or due to detection limits for substructures similar to the detection limits for SBMs \cite{Peixoto2014HierarchicalNetworks}. In the context of subgraph configuration models detection limits are likely to occur when the models only contain a few copies of irregular/non-symmetric and sparse atoms within an otherwise relatively dense random graph.

\subsection*{Model selection using posterior odds ratio}
We compare alternative models using posterior odds ratio. Given that all model variants have common parameter spaces of the same type i.e. atoms, counts and degree distributions setting up all models with the same prior allows for an objective comparison between models. Given two models $M_1$ and $M_2$ and MAP-configurations corresponding to these models $C_{\Sigma,1}$ and $C_{\Sigma,2}$ we consider the following posterior odds ration:
\begin{eqnarray}\label{modelselect}
 \Lambda_{C_{\Sigma,1},C_{\Sigma,2}}&=&\frac{P(C_{\Sigma,1},M_1|G)}{P(C_{\Sigma,2},M_2|G)}\\
 &=&\frac{P(G|C_{\Sigma,1})P(C_{\Sigma,1}|\theta(C_{\Sigma,1}))P(\theta(C_{\Sigma,1}))P(M_1)}{P(G|C_{\Sigma,2})P(C_{\Sigma,2}|\theta(C_{\Sigma,2}))P(\theta(C_{\Sigma,2}))P(M_2)}\\\
 &=&\exp(\Sigma_2(C_{\Sigma,2})-\Sigma_1(C_{\Sigma,1})),
\end{eqnarray}
where both models are assumed to have equal prior probability ($P(M_1)=P(M_2)=1/2$) and $P(G|C_{\Sigma,1})=P(G|C_{\Sigma,2})=1$ as both $C_1$ and $C_2$ are covers of $G$. Hence, $\Lambda_{C_{\Sigma,1},C_{\Sigma,2}}$ gives us a measure as to which configuration is more likely under the assumption that both models have equal prior probability. Note that this differs from direct model selection where the goal is to find which class of generative models is more likely given the data which would require the evaluation of $ \Lambda_{M_1,M_2}=\frac{P(M_1|G)}{P(M_2|G)}$. In general this requires the evaluation a sum over all covers of $G$ for both models. However, in the context where the goal is to select between subgraph configurations and corresponding generative models $\Lambda_{C_{\Sigma,1},C_{\Sigma,2}}$ is the more informative quantity.

\bibliographystyle{ieeetr}
\bibliography{mendeley,thesis}
\subsection*{Acknowledgements}
This work was supported in by the European Research Council under Grant CoG
2015-682172NETS, within the Seventh European Union Framework Program. AW also acknowledges support from the UK Food Standards Agency.
\subsection*{Author contributions}
AW and SO  conceived the study and wrote the manuscript. AW implemented the method and performed the numerical analysis.  
\subsection*{Code and data availability}
All data sets are publicly available at cited sources. An implementation of the main method along with the complete inferred subgraph configurations and the code required to replicate the presented analysis is publicly available at: \hyperlink{https://github.com/AnatolWegner/HigherOrderMotifs}{https://github.com/AnatolWegner/HigherOrderMotifs}  
\subsection*{Additional information}

\paragraph{Supplementary information:} Supplementary information accompanies this paper. 
\paragraph{Competing financial interests:} The authors declare no competing financial interests. 
\end{document}